\documentclass[journal,10pt]{IEEEtran}
\usepackage{cite}
\usepackage{amsmath,amssymb,amsfonts}
\usepackage{algorithm, algorithmic}
\usepackage{graphicx}
\usepackage{subfigure}
\usepackage{textcomp}
\usepackage{multirow}
\usepackage{stfloats}
\usepackage{url}
\usepackage{color}
\usepackage{bm}
\usepackage{makecell}
\usepackage{flushend}
\usepackage{float}
\usepackage{threeparttable}
\usepackage{setspace}
\usepackage{upgreek}
\usepackage{bbm}
\usepackage{pifont}

\usepackage[colorlinks,
linkcolor=blue,
anchorcolor=blue,
citecolor=blue, 
urlcolor=black,
]{hyperref}

\newcommand{\nop}[1]{}
\allowdisplaybreaks[4]

\begin{document}
	
	\title {Fine-Grained AI Model Caching and Downloading With Coordinated Multipoint Broadcasting\\in Multi-Cell Edge Networks
	}
	
	\author{~Yang~Fu,~Peng~Qin,~\IEEEmembership{Member,~IEEE}, Yueyue Zhang, Pao Cheng, Jun Lu, and Yifei Wang
		
		\thanks{This work was supported in part by 62201212, 62271201, 2023YFB2904700, and MPIS202409. \textit{(Corresponding author: Peng Qin)}}
		\thanks{
			Yang Fu, Peng Qin, Yifei Wang are with the State Key Laboratory of Alternate Electrical Power System with Renewable Energy Sources, School of Electrical and Electronic Engineering, North China Electric Power University, Beijing 102206, China (e-mail: qinpeng@ncepu.edu.cn).}
		\thanks{
			Yueyue Zhang is with the China Satellite Network Group Shanghai Research Institute, Shanghai 200131, China.}
		
		}

	\markboth{IEEE Transactions on Wireless Communications}
	{}

	\maketitle

	\begin{abstract}	
    6G networks are envisioned to support on-demand AI model downloading to accommodate diverse inference requirements of end users. By proactively caching models at edge nodes, users can retrieve the requested models with low latency for on-device AI inference. However, the substantial size of contemporary AI models poses significant challenges for edge caching under limited storage capacity, as well as for the concurrent delivery of heterogeneous models over wireless channels. To address these challenges, we propose a fine-grained AI model caching and downloading system that exploits parameter reusability, stemming from the common practice of fine-tuning task-specific models from a shared pre-trained model with frozen parameters. This system selectively caches model parameter blocks (PBs) at edge nodes, eliminating redundant storage of reusable parameters across different cached models. Additionally, it incorporates coordinated multipoint (CoMP) broadcasting to simultaneously deliver reusable PBs to multiple users, thereby enhancing downlink spectrum utilization. Under this arrangement, we formulate a model downloading delay minimization problem to jointly optimize PB caching, migration (among edge nodes), and broadcasting beamforming. To tackle this intractable problem, we develop a distributed multi-agent learning framework that enables edge nodes to explicitly learn mutual influence among their actions, thereby facilitating cooperation. Furthermore, a data augmentation approach is proposed to adaptively generate synthetic training samples through a predictive model, boosting sample efficiency and accelerating policy learning. Both theoretical analysis and simulation experiments validate the superior convergence performance of the proposed learning framework. Moreover, experimental results demonstrate that our scheme significantly reduces model downloading delay compared to benchmark methods. 
	\end{abstract}


	\IEEEpeerreviewmaketitle
	
	\section{Introduction}
	
	\subsection{Background}
	
	\IEEEPARstart{T}{he} rapid proliferation of edge computing resources, coupled with transformative breakthroughs in artificial intelligence (AI), has given rise to edge AI \cite{1}. As a pivotal enabler of 6G networks, edge AI supports ubiquitous inference services through the distributed deployment of AI models across edge infrastructure and user devices \cite{2,3}. For inference tasks characterized by high data sensitivity and stringent privacy requirements, such as mobile health and virtual assistants, on-device AI inference is often mandatory to preserve all data at the user side \cite{4}. However, the substantial storage demands of contemporary AI models (e.g., Apple’s on-device large language model (LLM) OpenELM, which comprises 3 billion parameters and requires 12 GB of storage) render it impractical for user devices with limited capacity to store all necessary models locally \cite{5}. A viable solution is to adaptively download AI models from the network\footnote{Accomplishing AI inference through task offloading to edge/cloud servers and downloading models to local devices represent two complementary rather than substitutable paradigms. This work focuses on the latter, motivated by its advantages in stable low-latency inference without repeated network dependency, enhanced data security, improved location and context awareness, and stronger sustainability for continuous tasks.}, accommodating real-time and diverse inference needs while preventing excessive local storage consumption. 
	
	Indeed, on-demand model downloading from an AI repository (which comprises a large collection of trained models and is typically deployed in cloud centers) has been identified as a key use case in the standardization of 6G \cite{6,7}. According to 3GPP technical specifications, such downloading operations should be completed within seconds for general inference tasks, and within 10-100 ms for time-sensitive applications including humanoid robot control and autonomous driving \cite{7,8}. To satisfy these stringent delay requirements, caching AI models at edge nodes near users is essential for eliminating the unpredictable delays inherent in cloud access \cite{9}. Meanwhile, advanced wireless transmission techniques (particularly broadcasting and coordinated multipoint (CoMP) considered in this paper) should be exploited to boost downlink throughput, thereby accelerating model delivery from edge nodes to end users. Building upon these two directions, research efforts have been dedicated to facilitating AI model caching and downloading in edge networks. 
	
	\subsection{Related Works}
	
    \textit{1) Works on Edge AI Model Caching:} Given the limited storage capacity of edge nodes, only a finite number of AI models can be cached from the AI repository. Therefore, existing works primarily focus on developing optimal model selection strategies at edge nodes to align with user requests. Literature \cite{10} investigated a single edge node scenario, where a subset of AI models was cached by the edge node to execute inference tasks. The authors proposed a deep reinforcement learning (DRL)-based approach to train the caching policy while introducing penalty mechanism to avoid frequent model change. Paper \cite{11} jointly optimized the number and type of AI models to be cached at an edge node, so as to provide scalable inference service. A rounding and relocation method was developed to handle the integer constraints of caching variables, thereby minimizing the holistic cost and accuracy loss. The authors of \cite{12,13} elaborated the cooperative caching in multi-cell scenarios, in which AI models could be migrated among edge nodes to enhance the cache hit ratio. Multi-agent DRL (MADRL) was adopted to make distributed caching decisions, while promoting model routing via backhaul links. In \cite{Fan2025Satellite}, different compressed versions of an AI model were deployed at cloud-edge-device nodes according to their computational capacities. The authors developed a DRL algorithm integrated with numerical optimization to derive the task offloading and resource allocation solution, thereby balancing inference delay, energy consumption, and accuracy loss. However, the aforementioned studies follow conventional principles of content caching, which directly stores entire AI models at edge nodes. Given that AI models typically require significantly greater storage space than traditional cached content, such coarse-grained caching methods are inefficient in utilizing the limited edge storage capacity.  
    
    Some works proposed storage-efficient caching strategies by exploiting the architectural properties of AI models, particularly deep neural networks. Layer splitting was considered in \cite{14} to determine the number of model layers to be cached at each edge node, then nodes storing different layers sequentially execute inference tasks through intermediate result exchange. Paper \cite{15} employed semantic splitting to partition the AI model into parallel disjoint fragments. Online learning was adopted to cache model fragments at different edge nodes, thereby reducing energy consumption and response time. Nevertheless, the methods proposed in \cite{14,15} are limited to caching a single AI model with a specific architecture. They cannot scale to the concurrent deployment of a large number of heterogeneous models from the AI repository, failing to meet diverse model downloading demands. To fill this gap, our work exploits not only the internal structure of individual AI models to partition them into fine-grained parameter sets, but also the interplay among different models by identifying their overlapping parameters for scalable edge caching. 

    \textit{2) Works on Broadcasting and CoMP:} By leveraging the broadcast nature of wireless channels, edge node can simultaneously deliver identical data to multiple users, thereby improving spectral efficiency. To mitigate inter-cell interference, multiple nodes can further form a CoMP cluster for joint data transmission, which effectively enhances the downlink throughput \cite{16}. Reference \cite{17} investigated the hybrid beamforming in a CoMP broadcasting scenario, in which the optimal precoder was derived by semi-definite program, then a maximum ratio combiner was employed to maximize the received signal power. In \cite{18}, a two-stage optimization approach was designed to maximize the quality of service (QoS) of data broadcasting, in which cells were first clustered to reduce inter-cell interference, then the service order was scheduled to decrease the total data reception delay. The authors of \cite{19} considered an air-ground communication system where multiple drones broadcast data to user groups using CoMP. To perform system optimization, QMIX was invoked to output the drone flight direction, then the CoMP beamforming was optimized using majorization minimization algorithm. Literature \cite{20} elaborated a multi-cell integrated sensing and communication system with CoMP, where the inter-cell reflections could be utilized to enhance the target estimation performance owing to the sharing of edge nodes’ data. The authors proposed robust CoMP beamforming approach taking into account imperfect channel state information (CSI). In the context of edge AI, broadcasting has been employed to distribute inference tokens across devices, enabling the parallel execution of mixture-of-experts foundation models \cite{Xie2025Mixture}. In \cite{Li2025JointCommunication}, user-offloaded tasks are jointly processed by multiple edge nodes that form a CoMP cluster for data exchange and result delivery, where the success probability is maximized to enhance both communication and computation efficiency. Although existing studies have demonstrated the effectiveness of broadcasting and CoMP techniques for enhancing data rates, their seamless integration with AI model downloading remains an open issue. Unlike traditional multimedia content, AI models are often trained for various downstream tasks, leading to low information overlap among user requests. Furthermore, the above works assume that all user data is readily available at edge nodes to enable CoMP, while overlooking the data exchange overhead. For AI models with large data volumes, excessive migration among edge nodes may incur severe downloading delays and backhaul congestion.
    
    \subsection{Motivations and Contributions}
    
    To address the limitations of previous studies, we propose a novel fine-grained AI model caching and downloading (FGAMCD) system that exploits parameter reusability among AI models. This property arises from the common practice of fine-tuning task-specific models from a shared pre-trained model, inherently maintaining a set of reusable parameters across different AI models \cite{21,22}. For instance, in convolutional neural networks (CNNs), the shallow layers responsible for extracting common visual features typically exhibit high parameter reusability across different tasks. This effect is even more pronounced in emerging LLMs, where parameter-efficient fine-tuning (PEFT) techniques (e.g., low-rank adaptation) typically freeze over 99\% of pre-trained parameters, resulting in a significant proportion of reused parameters \cite{23}. 
    
    In the proposed FGAMCD system, edge nodes selectively cache parameter blocks (PBs), which are fine-grained components of AI models, such as neural network layers or Transformer blocks. When multiple AI models are cached, the edge node only needs to store a single copy of their reusable PBs, thereby enhancing storage utilization. Additionally, it integrates broadcasting to simultaneously deliver PBs to multiple users, leveraging potential PB reuse across different requested models. To further reduce the downloading delay, FGAMCD optimizes PB migration within the backhaul network, facilitating the CoMP transmission of edge nodes while balancing coordinated gains and data exchange overhead. 
    
    Parameter reusability has also been explored in prior works to support AI model caching and delivery. The TrimCaching framework proposed in \cite{24} optimizes edge caching of AI models by considering parameter sharing among models, thereby improving cache hit ratio. However, this approach orthogonally allocates downlink spectrum to each user and relies on traditional unicasting for model delivery, leaving room for downloading delay reduction. Moreover, the proposed heuristic caching method incurs high computational complexity, limiting its adaptability to dynamic user requests and communication conditions. Work in \cite{25} developed a model broadcasting protocol to serve multiple users performing heterogeneous inference tasks, along with the joint optimization of parameter selection and power control. Although this protocol incorporates broadcasting of reused PBs, the proposed approach cannot be directly extended to multi-cell and multi-antenna communication systems, which require dedicated PB migration and beamforming design. To the best of our knowledge, this is the first work to jointly optimize AI model caching, migration, and delivery in multi-cell edge networks, through a backhaul-wireless co-design for communication-efficient model downloading. In particular, the combination of caching-aware PB migration with CoMP broadcasting unleashes the full potential of parameter reusability, offering simultaneous improvements in storage and spectrum efficiency, which remain only partially addressed in \cite{24,25}. The main contributions are summarized below. 
    
    \begin{enumerate}
		\item We propose an FGAMCD system to fulfill diverse downloading requests by strategically caching PBs from an AI model repository across multiple edge nodes, enabling efficient model retrieval for end users. Subsequently, the edge nodes cooperatively deliver the cached PBs to users through CoMP broadcasting, while PBs are migrated within backhaul links to facilitate coordinated transmission and enhance downlink rates. After receiving all the required PBs, each user reconstructs the complete model to perform on-device AI inference. With the objective to minimize the total model downloading delay, we formulate a joint optimization problem of PB caching, migration, and broadcasting beamforming, subject to user QoS requirements, edge storage capacity and transmission power constraints. 
		
		\item The formulated problem is an intractable mixed-integer nonlinear programming (MINLP) problem, exacerbated by the substantial computational complexity of centralized control across multiple edge nodes. To solve this problem, we develop a distributed MADRL framework, named multi-agent action semantics network with data augmentation (MAASN-DA), which learns PB caching and migration policies by explicitly characterizing the influence of each edge node’s action on others through a specialized actor network. In addition, MAASN-DA adaptively generates synthetic training samples through a predictive model to enrich the experience replay buffer, thereby accelerating policy learning. A robust optimization subroutine for deriving CoMP broadcasting beamformers under imperfect CSI is proposed and incorporated into the reward calculation, thus completing the MAASN-DA training loop. 
		
		\item We provide a theoretical convergence analysis for MAASN-DA, deriving a closed-form upper bound on the Q-value approximation error to offer theoretical guidelines for learning hyperparameter configuration. Furthermore, extensive experiments utilizing real-world dataset and AI models are conducted. The ablation study demonstrates that the proposed MAASN-DA achieves higher cumulative reward and faster convergence speed than existing MADRL methods. Moreover, experimental results reveal a two-fold performance gain achieved by the FGAMCD system: \textit{i) caching efficiency gain} by eliminating redundant storage of reusable PBs, and \textit{ii) downloading efficiency gain} through coordinated PB broadcasting that simultaneously serves multiple user requests. Compared with conventional coarse-grained caching and unicasting-based model delivery, FGAMCD reduces model downloading delay by 29.74\% to 67.86\%. 
	\end{enumerate}
    
    \textit{Notations:} Superscripts $\text{H}$ and $\text{T}$ denote the conjugate transpose and transpose, respectively. $\|\cdot\|$ indicates the norm of a vector or spectral norm of a matrix. ${{\mathbb{C}}^{M\times N}}$ specifies the space of $M\times N$ complex matrices, and ${\mathbb{R}}$ is the set of real numbers. $\{x\}^{+}$ is equal to $\max\{x,0\}$. $\mathbf{X}\succeq 0$ means that $\mathbf{X}$ is a semi-definite matrix. $\mathbf{O}$ denotes the all-zero matrix. 
    
	\section{System Model and Problem Formulation} \label{sec:model}
    
    \begin{figure}[t]
    	\begin{center}
    		\centerline{\includegraphics[width=8.5cm]{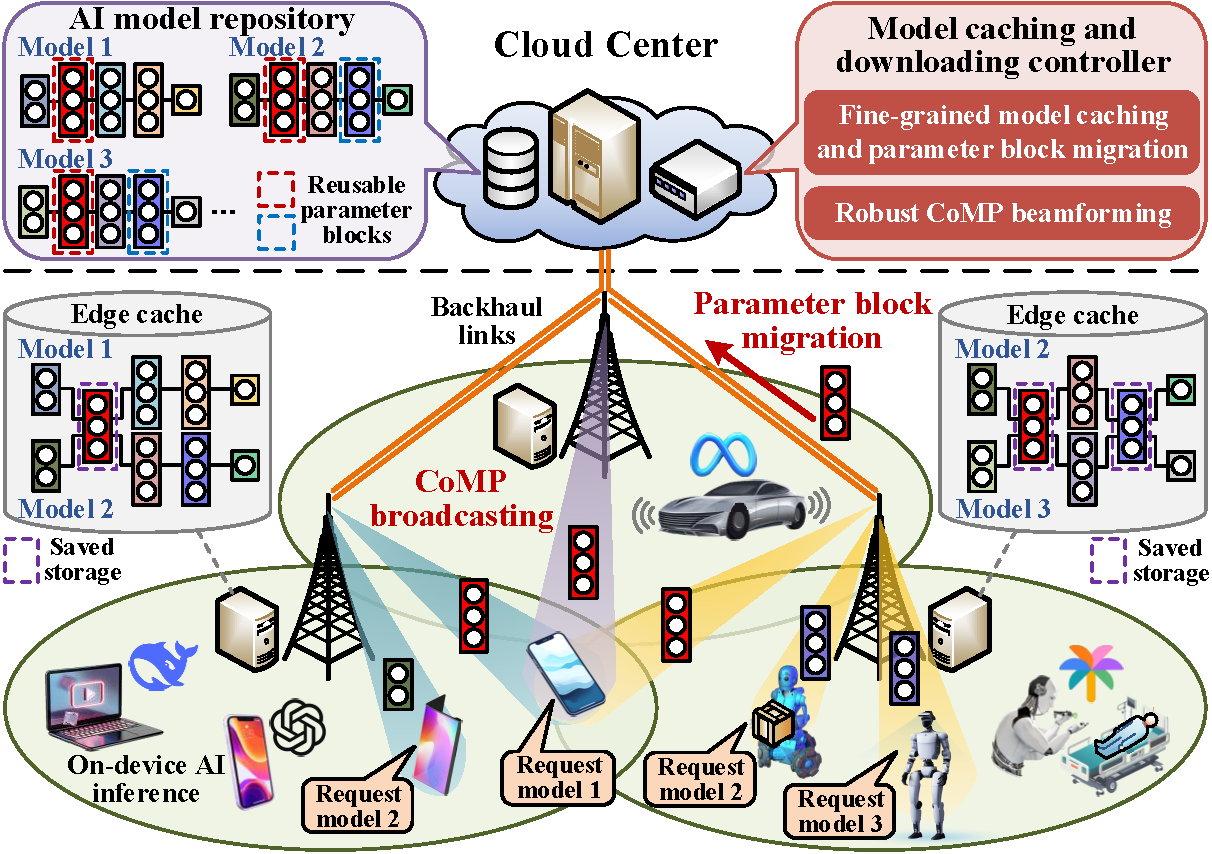}}
    	\end{center}
    	\vspace{-4mm}		
    	\caption{Fine-grained AI model caching and downloading system model.} 
    	\label{fig:1}
    \end{figure} 

	We consider a multi-cell network architecture as illustrated in Fig. \ref{fig:1}, designed to cache popular AI models\footnote{To mitigate the risk of sensitive information leakage or adversarial attacks arising from model distribution across edge nodes, the proposed FGAMCD framework can be integrated with ownership protection mechanisms such as model watermarking \cite{Wang2025ModelOwnership}. In this case, each edge-cached model is embedded with a watermark (e.g., a binary vector derived from model parameters), enabling users to verify the authenticity of the downloaded model.} across $N$ distributed edge nodes, which then provide model downloading service to $U$ end users. The sets of edge nodes and users are signified by $\mathcal{N}=\left\{ 1,\ldots ,n,\ldots ,N \right\}$ and $\mathcal{U}=\left\{ 1,\ldots ,u,\ldots ,U \right\}$, respectively. The cloud center maintains a comprehensive AI model repository $\mathcal{J}=\left\{ 1,\ldots ,j,\ldots ,J \right\}$ comprising $J$ models, which can be downloaded by users to facilitate on-device AI inference. All edge nodes are interconnected with the cloud server via backhaul links. Main notations to be used are summarized in Table \ref{tab:symbol}. The system operation encompasses the following phases:  
    
    \begin{table}[t]\footnotesize
    	\centering
    	\setlength{\abovecaptionskip}{0pt}    
    	\setlength{\belowcaptionskip}{10pt}
    	\caption{Summary of Main Notations} \label{tab:symbol}
    		\begin{threeparttable}
    			\begin{tabular}{p{1.5cm}<{\centering} p{6.5cm}}
    				\hline
    				\textbf{Notation}&\textbf{Description}\\
    				\hline
    				$N$, $\mathcal{N}$ & Number and set of edge nodes\\
    				$U$, $\mathcal{U}$ & Number and set of users\\
    				$J$, $\mathcal{J}$ & Number and set of AI models\\
    				$K$, $\mathcal{K}$ & Number and set of PBs\\
    				$r_u$& Target AI model of user $u$\\
    				${{\mathcal{K}}_{j}}$& Set of PBs that constitute AI model $j$\\
    				${{a}_{n}}\left( k \right)$& Binary variable indicating whether node $n$ caches PB $k$\\
    				$S\left( k \right)$&Data size of PB $k$\\
    				${{C}_{n}}$&Storage capacity of node $n$\\
    				${{b}_{n,m}}\left( k \right)$& Binary variable indicating whether to migrate PB $k$ from edge node $n$ to $m$\\
    				${{\lambda }_{n}}\left( k \right)$& Binary indicator representing whether edge node $n$ participates in the delivery of PB $k$\\
    				$M$& Number of antennas at each edge node\\
    				${{\mathbf{{h}}}_{n,u}}(k)$& Actual channel between node $n$ and user $u$ for PB $k$\\
    				${{\mathbf{\tilde{h}}}_{n,u}}(k)$&Estimated channel between node $n$ and user $u$ for PB $k$\\
    				${{\mathbf{w}}_{n}}\left( k \right)$&Beamforming at edge node $n$ for broadcasting PB $k$\\
    				${{R}_{u}}\left( k \right)$, $Q_u$&Downloading rate and QoS requirement for user $u$\\
    				$R_{n,m}^{\text{bac}}\left( k \right)$&Backhaul link rate between edge node $n$ and $m$\\
    				$T(k)$, $T$ & Downloading delay for PB $k$ and total downloading delay\\
    				${{\mathbf{o}}_{n}}\!\left( k \right)$, ${{\mathbf{s}}}\left( k \right)$&Observation of agent $n$ and global state in step $k$\\
    				${{\mathbf{d}}_{n}}\!\left( k \right)$, ${{r}\!}\left( k \right)$&Action of agent $n$ and reward in step $k$\\
    				${{\bm{\eta} }_{\text{in}}}$, $\!{{\bm{\eta} }_{\text{re}}}$, $\!{{\bm{\eta} }_{\text{out}}}$&Weighting parameters of ESN\\
    				${{\mathbf{q}}}\left( k \right)$&Reservoir state of ESN in step $k$\\
    				$\xi$, $\tau_0$&Selection threshold and proportion for synthetic samples\\ 
    				${{{\bm{\varphi} }}_{n}}$&Parameters of agent $n$'s action semantics actor network\\
    				${{{\bm{\theta} }}_{n}},{{{\bm{\theta} }}^{\text{mix}}}$&Parameters of agent $n$'s critic and mixing network\\
    				$\gamma$, $E$ & Discount factor and number of training episodes\\
    				\hline
    			\end{tabular}
    	\end{threeparttable}
    \end{table}

	\begin{itemize}
		\item \textbf{User Model Request:} Each user $u$ submits a model request to its associated edge nodes, specifying both the target AI model ${{r}_{u}}\in \mathcal{J}$ from the repository and the corresponding QoS requirement ${{Q}_{u}}$ for model downloading. 
		\item \textbf{Fine-Grained AI Model Caching:} Edge nodes strategically cache a selected set of AI models from the cloud center, enabling collaborative fulfillment of user requests. Particularly, the proposed system exploits the parameter shareability, where distinct AI models may contain reusable PBs due to the prevalent adoption of parameter-efficient fine-tuning techniques. Accordingly, a fine-grained model caching mechanism is implemented to cache specific PBs instead of entire AI models at each edge node. This design ensures that the reusable PB is stored only once in the edge cache, thereby significantly enhancing storage efficiency. More details will be elaborated in Section II-A. 
		\item \textbf{Model Downloading With CoMP Broadcasting:} During this phase, users retrieve the requested AI models from edge nodes through a coordinated transmission process. Specifically, edge nodes utilize backhaul links to exchange their cached PBs, promoting the execution of CoMP techniques. Subsequently, they cooperatively broadcast PBs to the requesting users, exploiting potential PB reuse across different requested models to improve downloading efficiency. We will detail the model downloading phase in Section II-B to D. 
		\item \textbf{On-Device AI inference:} End users reconstruct AI models from the received PBs to perform various inference tasks, e.g., text generation for virtual assistance applications and object recognition for humanoid robot control\footnote{After downloading, model reconstruction simply loads the PBs into their designated positions within the model architecture, without requiring retraining or fine-tuning. The associated overhead is negligible compared to PB downloading \cite{Geng2025Layer}. Moreover, this process does not affect inference accuracy, as it restores the original model exactly without modifying its architecture or parameter values.}. This local inference fashion inherently guarantees data privacy and security by eliminating the need for sensitive data/feature uploading. 
	\end{itemize}

	\subsection{Fine-Grained AI Model Caching}
	
	For each AI model $j$ in the repository, we define ${{\mathcal{K}}_{j}}$ as the set of PBs that constitute the model. $\mathcal{K}=\bigcup\nolimits_{j\in \mathcal{J}}{{{\mathcal{K}}_{j}}}=\left\{ 1,\ldots ,k,\ldots ,K \right\}$ is the collection of all PBs\footnote{The integration of FGAMCD with model compression is straightforward. Diverse sub-models can be derived from an original model through techniques such as quantization, pruning, and knowledge distillation \cite{6}. The PBs of these sub-models can be also incorporated into $\mathcal{K}$ without affecting our subsequent modeling and solution, thereby allowing users to flexibly select appropriate model versions to improve downloading efficiency.}, and we have $\left| \mathcal{K} \right|\le \sum\nolimits_{j\in \mathcal{J}}{\left| {{\mathcal{K}}_{j}} \right|}$ due to the parameter shareability. The fine-grained caching decision is represented by binary variable ${{a}_{n}}\left( k \right)\in \left\{ 0,1 \right\}$, where ${{a}_{n}}\left( k \right)=1$ if PB $k$ is cached by edge node $n$, otherwise ${{a}_{n}}\left( k \right)=0$. Therefore, $\prod\nolimits_{k\in {{\mathcal{K}}_{j}}}{{{a}_{n}}}\left( k \right)=1$ implies that model $j$ is completely cached by node $n$. Consider that the total cached PBs cannot exceed the storage capacity of each edge node, we have constraint\footnote{In practice, each edge node maintains an index table of PB identifiers, sizes, and locations to support lookup according to user requests. Since such information requires only minimal storage and thus negligible compared to cached PBs themselves, the indexing overhead is omitted \cite{10,11}. }
	\begin{align}
		\sum\limits_{k\in \mathcal{K}}{{{a}_{n}}\left( k \right)S}\left( k \right)\le {{C}_{n}},\ \forall n, \label{eq:1}
	\end{align}
	where $S\left( k \right)$ is the size of PB $k$, ${{C}_{n}}$ signifies the storage capacity of node $n$. 
	
	\begin{figure}[t]
		\begin{center}
			\centerline{\includegraphics[width=8.5cm]{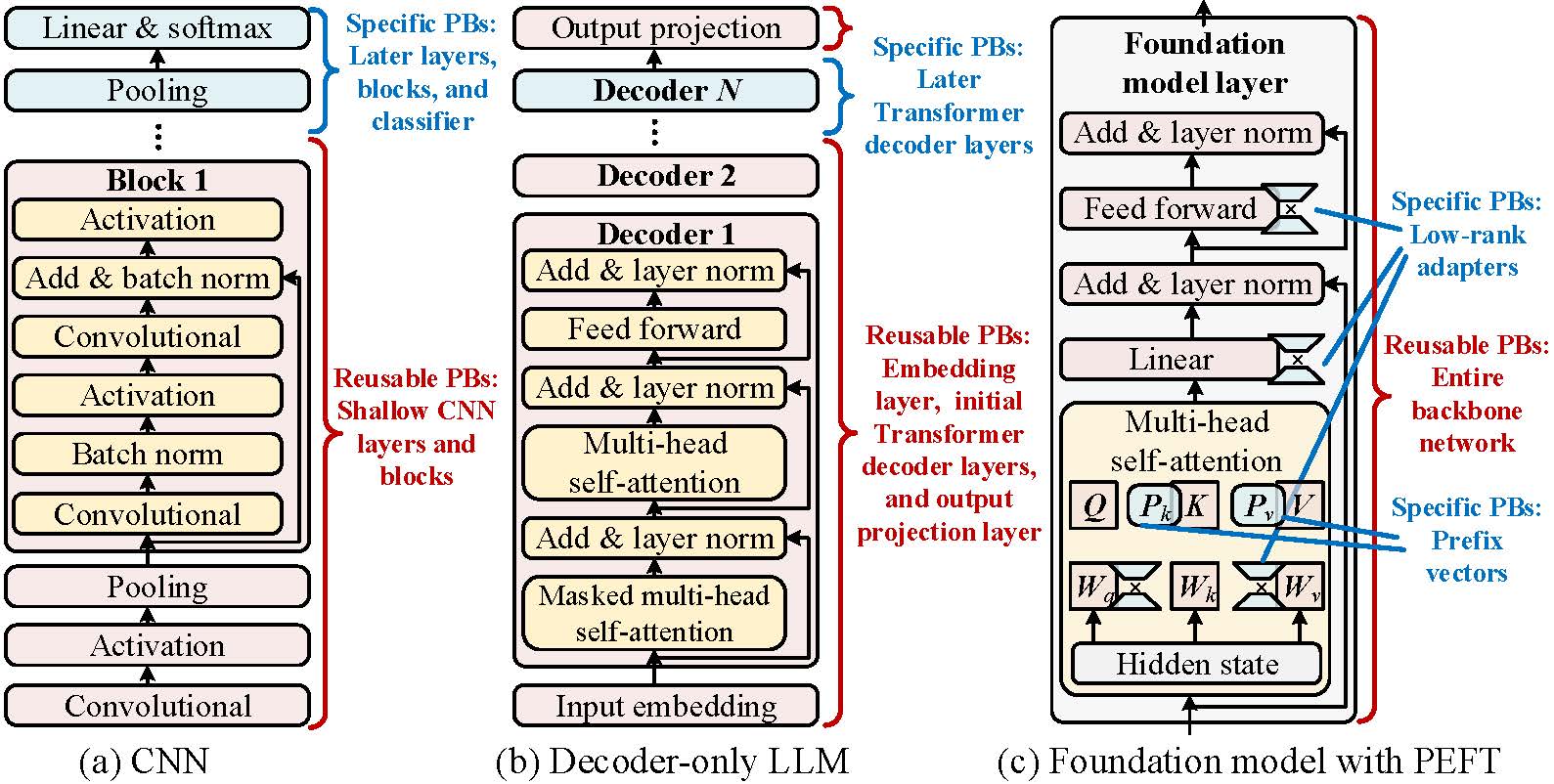}}
		\end{center}
		\vspace{-4mm}		
		\caption{Identification of PBs across AI model architectures.}
		\label{fig:PB_definition}
	\end{figure}  
	
	\textit{Remark 1 (Identification of PBs Across AI Model Architectures):} The definition of a PB depends on the underlying model architecture and parameter reuse conditions. As illustrated in Fig. \ref{fig:PB_definition}, in CNNs, a PB can correspond to a single layer (e.g., convolution, activation, or linear) or a block containing residual connections or inception modules. Reusable PBs are often shallow layers or blocks that capture generic visual features. For decoder-only LLMs, a PB may refer to the input embedding, a Transformer decoder layer, or the output projection (which typically shares parameters with the embedding). In this case, reusable PBs are usually the embedding/output layers and the initial decoder layers that extract common linguistic knowledge. In foundation models with PEFT, the backbone network itself can serve as a reusable PB across different downstream tasks, while the inserted fine-tuning parameters (e.g., low-rank adapters or prefix vectors) are treated as task-specific PBs. For models without clear parameter sharing, we directly treat them as individual PBs. 
	
	\subsection{Parameter Block Migration}
	
	Prior to model delivery over the wireless edge network, edge nodes can exchange the cached PBs via backhaul links. This exchange enables multiple nodes to cooperatively transmit PBs to users, thereby enhancing the CoMP performance at the expense of increased backhaul delay. To achieve an optimal tradeoff, we introduce PB migration variable ${{b}_{n,m}}\left( k \right)\in \left\{ 0,1 \right\}$, where ${{b}_{n,m}}\left( k \right)=1$ indicates that PB $k$ cached by edge node $n$ is migrated to node $m\in \mathcal{N}\backslash \left\{ n \right\}$, otherwise ${{b}_{n,m}}\left( k \right)=0$. Obviously, a PB can be migrated from edge node $n$ to other nodes only when it is cached by node $n$, thus we impose the following restraint: 
	\begin{align}
		{{a}_{n}}\left( k \right)\ge \underset{m\in \mathcal{N}\backslash \left\{ n \right\}}{\mathop{\max }}\,\left\{ {{b}_{n,m}}\left( k \right) \right\},\ \forall k,n. \label{eq:2}
	\end{align}
    
    As a result, edge node $n$ can participate in the delivery of PB $k$ if at least one condition is met: i) PB $k$ is already in node $n$’s cache; ii) PB $k$ is migrated to node $n$ from another node. Formally, we define binary indicator ${{\lambda }_{n}}\left( k \right)\in \left\{ 0,1 \right\}$ with ${{\lambda }_{n}}\left( k \right)=1$ implying node $n$’s participation in the delivery of PB $k$, then ${{\lambda }_{n}}\left( k \right)$ is calculated by
    \begin{align}
       {{\lambda }_{n}}\left( k \right)=\min \bigg\{ {{a}_{n}}\left( k \right)+\sum\limits_{m\in \mathcal{N}\backslash \left\{ n \right\}}{{{b}_{m,n}}\left( k \right)},1 \bigg\}. \label{eq:3}
    \end{align}
    Afterwards, ${{\lambda }_{n}}\left( k \right)$ will be leveraged to formulate the received signal for users during PB broadcasting. 
    
    \subsection{Parameter Block Broadcasting}
    
    Each edge node equips with $M$ antennas to deliver PBs to single-antenna users. Denote ${{\mathbf{h}}_{n,u}}\left( k \right)=\sqrt{\upsilon d_{n,u}^{-\alpha }\left( k \right)}{{\mathbf{\bar{h}}}_{n,u}}\left( k \right)$ as the wireless channel from edge node $n$ to user $u$ when broadcasting PB $k$, $\upsilon$ is the channel gain at reference distance of 1 m, ${{d}_{n,u}}\left( k \right)$ represents the transmission distance, $\alpha $ indicates the pathloss exponent, and ${{\mathbf{\bar{h}}}_{n,u}}\left( k \right)$ is the small-scale fading component. In practice, edge nodes face challenges in acquiring perfect CSI. Hence, we model the relationship between actual channel ${{\mathbf{h}}_{n,u}}\left( k \right)$ and estimated channel ${{\mathbf{\tilde{h}}}_{n,u}}\left( k \right)$ considering CSI uncertainty as follows
    \begin{align}
    	&\!\!{{\mathbf{h}}_{n,u}}\left( k \right)={{\mathbf{\tilde{h}}}_{n,u}}\left( k \right)+{{\mathbf{e}}_{n,u}}\left( k \right),\nonumber\\
    	&\qquad\qquad\qquad\quad {{\mathbf{e}}_{n,u}}\left( k \right)\in{\mathcal{E}}_{n,u}=\left\{ \mathbf{e}|{{\mathbf{e}}^{\text{H}}}{{\mathbf{C}}_{n,u}}\mathbf{e}\le\! 1 \right\}, \label{eq:4}
    \end{align}
    where ${{\mathbf{\tilde{h}}}_{n,u}}\left( k \right)\in {{\mathbb{C}}^{M\times 1}}$ denotes the estimated channel vector, which can be acquired either through pilot signal reception from the user or generated by a diffusion model \cite{Gong2025Digital}. This novel approach in \cite{Gong2025Digital} exploits the user location as conditional input and infers the statistical CSI from random noise via a reverse denoising process. ${{\mathbf{e}}_{n,u}}\left( k \right)$ represents the channel estimation error, which is restricted within a spherical set with ${{\mathbf{C}}_{n,u}}$ determining the shape and size of the spherical set. 
    
    PBs are sequentially broadcasted from edge nodes to users through CoMP transmission. Consider the broadcasting of an arbitrary PB, say PB $k$, the transmitted symbol of edge node $n$ is denoted by ${{\lambda }_{n}}\left( k \right)x\left( k \right)$ with $\mathbb{E}\{ {{\left| x\left( k \right) \right|}^{2}} \}=1$. Then, the signal received at user $u$ can be written as
    \begin{align}
    	{{y}_{u}}\left( k \right)=\sum\limits_{n\in \mathcal{N}}{\mathbf{h}_{n,u}^{\text{H}}(k){{\mathbf{w}}_{n}}\left( k \right){{\lambda }_{n}}\left( k \right)x\left( k \right)}+{{z}_{u}}\left( k \right), \label{eq:5}
    \end{align}
    where ${{\mathbf{w}}_{n}}\left( k \right)\in {{\mathbb{C}}^{M\times 1}}$ signifies the transmission beamforming at edge node $n$, ${{z}_{u}}\left( k \right)$ is the channel noise, which obeys a complex Gaussian distribution with zero mean and variance $\sigma _{u}^{2}$. Consequently, the downloading rate for user $u$ during the broadcasting of PB $k$ is given by
	\begin{align}
		\!\!\!{{R}_{u}}\left( k \right)\!=\!B{{\log }_{2}}\left( \!1\!+\!\frac{{{\left| \sum\nolimits_{n\in \mathcal{N}}{{{\lambda }_{n}}\left( k \right)\mathbf{h}_{n,u}^{\text{H}}\left( k \right){{\mathbf{w}}_{n}}\left( k \right)} \right|}^{2}}}{\sigma _{u}^{2}} \right), \label{eq:6}
	\end{align}
	where $B$ is the network bandwidth shared by all edge nodes. 
	
	\subsection{Model Downloading Delay}
	
	The model downloading delay measures the total time required for all users to successfully retrieve the PBs necessary for assembling their requested AI models. Given that all PBs are delivered sequentially, we first derive the downloading delay for a specific PB $k$ (with data size $S\left( k \right)$) in the sequel\footnote{Similar to \cite{11,24}, we do not account for the model fetching delay from the cloud to edge nodes for two reasons. First, model placement at the edge typically occurs at a much longer timescale than user downloading. Once cached, user requests only trigger PB migration and wireless delivery, which dominate the downloading delay. Second, our work focuses on users served by collaborative edge nodes to explore the performance gain of fine-grained caching, while the traditional cloud-edge-device model delivery lies beyond the scope of this study.}
	\begin{align}
		T\left( k \right)=&\sum\limits_{n\in \mathcal{N}}{\sum\limits_{m\in \mathcal{N}\backslash \left\{ n \right\}}{\frac{{{b}_{n,m}}\left( k \right)S\left( k \right)}{R_{n,m}^{\text{bac}}\left( k \right)}}}\nonumber\\
		&\qquad\quad+\underset{u\in \mathcal{U}}{\mathop{\max }}\,\frac{\mathbb{I}\left\{ k\in {{\mathcal{K}}_{{{r}_{u}}}} \right\}S\left( k \right)}{\underset{{{\mathbf{e}}_{n,u}}\left( k \right)\in {{\mathcal{E}}_{n,u}},\forall n}{\mathop{\min }}\,{{R}_{u}}\left( k \right)},  \label{eq:7}
	\end{align}
	where the first term of the right-hand-side quantifies the migration delay, and $R_{n,m}^{\text{bac}}\left( k \right)$ denotes the backhaul link rate\footnote{Note that $R_{n,m}^{\text{bac}}\left( k \right)$ is finite and varies across different $k$ due to dynamic backhaul bandwidth availability and congestion conditions, introducing additional delay for PB migration. Our subsequent goal is to jointly design caching and migration strategies to minimize the overall downloading latency.} from edge node $n$ to $m$ when delivering PB $k$. The second term is the worst case broadcasting delay, where $\mathbb{I}\left\{ x \right\}=1$ if $x$ is true, otherwise $\mathbb{I}\left\{ x \right\}=0$, thus $\mathbb{I}\left\{ k\in {{\mathcal{K}}_{{{r}_{u}}}} \right\}$ indicates whether $u$ is the requesting user for PB $k$. On this basis, we express the total model downloading delay as 
	\begin{align}
		T=\sum\limits_{k\in \mathcal{K}}{T\left( k \right)}. \label{eq:8}
	\end{align}
    
    \subsection{Problem Formulation}
    
    Our objective is to minimize the model downloading delay while satisfying both edge storage capacity constraints and user QoS requirements, thereby boosting timely on-device AI inference. The designed variables incorporate fine-grained model caching $\mathbf{a}=\left[ {{a}_{n}}\left( k \right):\forall k,n \right]$, PB migration $\mathbf{b}=\left[ {{b}_{n,m}}\left( k \right):\forall k,n,m \right]$, and broadcasting beamforming $\mathbf{w}=\left[ {{\mathbf{w}}_{n}}\left( k \right):\forall k,n \right]$. As a consequence, the optimization problem is formulated as
    \begin{subequations}
    	\begin{equation}
    		\begin{aligned}
    			\textbf{P1}: \underset{\mathbf{a},\mathbf{b},\mathbf{w}}{\mathop{\min }}\,T, \label{eq:9a}
    		\end{aligned}
    	\end{equation}
    	\vspace{-5mm}
    	\begin{align}
    		\mbox{s.t.}\ 
    		&\underset{{{\mathbf{e}}_{n,u}}\left( k \right)\in {{\mathcal{E}}_{n,u}},\forall n}{\mathop{\min }}\,{{R}_{u}}\left( k \right)\ge \mathbb{I}\left\{ k\in {{\mathcal{K}}_{{{r}_{u}}}} \right\}{{Q}_{u}},\ \forall k,u, \label{eq:9b}\\
    		&{{a}_{n}}\left( k \right)\in \left\{ 0,1 \right\},\text{ }{{b}_{n,m}}\left( k \right)\in \left\{ 0,1 \right\},\text{ }\forall k,n,m, \label{eq:9c}\\
    		&{{\left\| {{\mathbf{w}}_{n}}\left( k \right) \right\|}^{2}}\le {{P}^{\max }},\ \forall k,n, \label{eq:9d}\\  
    		&\text{(\ref{eq:1}), (\ref{eq:2})}\nonumber
    	\end{align} 
    \end{subequations}
    where (\ref{eq:9b}) ensures that the downloading rate of the requested PB remains above the user QoS threshold for all potential channel estimation errors. (\ref{eq:9c}) indicates that the caching and migration variables are binary. In (\ref{eq:9d}), the transmission power of each edge node is restricted by power budget ${{P}^{\max }}$. (\ref{eq:1}) and (\ref{eq:2}) specifies edge storage capacity and PB migration constraints, respectively. 
    
    However, \textbf{P1} is an intractable MINLP problem exacerbated by CSI uncertainty, where the number of possible channel errors is infinite. Besides, the centralized control of PB caching and migration results in an extensive decision space as well as significant information exchange overhead, which imposes substantial computational complexity. To efficiently address \textbf{P1}, we develop an MADRL-based approach in Section III to enable collaborative decision-making among edge nodes in a distributed manner. 
    
    \section{Fine-Grained AI Model Caching and Downloading Solution} 
    
    In this section, we present the proposed solution for FGAMCD. First, we analyze the limitations of existing MADRL methods in addressing \textbf{P1}, and provide an overview of our designed multi-agent action semantics network with data augmentation (MAASN-DA) framework, along with its specialized enhancements. Next, we detail the individual components of MAASN-DA framework, followed by the development of the overall training algorithm. 
    
    \subsection{Overview of MAASN-DA Framework}
    
    In \textbf{P1}, each edge node serves as an agent responsible for determining its caching, migration, and beamforming decisions. Meanwhile, the decisions made by different agents have mutual influence, and jointly affect the model downloading delay, which align with the MADRL framework \cite{26,27,Fan2025MADRL}. For instance, \cite{Fan2025MADRL} exploited multi-agent deep deterministic policy gradient (MADDPG) to optimize model partitioning, tolerance latency, and updating frequency for split federated learning, thereby coordinating the global model convergence across multiple edge nodes. However, none of the existing MADRL methods can be directly applied to solve \textbf{P1} due to the following reasons: \textit{1) Implicitly Characterization of Actions’ Mutual Influence:} Existing methods generally employ a single black-box neural network to output each agent’s action, neglecting the fact that different action dimensions may have distinct impacts on other agents, which is inefficient for capturing the complex coupling among the caching and migration policies of multiple edge nodes. \textit{2) Low Sample Efficiency:} MADRL methods suffer from inefficient sample collection due to the costly agent-environment interactions, which can significantly impede the training speed and hinder the model convergence. \textit{3) Multi-Agent Credit Assignment:} MADRL methods often rely on a centralized training stage that leverages a global critic function to learn distributed policies, which leads to challenges in fairly evaluating the contributions of individual agents \cite{28}. 
    
    \begin{figure}[t] \centering
    	\begin{center}
    		\centerline{\includegraphics[width=7.8cm]{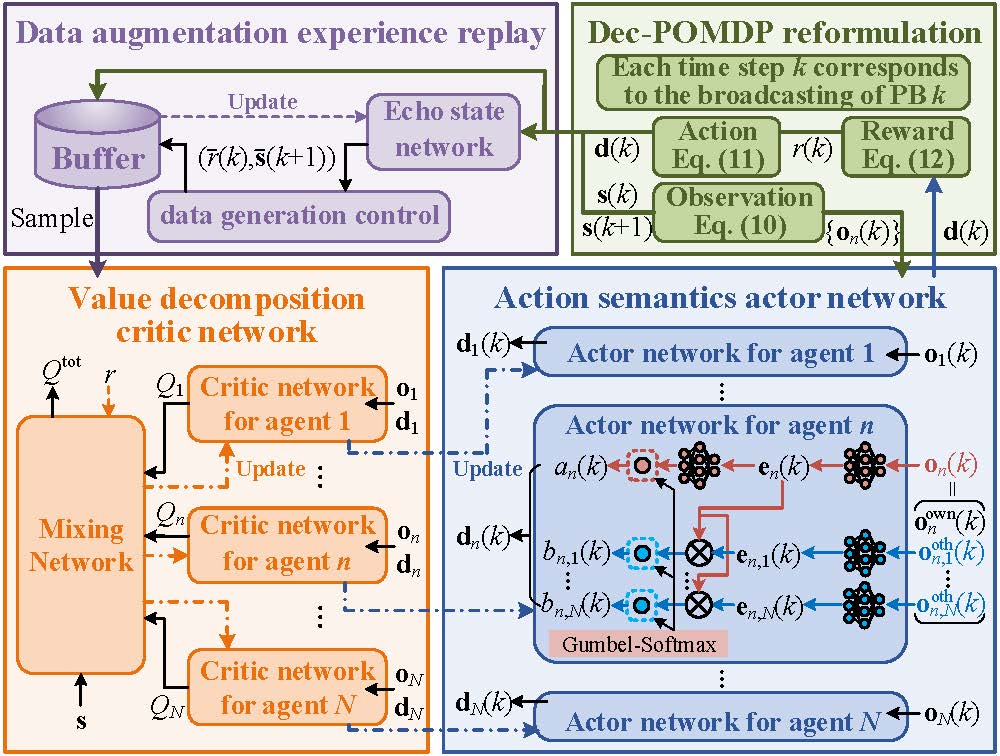}}
    	\end{center}
    	\vspace{-4mm}		
    	\caption{The proposed MAASN-DA framework.}
    	\label{fig:2}     
    \end{figure}

    To tackle the above limitations, we propose MAASN-DA framework as depicted in Fig. \ref{fig:2}, which incorporates four main components:
    \begin{itemize}
    	\item \textbf{Dec-POMDP Reformulation:} The original \textbf{P1} is first reformulated as a decentralized partially observable Markov decision process (Dec-POMDP), specifically designed to enable MADRL agents to make sequential decisions for PB caching and delivery.
    	\item \textbf{Action Semantics Actor Network:} A specialized actor network is designed for each agent to explicitly capture the effects of different action dimensions on other agents (i.e., action semantics). This network consists of several sub-modules, each processing distinct parts of the agent’s observation and producing corresponding action dimensions based on the action semantics. 
    	\item \textbf{Data Augmentation Experience Replay:} After collecting experience tuples through interactions with the environment, we utilize an echo state network (ESN) to generate additional training samples by leveraging its prediction capabilities, thereby enriching the replay buffer and enhancing sample efficiency. Furthermore, both the quality and quantity of the generated samples are carefully controlled to ensure the convergence performance.
    	\item \textbf{Value Decomposition Critic Network:} Each agent employs a local critic to estimate its individual Q-value, which is subsequently combined to obtain the global Q-value via a mixing network. During the training phase, the global Q-value is decomposed to compute the policy gradients of individual agents, thus addressing the credit assignment problem. 
    \end{itemize}
    
    We present detailed elucidation of these components in the following subsections. 
    
    \subsection{Dec-POMDP Reformulation}
    
    Dec-POMDP features the distributed decision-makings for multiple agents (i.e., edge nodes) with partially observations, which can be defined by $\left( \mathcal{O},\mathcal{S},\mathcal{D},\pi ,r,\gamma  \right)$, where $\mathcal{O}$, $\mathcal{S}$, and $\mathcal{D}$ are observation, state, and action space, respectively. In this work, we optimize the caching and downloading of each PB sequentially, so the index of PB $k$ corresponds to the time step in MADRL. In an arbitrary step $k$, each agent $n$ obtains an observation ${{\mathbf{o}}_{n}}\left( k \right)\in \mathcal{O}$, which is a part of global state $\mathbf{s}\left( k \right)\in \mathcal{S}$, then takes action ${{\mathbf{d}}_{n}}\left( k \right)\in \mathcal{D}$ based on distributed policy $\pi$. Given reward function $r\left( k \right)$ and discount factor $\gamma$, all agents collaborate to learn an optimal $\pi$ that maximizes the accumulative discounted reward. We specify these elements in the sequel. 
    
    \textit{1) Observation:} ${{\mathbf{o}}_{n}}\left( k \right)$ can be intuitively divided into two parts: $\mathbf{o}_{n}^{\text{own}}\left( k \right)$ that contains environmental information and agent $n$’s own properties, as well as $\mathbf{o}_{n,m}^{\text{oth}}\left( k \right),m\in \mathcal{N}\backslash \left\{ n \right\}$ that represents the observations of agent $n$ on other agents. They are written as
    \begin{subequations}
    	\begin{align}
    		&\mathbf{o}_{n}^{\text{own}}\left( k \right)=\left[ S\left( k \right),\left\{ \mathbb{I}\left\{ k\in {{\mathcal{K}}_{{{r}_{u}}}} \right\}:u\in {{\mathcal{U}}_{n}} \right\},{{{\tilde{C}}}_{n}}\left( k \right) \right], \label{eq:10a}\\
    		&\mathbf{o}_{n,m}^{\text{oth}}\left( k \right)={{\varpi }_{n,m}}\nonumber\\
    		&\quad\ \times \left[ R_{n,m}^{\text{bac}}\left( k \right),\left\{ \mathbb{I}\left\{ k\in {{\mathcal{K}}_{{{r}_{u}}}} \right\}:u\in {{\mathcal{U}}_{m}} \right\},{{{\tilde{C}}}_{m}}\left( k \right) \right], \label{eq:10b}\\
    		&{{\mathbf{o}}_{n}}\left( k \right)=\left[ \mathbf{o}_{n}^{\text{own}}\left( k \right),\left\{ \mathbf{o}_{n,m}^{\text{oth}}\left( k \right):m\in \mathcal{N}\backslash \left\{ n \right\} \right\} \right], \label{eq:10c}
    	\end{align}
    \end{subequations}
    where ${{\mathcal{U}}_{n}}$ denotes the set of users associated with node $n$. ${{\tilde{C}}_{n}}\left( k \right)$ signifies the remaining storage capacity before caching PB $k$ with ${{\tilde{C}}_{n}}\left( k+1 \right)={{\{ {{{\tilde{C}}}_{n}}\left( k \right)-{{a}_{n}}\left( k \right)S\left( k \right) \}}^{+}}$ and ${{\tilde{C}}_{n}}\left( 1 \right)={{C}_{n}}$. Binary indicator ${{\varpi }_{n,m}}=1$ implies that the information of node $m$ can be observed by $n$, otherwise ${{\varpi }_{n,m}}=0$. Note that ${{\varpi }_{n,m}}$ is determined by the distribution of edge nodes and the allowed information exchange overhead. Moreover, the global state in step $k$ is $\mathbf{s}\left( k \right)=\left[ {{\mathbf{o}}_{n}}\left( k \right):n\in \mathcal{N} \right]$. 
    
   \textit{2) Action:} Theoretically, all optimization variables of \textbf{P1} $\mathbf{a},\mathbf{b},\mathbf{w}$ can be regarded as the agents’ actions. Nonetheless, the optimality of MADRL degrades and the training complexity increases with the growth of action dimension. As a remedy, we propose to decouple the variables into two groups, where the PB caching and migration are optimized by MADRL, i.e., the action of agent $n$ in step $k$ is 
    \begin{align}
    	{{\mathbf{d}}_{n}}\left( k \right)=\left[ {{a}_{n}}\left( k \right),\left\{ {{b}_{n,m}}\left( k \right):m\in \mathcal{N}\backslash \left\{ n \right\} \right\} \right], \label{eq:11}
    \end{align}
    then the high-dimensional beamforming $\left[ {{\mathbf{w}}_{n}}\left( k \right):\forall n \right]$ is derived from an optimization subroutine with given joint action $\mathbf{d}\left( k \right)=\left[ {{\mathbf{d}}_{n}}\left( k \right):\forall n \right]$, which will be detailed in Section III-F. Accordingly, the distributed policy that maps local observation to action is expressed as ${{\mathbf{d}}_{n}}\left( k \right)=\pi \left( {{\mathbf{o}}_{n}}\left( k \right) \right)$. 
    
    \textit{3) Reward:} According to \textbf{P1}, the reward design targets to minimize the downloading delay $T\left( k \right)$ of each PB $k$ while meeting the constraints, i.e.,  
    \begin{align}
    	r\left( k \right)=
    	\left\{\begin{array}{cl}
    		&\!\!\!\!\!\!\!\!\!-T\left( k \right)-{{r}_{1}}\Lambda \left( k \right),\\
    		&\,\quad\sum\nolimits_{u\in \mathcal{U}}{\mathbb{I}\left\{ k\in {{\mathcal{K}}_{{{r}_{u}}}} \right\}}>0,\sum\nolimits_{n\in \mathcal{N}}{{{\lambda }_{n}}\left( k \right)}>0, \\
    		&\!\!\!\!\!\!\!\!\!-{{r}_{2}},\text{ }\sum\nolimits_{u\in \mathcal{U}}{\mathbb{I}\left\{ k\in {{\mathcal{K}}_{{{r}_{u}}}} \right\}}>0,\sum\nolimits_{n\in \mathcal{N}}{{{\lambda }_{n}}\left( k \right)}=0, \\
    		&\!\!\!\!\!\!\!\!\!0,\text{ otherwise},
    	\end{array}\right.  \label{eq:12}
    \end{align}
    where ${{r}_{1}},{{r}_{2}}>0$ denote penalties for violating the constraints. To be specific, $\Lambda \left( k \right)=1$ if the optimization subroutine returns an infeasible solution for QoS or transmission power constraints, and all agents receive penalty ${{r}_{1}}$, otherwise $\Lambda \left( k \right)=0$. If PB $k$ is requested by some users whereas no agent can deliver PB $k$, then agents are punished\footnote{Two factors may lead to insufficient edge storage capacity for caching a PB requested by certain users. First, the caching decisions may fail to fully utilize the available storage, in which case the reward penalty guides the agents to adjust their policies. Second, the total system resources may be inadequate to satisfy all user requests, rendering problem \textbf{P1} mathematically infeasible. In such situations, users can either defer to the next request period or directly retrieve the models from the cloud.} by ${{r}_{2}}$. If PB $k$ is not required by any users, we set $r\left( k \right)=0$. Note that the other constrains in \textbf{P1} can be ensured by our actor network design. 
    
    \subsection{Action Semantics Actor Network}
    
    It is observed from (11) that ${{a}_{n}}\left( k \right)$ only affects the agent $n$’s own property, i.e., the remaining storage capacity, while the PB migration ${{b}_{n,m}}\left( k \right)$ directly impacts the caching and downloading of other agents. This motivates us to explicitly extract the action semantics by dividing the black-box actor network into multiple sub-modules, each processing a distinct part of the agent’s observation \cite{29}. As shown in Fig. \ref{fig:2}, the actor network of each agent $n$ consists of $N$ sub-modules. The first one takes the full observation ${{\mathbf{o}}_{n}}\left( k \right)$ as input, utilizes two neural networks to yield observation embedding ${{\mathbf{e}}_{n}}\left( k \right)$ and ${{a}_{n}}\left( k \right)$, respectively. The rest $N-1$ sub-modules contain $N-1$ neural networks to generate the action dimensions related with other influenced agents. Each of the $N-1$ sub-modules (say the $m$-th sub-module) uses a part of observation $\mathbf{o}_{n,m}^{\text{oth}}\left( k \right)$ as input to determine embedding ${{\mathbf{e}}_{n,m}}\left( k \right)$ related with influenced agent $m$, then combines ${{\mathbf{e}}_{n}}\left( k \right)$ and ${{\mathbf{e}}_{n,m}}\left( k \right)$ via inner product to output ${{b}_{n,m}}\left( k \right)$. 
    
    Following the above structure, the action semantics actor network of agent $n$ is represented by $\pi \left( {{\mathbf{o}}_{n}}\left( k \right);{{\varphi }_{n}} \right)$ with ${{\varphi }_{n}}$ being the network parameters. For the sake of satisfying constraints (\ref{eq:1}), (\ref{eq:2}), and (\ref{eq:9c}), we further adapt the output layer of $\pi \left( {{\mathbf{o}}_{n}}\left( k \right);{{\varphi }_{n}} \right)$ as below. Given an output value $\tilde{d}$, we leverage the Gumbel-Softmax reparameterization to yield differentiable binary variable as
	\begin{align}
		d=GS\left( {\tilde{d}} \right)=\text{Sigmoid}\bigg( \frac{\tilde{d}+\ln \varsigma -\ln \left( 1-\varsigma  \right)}{{{s}_{\text{temp}}}} \bigg), \label{eq:13}
	\end{align}
    where $\text{Sigmoid}\left( x \right)=\frac{{{\text{e}}^{x}}}{1+{{\text{e}}^{x}}}$, $\varsigma$ is a random variable obeying uniform distribution within [0, 1], ${{s}_{\text{temp}}}$ denotes the Softmax coefficient, and smaller ${{s}_{\text{temp}}}$ makes $d$ closer to a binary variable. Therefore, given the variables output by the sub-modules of $\pi \left( {{\mathbf{o}}_{n}}\left( k \right);{{\bm{\varphi} }_{n}} \right)$, denoted by ${{\tilde{a}}_{n}}\left( k \right)$ and ${{\tilde{b}}_{n,m}}\left( k \right)$, we invoke the following operations: 
    \begin{align}
    	\!\!{{a}_{n}}\left( k \right)\!=\!GS\left( {{{\tilde{a}}}_{n}}\left( k \right) \right),\ {{b}_{n,m}}\left( k \right)\!=\! GS( {{{\tilde{b}}}_{n,m}}\left( k \right) ),\ \forall m, \label{eq:14}
    \end{align}
     In addition, if the edge node’s storage capacity is not adequate for caching PB $k$, i.e., ${{\tilde{C}}_{n}}\left( k \right)<S\left( k \right)$, ${{a}_{n}}\left( k \right)$ is forced to be 0. For ${{a}_{n}}\left( k \right)=0$, we set ${{b}_{n,m}}\left( k \right)=0,m\in \mathcal{N}\backslash \left\{ n \right\}$ to prohibit PB migration from node $n$.\footnote{It is worth noting that an edge node with fully occupied storage can still relay PBs received from other nodes to users. This is because such PBs are only temporarily buffered in volatile memory (e.g., DRAM/VRAM) rather than persistently cached, and thus do not consume the node’s storage capacity.}
     
     \subsection{Data Augmentation Experience Replay}
     
    In each time step $k$, all agents interact with the environment and collect a new experience tuple $\left( \mathbf{s}\left( k \right),\mathbf{d}\left( k \right),r\left( k \right),\mathbf{s}\left( k+1 \right) \right)$. In the meantime, we adopt an ESN to predict the reward and next state $\left( \bar{r}\left( k \right),\mathbf{\bar{s}}\left( k+1 \right) \right)$, thereby synthesizing training data and boosting sample efficiency. The reason for choosing an ESN over other predictive models, such as recurrent neural networks (RNNs) or transformers, is that it only requires training the output weights while keeping the high-dimensional hidden weights fixed. This significantly reduces training complexity and mitigate gradient vanishing/exploding issues.
    
	Specifically, ESN consists of input layer, reservoir layer, and output layer with weighting parameters ${{\bm{\eta} }_{\text{in}}}$, ${{\bm{\eta} }_{\text{re}}}$, and ${{\bm{\eta} }_{\text{out}}}$, respectively. Taking sequence $\mathbf{v}\left( 1 \right),\ldots ,\mathbf{v}\left( k \right)$ as input, where $\mathbf{v}\left( k \right)=\left( \mathbf{s}\left( k \right),\mathbf{d}\left( k \right) \right)$, ESN recurrently updates the reservoir states $\mathbf{q}\left( 1 \right),\ldots ,\mathbf{q}\left( k \right)$ and yields the prediction output as follows \cite{30}:
	\begin{align}
		&\mathbf{q}\left( k \right)=\tanh \left( {{\bm{\eta} }_{\text{in}}}\mathbf{v}\left( k \right)+{{\bm{\eta} }_{\text{re}}}\mathbf{q}\left( k-1 \right) \right),\nonumber\\
		&\left( \bar{r}\left( k \right),\mathbf{\bar{s}}\left( k+1 \right) \right)={{\bm{\eta} }_{\text{out}}}\mathbf{q}\left( k \right). \label{eq:15}
	\end{align}
	
	We fix ${{\bm{\eta} }^{\text{in}}}$ and ${{\bm{\eta} }^{\text{re}}}$ after random initialization, and only update ${{\bm{\eta} }^{\text{out}}}$ during the tuning of ESN. The loss function to be minimized is given by
	\begin{align}
		{{L}^{\text{e}}}\left( {{\bm{\eta} }_{\text{out}}} \right)={{\left\| {{\bm{\eta} }_{\text{out}}}\mathbf{q}\left( k \right)-\left( r\left( k \right),\mathbf{s}\left( k+1 \right) \right) \right\|}^{2}}, \label{eq:16}
	\end{align}
	where real experience tuples from the environment $\left( r\left( k \right),\mathbf{s}\left( k+1 \right) \right),\forall k$ serve as labels for tuning the ESN. 
	
	Considering that the extensive inclusion of synthetic training samples may degrade the convergence performance, we design a data generation control strategy to filter out low-quality data while adaptively adjusting the number of synthetic samples. Concretely, we invoke the following condition to identify satisfactory data predicted by ESN: 
	\begin{align}
		\left\| \left( \bar{r}\left( k \right),\mathbf{\bar{s}}\left( k+1 \right) \right)-\left( r\left( k \right),\mathbf{s}\left( k+1 \right) \right) \right\|\le \xi , \label{eq:17}
	\end{align}
    where $\xi$ denotes the data selection threshold. Moreover, the ESN can generate at most ${{\tau }_{0}}K$ synthetic samples per episode in the initial training stage with ${{\tau }_{0}}\in \left[ 0,1 \right]$ being a tunable proportion. The number of synthetic samples should decrease with the training episodes, thus we have
    \begin{align}
    	{{\tau }_{e}}=\left\lfloor {{\tau }_{0}}K{{\Lambda }^{\left\lfloor {e}/{{\bar{E}}}\; \right\rfloor }} \right\rfloor, \label{eq:18}
    \end{align}
	where ${{\tau }_{e}}$ represents the maximum number of synthetic samples in the $e$-th episode, $\Lambda \in \left( 0,1 \right)$ is the attenuation factor, and the number of synthetic samples declines every $\bar{E}$ episodes. Thereafter, both real experience tuples and synthetic samples are stored in the replay buffer. 
	
	\subsection{Value Decomposition Critic Network}
	
	The critic network of MAASN-DA is designed based on the value decomposition architecture, allowing for the evaluation of each agent’s contribution to the overall system. Each agent $n$ possesses a local critic $Q\left( {{\mathbf{o}}_{n}}\left( k \right),{{\mathbf{d}}_{n}}\left( k \right);{{\bm{\theta} }_{n}} \right)$ with parameters ${{\bm{\theta} }_{n}}$ to estimate its Q-value ${{Q}_{n}}$, then a mixing network is adopted to combine the individual Q-value of each agent, and produce a global Q-value as
	\begin{align}
		{{Q}^{\text{tot}}}={{g}^{\text{mix}}}\left( \mathbf{s},{{Q}_{1}},\ldots ,{{Q}_{N}};{{\bm{\theta} }^{\text{mix}}} \right), \label{eq:19}
	\end{align}
	where ${{g}^{\text{mix}}}$ indicates the mixing function, and ${{\bm{\theta} }^{\text{mix}}}$ is the parameters of the mixing network. Inspired by the well-known QMIX algorithm \cite{31}, we design ${{g}^{\text{mix}}}$ by forcing monotonicity between ${{Q}^{\text{tot}}}$ and each ${{Q}_{n}}$, i.e.,
	\begin{align}
		{\partial {{Q}^{\text{tot}}}}/{\partial {{Q}_{n}}}\ge 0,\ \forall n\in \mathcal{N}, \label{eq:20}
	\end{align}
    such that any action ${{\mathbf{d}}_{n}}\left( k \right)$ that increases the individual Q-value ${{Q}_{n}}$ also improves the global Q-value ${{Q}^{\text{tot}}}$. To this end, the parameter ${{\bm{\theta} }^{\text{mix}}}$ is yielded by separate hypernetworks, each of which employs an absolute activation function to output the weights for one layer of the mixing network, thereby guaranteeing the nonnegativity. 
	
	During the training stage, we randomly sample a mini-batch of samples in the replay buffer, and perform forward propagation through the neural networks in MAASN-DA to compute the loss functions. To be specific, the loss for the value decomposition critic network is given by 
	\begin{align}
		{{L}^{\text{c}}}\left( \left\{ {{\bm{\theta} }_{n}} \right\},{{\bm{\theta} }^{\text{mix}}} \right)\!=\!\mathbb{E}\!\left[ {{Q}^{\text{tar}}}\!\left( k \right)\!-\!{{Q}^{\text{tot}}}\!\left( \mathbf{s}\left( k \right),\mathbf{d}\left( k \right);\left\{ {{\bm{\theta} }_{n}} \right\}\!,{{\bm{\theta} }^{\text{mix}}} \right) \right]\!, \label{eq:21}
	\end{align}
	where \[{{Q}^{\text{tar}}}\!\left( k \right)\!=\!r\!\left( k \right)\!+\!\gamma {{Q}^{\text{tot}}}\big( \mathbf{s}\!\left( k\!+\!1 \right)\!,\!\left\{ \pi\! \left( {{\mathbf{o}}_{n}}\!\left( k\! +\! 1 \right)\!;{{{\hat{\bm{\varphi} }}}_{n}}\!\right) \right\}\!;\!\{ {{{\hat{\bm{\theta} }}}_{n}}\},\!{{{\hat{\bm{\theta} }}}^{\text{mix}}} \big),\]${{\hat{\bm{\varphi} }}_{n}}$ and ${{\hat{\bm{\theta} }}_{n}},{{\hat{\bm{\theta} }}^{\text{mix}}}$ are the parameters of the target actor and critic networks, respectively. These target networks share the same structure as the main actor and critic networks, while their parameters are slowly updated to stabilize training. Additionally, the global Q-value is decomposed to evaluate the contribution of each agent $n$, thereby calculating its actor loss as follows:
    \begin{align}
    	{{L}^{\text{a}}}\left( {{\bm{\varphi} }_{n}} \right)=\mathbb{E}\left[ -Q\left( {{\mathbf{o}}_{n}}\left( k \right),\pi \left( {{\mathbf{o}}_{n}}\left( k \right);{{\bm{\varphi} }_{n}} \right);{{\bm{\theta} }_{n}} \right) \right],\ \forall n. \label{eq:22}
    \end{align}

    Consequently, the actor and critic networks of MAASN-DA are updated using gradient descent based on the above loss functions. 
    
    \subsection{Robust CoMP Beamforming Subroutine}
    
    To proceed with, we propose a robust CoMP beamforming subroutine to optimize $\left[ {{\mathbf{w}}_{n}}\left( k \right):\forall n \right]$ under CSI uncertainty. It is notable that with given PB caching and migration decisions $\mathbf{d}\left( k \right)$, the participation of each edge node, i.e., ${{\lambda }_{n}}\left( k \right)$, can be determined by (\ref{eq:3}). We define $\mathbf{w}\left( k \right)={{[ \mathbf{w}_{1}^{\text{T}}\left( k \right),\ldots ,\mathbf{w}_{N}^{\text{T}}\left( k \right) ]}^{\text{T}}}\in {{\mathbb{C}}^{MN\times 1}}$ and $\mathbf{W}\left( k \right)=\mathbf{w}\left( k \right){{\mathbf{w}}^{\text{H}}}\left( k \right)\in {{\mathbb{C}}^{MN\times MN}}$, then the beamforming problem for broadcasting PB $k$ can be recast as
	\begin{subequations}
		\begin{equation}
			\begin{aligned}
				\textbf{P2}(k): \underset{\mathbf{W}\left( k \right),\zeta }{\mathop{\min }}\,\frac{1}{\zeta }, \label{eq:23a}
			\end{aligned}
		\end{equation}
		\vspace{-5mm}
		\begin{align}
			\mbox{s.t.}\ 
			&\underset{{{\mathbf{e}}_{n,u}}\left( k \right)\in {{\mathcal{E}}_{n,u}},\forall n}{\mathop{\min }}\,{{R}_{u}}\left( k \right)\ge \mathbb{I}\left\{ k\in {{\mathcal{K}}_{{{r}_{u}}}} \right\}{{Q}_{u}},\ \forall u, \label{eq:23b}\\
			&\sum\nolimits_{l=\left( n-1 \right)M+1}^{nM}{{{\left[ \mathbf{W}\left( k \right) \right]}_{l,l}}}\le {{P}^{\max }},\ \forall n, \label{eq:23c}\\
			&\frac{1}{\zeta }\ge \frac{\mathbb{I}\left\{ k\in {{\mathcal{K}}_{{{r}_{u}}}} \right\}S\left( k \right)}{\underset{{{\mathbf{e}}_{n,u}}\left( k \right)\in {{\mathcal{E}}_{n,u}},\forall n}{\mathop{\min }}\,{{R}_{u}}\left( k \right)},\ \forall u, \label{eq:23d}\\  
			&\mathbf{W}\left( k \right)\succeq 0, \label{eq:23e}\\ 
			&\text{rank}\left[ \mathbf{W}\left( k \right) \right]=1, \label{eq:23f}
		\end{align} 
	\end{subequations}
    where $\zeta$ is an auxiliary variable with $\frac{1}{\zeta }$ being the maximum worst case broadcasting delay among users, and ${{R}_{u}}\left( k \right)$ can be rewritten as
    \begin{align}
    	&\!\!{{R}_{u}}\left( k \right)=B\times\nonumber\\
    	&\!\!{{\log }_{2}}\big( 1\!+\!{{{\big[ {{{\mathbf{\tilde{h}}}}_{u}}\left( k \right)\!+\!{{\mathbf{e}}_{u}}\left( k \right) \big]}^{\text{H}}}\mathbf{W}\!\left( k \right)\!\big[ {{{\mathbf{\tilde{h}}}}_{u}}\left( k \right)\!+\!{{\mathbf{e}}_{u}}\left( k \right) \!\big]}\!/{\sigma _{u}^{2}} \big), \label{eq:24}
    \end{align}
    where ${{\mathbf{\tilde{h}}}_{u}}\left( k \right)={{[ {{\lambda }_{1}}\left( k \right)\mathbf{\tilde{h}}_{1,u}^{\text{T}}\left( k \right),\ldots ,{{\lambda }_{N}}\left( k \right)\mathbf{\tilde{h}}_{N,u}^{\text{T}}\left( k \right) ]}^{\text{T}}}$ and ${{\mathbf{e}}_{u}}\left( k \right)={{[ {{\lambda }_{1}}\left( k \right)\mathbf{e}_{1,u}^{\text{T}}\left( k \right),\ldots ,{{\lambda }_{N}}\left( k \right)\mathbf{e}_{N,u}^{\text{T}}\left( k \right) ]}^{\text{T}}}$ are the stacked imperfect channel and estimation error for user $u$, respectively. To handle the infinite possibilities of channel errors in constraints (\ref{eq:23b}) and (\ref{eq:23d}), we invoke S-Procedure as illustrated in the following lemma \cite{32}. 
    
    \textit{Lemma 1:} Let ${{\mathbf{A}}_{1}},{{\mathbf{A}}_{2}}\in {{\mathbb{H}}^{MN\times MN}}$, ${{\mathbf{b}}_{1}},{{\mathbf{b}}_{2}}\in {{\mathbb{C}}^{MN\times 1}}$, ${{\kappa }_{1}},{{\kappa }_{2}}\in \mathbb{R}$, for $\forall \mathbf{e}\in {{\mathbb{C}}^{MN\times 1}}$, the implication 
    \begin{subequations}
    	\begin{align}
    		&{{\mathbf{e}}^{\text{H}}}{{\mathbf{A}}_{1}}\mathbf{e}+2\operatorname{Re}\left\{ \mathbf{b}_{1}^{\text{H}}\mathbf{e} \right\}+{{\kappa }_{1}}\le 0 \label{eq:25a}\\
    		\Rightarrow &{{\mathbf{e}}^{\text{H}}}{{\mathbf{A}}_{2}}\mathbf{e}+2\operatorname{Re}\left\{ \mathbf{b}_{2}^{\text{H}}\mathbf{e} \right\}+{{\kappa }_{2}}\le 0 \label{eq:25b}
    	\end{align}
    \end{subequations}
    holds if and only if there exists a $\varepsilon \ge 0$ such that 
    \begin{align}
    	\left[ \begin{matrix}
    		\varepsilon {{\mathbf{A}}_{1}}-{{\mathbf{A}}_{2}} & \varepsilon {{\mathbf{b}}_{1}}-{{\mathbf{b}}_{2}}  \\
    		\varepsilon \mathbf{b}_{1}^{\text{H}}-\mathbf{b}_{2}^{\text{H}} & \varepsilon {{\kappa }_{1}}-{{\kappa }_{2}}  \\
    	\end{matrix} \right]\succeq \mathbf{0}.
    	 \label{eq:26}
    \end{align}
    
    Take (\ref{eq:23b}) as an example, we first rewrite it as 
	\begin{align}
		&\underset{{{\mathbf{e}}_{n,u}}\left( k \right)\in {{\mathcal{E}}_{n,u}},\forall n}{\mathop{\max }}\,\mathbb{I}\left\{ k\in {{\mathcal{K}}_{{{r}_{u}}}} \right\}\cdot \big[ -\mathbf{e}_{u}^{\text{H}}\left( k \right)\mathbf{W}\left( k \right){{\mathbf{e}}_{u}}\left( k \right)\nonumber\\
		&+2\operatorname{Re}\left\{ -\mathbf{\tilde{h}}_{u}^{\text{H}}\left( k \right)\mathbf{W}\left( k \right){{\mathbf{e}}_{u}}\left( k \right) \right\}\!+\!\kappa _{u}^{\left\langle 1 \right\rangle }\left( k \right) \big]\le 0,\ \forall u,  \label{eq:27}
	\end{align}
    where $\kappa _{u}^{\left\langle 1 \right\rangle }\left( k \right)=\sigma _{u}^{2}\left( {{2}^{{{{Q}_{u}}}/{B}\;}}-1 \right)-\mathbf{\tilde{h}}_{u}^{\text{H}}\left( k \right)\mathbf{W}\left( k \right){{\mathbf{\tilde{h}}}_{u}}\left( k \right)$. According to (\ref{eq:4}), we also have 
    \setlength{\arraycolsep}{0pt}
    \begin{align}
    	\!\!\!\mathbf{e}_{u}^{\text{H}}\left( k \right){{\mathbf{C}}_{u}}{{\mathbf{e}}_{u}}\left( k \right)-N\le 0,\text{ }{{\mathbf{C}}_{u}}=\left[ \begin{matrix}
    		{{\mathbf{C}}_{1,u}} & \ldots  & \mathbf{O}  \\
    		\vdots  & \ddots  & \vdots   \\
    		\mathbf{O} & \ldots  & {{\mathbf{C}}_{N,u}}  \\
    	\end{matrix} \right]\!,\forall u.\!
    	  \label{eq:28}
    \end{align}
    Then, by substituting (\ref{eq:28}) into (\ref{eq:25a}) and (\ref{eq:27}) into (\ref{eq:25b}), constraint (\ref{eq:23b}) can be transformed into 
    \begin{align}
    	\!\!\!\mathbb{I}\!\left\{ k\!\in\! {{\mathcal{K}}_{{{r}_{u}}}}\! \right\}\!\cdot\!\! \left[ \begin{matrix}
    		\varepsilon _{u}^{\left\langle 1 \right\rangle }{{\mathbf{C}}_{u}}+\mathbf{W}\left( k \right) & {{\mathbf{W}}^{\text{H}}}\left( k \right){{{\mathbf{\tilde{h}}}}_{u}}\left( k \right)  \\
    		\mathbf{\tilde{h}}_{u}^{\text{H}}\left( k \right)\mathbf{W}\left( k \right) & -\varepsilon _{u}^{\left\langle 1 \right\rangle }N-\kappa _{u}^{\left\langle 1 \right\rangle }\left( k \right)  \\
    	\end{matrix} \right]\!\!\succeq\!\mathbf{0},\forall u,\!\!\! \label{eq:29}
    \end{align}
    where $\varepsilon _{u}^{\left\langle 1 \right\rangle }$ is a non-negative auxiliary variable. Analogously, we convert (\ref{eq:23d}) into 
    \begin{align}
    	\!\!\!\mathbb{I}\left\{ k \!\in\! {{\mathcal{K}}_{{{r}_{u}}}} \right\}\!\cdot\!\! \left[ \begin{matrix}
    		\varepsilon _{u}^{\left\langle 2 \right\rangle }{{\mathbf{C}}_{u}}\!+\!\mathbf{W}\left( k \right) & {{\mathbf{W}}^{\text{H}}}\left( k \right){{{\mathbf{\tilde{h}}}}_{u}}\left( k \right)  \\
    		\mathbf{\tilde{h}}_{u}^{\text{H}}\left( k \right)\mathbf{W}\left( k \right) & -\varepsilon _{u}^{\left\langle 2 \right\rangle }N\!-\!\kappa _{u}^{\left\langle 2 \right\rangle }\left( k \right)  \\
    	\end{matrix} \right]\!\!\succeq\!\mathbf{0},\forall u,\!\!\!
    	 \label{eq:30}
    \end{align}
    where $\kappa _{u}^{\left\langle 2 \right\rangle }\left( k \right)=\sigma _{u}^{2}\left( {{2}^{{\zeta S\left( k \right)}/{B}\;}}-1 \right)-\mathbf{\tilde{h}}_{u}^{\text{H}}\left( k \right)\mathbf{W}\left( k \right){{\mathbf{\tilde{h}}}_{u}}\left( k \right)$. 
    
    Subsequently, we tackle non-convex rank constraint (\ref{eq:23f}) by constructing an equivalent difference-of-convex (DC) expression. Specifically, $\text{rank}\left[ \mathbf{W}\left( k \right) \right]=1$ if and only if $\mathbf{W}\left( k \right)$ possesses only one nonzero eigenvalue, i.e., ${{\sigma }_{1}}\left[ \mathbf{W}\left( k \right) \right]>0,{{\sigma }_{m}}\left[ \mathbf{W}\left( k \right) \right]=0,m\in \left\{ 2,\ldots ,MN \right\}$, thus we have 
    \begin{align}
    	\text{tr}\left[ \mathbf{W}\left( k \right) \right]\!=\!\sum\limits_{m=1}^{MN}{{{\sigma }_{m}}\left[ \mathbf{W}\left( k \right) \right]}\!=\!{{\sigma }_{1}}\left[ \mathbf{W}\left( k \right) \right]\!=\!\left\| \mathbf{W}\left( k \right) \right\|.
    	\label{eq:31}
    \end{align}
    As a consequence, define $\bm{\varepsilon} =\left[ \varepsilon _{u}^{\left\langle 1 \right\rangle },\varepsilon _{u}^{\left\langle 2 \right\rangle }:\forall u \right]$, we can transform \textbf{P2} into the following DC program: 
    \begin{equation}
    	\begin{aligned}
    		\textbf{P2.1}(k): \underset{\mathbf{W}\left( k \right),\zeta ,\bm{\varepsilon} }{\mathop{\min }}\,\frac{1}{\zeta }+\mu \left( \text{tr}\left[ \mathbf{W}\left( k \right) \right]-\left\| \mathbf{W}\left( k \right) \right\| \right), \label{eq:32}
    	\end{aligned}
    \end{equation}
    \vspace{-5mm}
    \begin{align}
    	\mbox{s.t.}\ 
    	&\text{(\ref{eq:23c}), (\ref{eq:23e}), (\ref{eq:29}), (\ref{eq:30})}\nonumber
    \end{align} 
    where $\mu$ is the regularization parameter \cite{33}. Then, we linearize the concave part $-\left\| \mathbf{W}\left( k \right) \right\|$ in (\ref{eq:32}) and reach a stationary solution by iteratively solving \textbf{P2.2}$\left( k \right)$ in the sequel
    \begin{equation}
    	\begin{aligned}
    		\!\textbf{P2.2}(k): \underset{\mathbf{W}\left( k \right),\zeta ,\bm{\varepsilon}}{\mathop{\min }}\,\frac{1}{\zeta }\!+\!\mu \left( \text{tr}\left[ \mathbf{W}\left( k \right) \right]\!-\!\text{tr}\left[ {{{\mathbf{\bar{u}}}}_{1}}\mathbf{\bar{u}}_{1}^{\text{H}}\mathbf{W}\left( k \right) \right] \right)\!, \label{eq:33}
    	\end{aligned}
    \end{equation}
    \vspace{-5mm}
    \begin{align}
    	\mbox{s.t.}\ 
    	&\text{(\ref{eq:23c}), (\ref{eq:23e}), (\ref{eq:29}), (\ref{eq:30})}\nonumber
    \end{align} 
    where ${{\left. {{{\mathbf{\bar{u}}}}_{1}}\mathbf{\bar{u}}_{1}^{\text{H}}=\frac{\partial \left\| \mathbf{W} \right\|}{\partial \mathbf{W}} \right|}_{\mathbf{W}=\mathbf{\bar{W}}\left( k \right)}}$ with ${{\mathbf{\bar{u}}}_{1}}$ denoting the eigenvector corresponding to the maximum eigenvalue of $\mathbf{\bar{W}}\left( k \right)$, and $\mathbf{\bar{W}}\left( k \right)$ is the solution obtained from the previous iteration. Notice that \textbf{P2.2}$\left( k \right)$ pertains to convex optimization issue that can be solved by interior-point method (IPM). Finally, the proposed robust CoMP beamforming subroutine converges to a rank-one solution $\mathbf{W}\left( k \right)$, then we reconstruct beamforming vector $\mathbf{w}\left( k \right)$ through eigenvalue decomposition. 
    
    \textit{Remark 2 (Solution Quality of the Beamforming Subroutine):} The proposed approach leverages the S-Procedure to handle channel uncertainty and reformulates the rank-1 constraint via DC programming. The introduction of auxiliary variables, each with a properly defined feasible region according to S-Procedure theory, does not affect optimality \cite{32}. The DC program is iteratively solved using a first-order approximation of the spectral norm, which may lead to sub-optimal solutions. Nevertheless, as reported in \cite{33}, this approach achieves superior solution quality and computational efficiency compared to the conventional Gaussian randomization method for recovering rank-1 solutions. 
    
    \begin{algorithm}[t]
    	\caption{Overall Training Algorithm for MAASN-DA} \label{alg:1}
    	\begin{algorithmic}[1]
    		\STATE \textbf{Input:} AI model repository $J,K,{{\mathcal{K}}_{j}},\forall j$, user requests $U,{{r}_{u}},{{Q}_{u}},\forall u$, FGAMCD environment $N,M,B,{{P}^{\max }},{{C}_{n}},{{\mathbf{\tilde{h}}}_{n,u}}\left( k \right),{{\mathbf{C}}_{n,u}},R_{n,m}^{\text{bac}}\left( k \right),\forall n,u,m$, and learning hyperparameters. 
    		\STATE \textbf{Initialize:} Parameter of ESN ${{\bm{\eta} }^{\text{in}}},{{\bm{\eta} }^{\text{re}}},{{\bm{\eta} }^{\text{out}}}$, actor network ${{\bm{\varphi} }_{n}},\forall n$, critic network ${{\bm{\theta} }_{n}},{{\bm{\theta} }^{\text{mix}}},\forall n$, and target networks ${{\hat{\bm{\varphi} }}_{n}},{{\hat{\bm{\theta} }}_{n}},{{\hat{\bm{\theta} }}^{\text{mix}}},\forall n$, replay buffer $\mathcal{B}$. 
    		\FOR{$e\in \mathcal{E}=\left\{ 1,\ldots ,E \right\}$} 
    		\FOR{$k\in \mathcal{K}$}
    		\STATE Each agent $n$ observes ${{\mathbf{o}}_{n}}\left( k \right)$ and takes action ${{\mathbf{d}}_{n}}\left( k \right)=\pi \left( {{\mathbf{o}}_{n}}\left( k \right);{{\bm{\varphi} }_{n}} \right)$. 
    		\STATE Iteratively solve \textbf{P2.2}$\left( k \right)$ to optimize CoMP beamforming $\mathbf{w}\left( k \right)$. 
    		\STATE Calculate reward $r\left( k \right)$ and transit to the next state $\mathbf{s}\left( k+1 \right)$. 
    		\STATE Store $\left( \mathbf{s}\left( k \right),\mathbf{d}\left( k \right),r\left( k \right),\mathbf{s}\left( k+1 \right) \right)$ in $\mathcal{B}$. 
    		\ENDFOR
    		\WHILE{1}
    		\STATE Sample experience $\left( \mathbf{s}\left( k \right),\mathbf{d}\left( k \right),r\left( k \right),\mathbf{s}\left( k+1 \right) \right)$ from $\mathcal{B}$.
    		\STATE Invoke the ESN to predict $\left( \bar{r}\left( k \right),\mathbf{\bar{s}}\left( k+1 \right) \right)$.
    		\STATE Tune the ESN by minimizing ${{L}^{\text{e}}}\left( {{\bm{\eta} }_{\text{out}}} \right)$ in (\ref{eq:16}).
    		\IF{$\left\| \left( \bar{r}\left( k \right),\mathbf{\bar{s}}\left( k+1 \right) \right)-\left( r\left( k \right),\mathbf{s}\left( k+1 \right) \right) \right\|\le \xi$}
    		\STATE Store $\left( \mathbf{s}\left( k \right),\mathbf{d}\left( k \right),\bar{r}\left( k \right),\mathbf{\bar{s}}\left( k+1 \right) \right)$ in $\mathcal{B}$. 
    		\ELSIF{the number of synthetic samples reaches ${{\tau }_{e}}$}
    		\STATE \textbf{break}
    		\ENDIF
    		\ENDWHILE
    		\STATE Sample a mini-batch of samples from $\mathcal{B}$.
    		\STATE Calculate the global Q-value using (\ref{eq:19}). 
    		\STATE Update the critic network by minimizing ${{L}^{\text{c}}}\left( \left\{ {{\bm{\theta} }_{n}} \right\},{{\bm{\theta} }^{\text{mix}}} \right)$ in (\ref{eq:21}). 
    		\STATE Update the actor network by minimizing ${{L}^{\text{a}}}\left( {{\bm{\varphi} }_{n}} \right),\forall n$ in (\ref{eq:22}). 
    		\STATE Slowly update the target networks. 
    		\ENDFOR
    		\STATE \textbf{Output:} Well-trained edge node agents for optimizing $\mathbf{a},\mathbf{b},\mathbf{w}$. 
    	\end{algorithmic}
    \end{algorithm}

	\subsection{Overall Training Algorithm for MAASN-DA}

	The developed training algorithm for MAASN-DA is outlined in \textbf{Algorithm 1}. In each time step $k$, the edge node agents make PB caching and migration decisions exploiting the action semantics actor network, then the CoMP beamforming is derived from the embedded optimization subroutine (Lines 4-9). After collecting experience tuples via agent-environment interactions, we invoke the ESN to synthesize additional training samples while tuning the ESN, followed by performing data generation control (Lines 10-19). Afterwards, the global Q-value is obtained based on the value decomposition critic network, then the actor and critic network parameters are updated by minimizing the loss functions (Lines 20-24). Finally, the well-trained agents can efficiently coordinate AI model caching and delivery processes, responding to diverse user requests and uncertain channel estimations, thus ensuring low-latency AI model downloading. 
	
	\section{Theoretical Analysis for MAASN-DA}
	
	In this section, we present theoretical analysis for the proposed MAASN-DA approach. To begin with, we analyze the convergence performance of critic network, characterizing the approximation error of Q-value and offering guidelines for configuring the hyperparameters of data augmentation experience replay. Afterwards, the convergence of actor network is analyzed, ensuring the achievement of local-optimal PB caching and downloading decisions. Lastly, we present complexity analysis for MAASN-DA. 
	
	\subsection{Convergence Analysis for Critic Network}
	
	We commence by making some assumptions that are commonly utilized in neural network analysis \footnote{Please note that these assumptions are only imposed to derive the closed-form convergence bound and extract theoretical insights; the actual implementation of the proposed algorithm does not depend on them. Violating these assumptions may lead to inaccurate bound characterization and sub-optimal hyperparameter selection, thereby degrading training convergence.}. 
	
	\textit{Assumption 1:} The state, action, and reward in each time step are bounded by $\left\| \mathbf{s}\left( k \right) \right\|\le {{B}_{\mathbf{s}}},\left\| \mathbf{d}\left( k \right) \right\|\le {{B}_{\mathbf{d}}},\left| r\left( k \right) \right|\le {{B}_{r}},\forall k$. This assumption is typically met due to all elements in $\mathbf{s}\left( k \right)$, $\mathbf{d}\left( k \right)$ and $r\left( k \right)$ possess explicit physical meanings. 
	
	\textit{Assumption 2:} The ESN for generating synthetic samples satisfies the echo state property, i.e., the spectral norms of weight matrices satisfy $\left\| {{\bm{\eta} }_{\text{in}}} \right\|\le \psi _{\text{in}}^{\max }$, $\left\| {{\bm{\eta} }_{\text{re}}} \right\|\le \psi _{\text{re}}^{\max }<1$, and ${{\bm{\eta} }_{\text{out}}}\le \psi _{\text{out}}^{\max }$. This assumption can be met by reasonably initializing the weights of the ESN \cite{30}.
	
	\textit{Assumption 3:} The value decomposition critic network can be fitted by a deep recurrent Q network (DRQN) with input, recurrent, and output weights ${{\bm{\omega} }_{\text{in}}}$, ${{\bm{\omega} }_{\text{re}}}$, and ${{\bm{\omega} }_{\text{out}}}$, respectively. This RNN is able to estimate the Q-value as: $\mathbf{p}\left( k \right)=\tanh \left( {{\bm{\omega} }_{\text{in}}}\mathbf{v}\left( k \right)+{{\bm{\omega} }_{\text{re}}}\mathbf{p}\left( k-1 \right) \right),Q\left( \mathbf{s}\left( k \right),\mathbf{d}\left( k \right) \right)={{\bm{\omega} }_{\text{out}}}\mathbf{p}\left( k \right)$, with given the state sequence up to step $k$. This assumption stems from the wide adaptation of DRQN in value decomposition MADRL \cite{31}. 
	
	\textit{Assumption 4:} The training of the critic network can be simplified as the fitted-Q iteration algorithm \cite{34}, in which the Q-value function is updated $E$ episodes, and the experience relay is deemed as a sampling distribution $\Omega $ over $\mathcal{S}\times \mathcal{D}$. Since the actor network is updated to maximize the Q-value estimated by the critic, this yields the greedy policy ${{\pi }_{E}}$ after $E$ iterations. 
	
	Under Assumption 1-4, we establish the following theorem to illustrate the convergence of the critic network. 
	
	\textit{Theorem 1:} Let ${{Q}_{{{\pi }_{E}}}}$ be the Q-value corresponding to ${{\pi }_{E}}$ and ${{Q}^{*}}={{\sup }_{\pi }}{{Q}_{\pi }}$ be the optimal Q-value. The Q-value approximation error is upper-bounded by 
	\begin{align}
		&{{\mathbb{E}}_{\Xi }}\left[ \left| {{Q}_{{{\pi }_{E}}}}-{{Q}^{*}} \right| \right]\le \underbrace{\frac{4{{\gamma }^{E+1}}}{{{\left( 1-\gamma  \right)}^{2}}}{{B}_{r}}}_{\text{Algorithmic error}}\nonumber\\
		&\quad+\underbrace{{{\vartheta }_{\Xi ,\Omega }}\left[ \frac{2\gamma }{{{\left( 1-\gamma  \right)}^{2}}}\sqrt{{{\nu }^{\max }}}+\xi \left( 1+\gamma \phi _{\text{out}}^{\max }\phi _{\text{in}}^{\max } \right) \right]}_{\text{Statistical error}}, \label{eq:34}
	\end{align}
	where $\Xi $ signifies the distribution corresponding to the Dec-POMDP in Section III-B, and ${{\vartheta }_{\Xi ,\Omega }}$ denotes the concentration coefficient determined by $\Xi $ and $\Omega $. $\phi _{\text{in}}^{\max }$ and $\phi _{\text{out}}^{\max }$ are the bounds of $\left\| {{\bm{\omega} }_{\text{in}}} \right\|$ and $\left\| {{\bm{\omega} }_{\text{out}}} \right\|$, respectively. ${{\nu }^{\max }}$ represents the maximum one-step error, given by
	\begin{align}
		{{\nu }^{\max }}\le &\underbrace{4\max {{\left\{ {{{B}_{r}}}/{\left( 1-\gamma  \right)}\;-\varsigma L_{\varsigma }^{\text{DRQN}},0 \right\}}^{2}}}_{\text{Estimation bias}}\nonumber\\
		&+\underbrace{{{{D}_{1}}{{V}^{2}}\ln U_{\varsigma }^{\text{DRQN}}}/{{{K}'}}\;+{{D}_{2}}{{V}^{2}}\varsigma }_{\text{Estimation variance}}, \label{eq:35}
	\end{align}
	where
	\begin{align}
		{K}'=K(& 1+{{\tau }_{0}}-\frac{{{\tau }_{0}}}{\xi }\{ \psi _{\text{out}}^{\max }\psi_{\text{in}}^{\max }\sqrt{B_{\mathbf{s}}^{2}+B_{\mathbf{d}}^{2}}\frac{1-{{\left( \psi _{\text{re}}^{\max } \right)}^{K}}}{1-\psi _{\text{re}}^{\max }}\nonumber\\
		&+\sqrt{B_{r}^{2}+B_{\mathbf{s}}^{2}} \})\nonumber
	\end{align}
    $\!\!$denotes the number of training samples in each episode, and $\varsigma$ is a positive constant. ${{D}_{1}}=8\sqrt{2{K}'}+{256}/{V}$, ${{D}_{2}}=4\sqrt{2{K}'}+52$, $V={{{B}_{r}}}/{\left( 1-\gamma  \right)}$. $L_{\varsigma }^{\text{DRQN}}$ and $U_{\varsigma }^{\text{DRQN}}$ are the upper- and lower-bounds of the exterior $\varsigma$-number of the DRQN, respectively. 
	
	\textit{Proof:} Please refer to Appendix A. $\qquad\qquad\qquad\qquad\ \ \blacksquare$
	
	Theorem 1 reveals that the Q-value approximation error in (\ref{eq:34}) is composed by an algorithmic error and a statistical error. The algorithmic error asymptotically diminishing to zero as the training proceeds. The statistical error is caused by the bias and variance that arise in estimating the Q-value function using the critic network (characterized by (\ref{eq:35})), as well as the prediction inaccuracies of the ESN. 
	
	On the other side, ${K}'$ encompasses both real and synthetic samples. Different from traditional experience replay that only collect $K$ samples in each episode, the proposed data augmentation trick enlarges ${K}'$ to boost sample efficiency, thus reducing the variance in Q-value estimation. Furthermore, it is evident that ${K}'$ is influenced by ${{\tau }_{0}}$ and $\xi$, and the judicious configure of these hyperparameters can mitigate the Q-value approximation error. Consequently, ${{\tau }_{0}}$ and $\xi$ are obtained by numerically minimizing the upper bound in (\ref{eq:34}) via a two-dimensional search. A practical example will be provided in Section V-C, where we traverse all possible combinations of ${{\tau}_{0}}$ and $\xi$ to find the one achieving the minimum upper bound value. Note that this search is performed only once prior to training, incurring negligible overhead. 
	
	\subsection{Convergence Analysis for Actor Network}
	
	We establish the following theorem to guarantee the convergence of the actor network, thereby ensuring the attainment of stable decisions of PB caching, migration, and broadcasting beamforming.
	
	\textit{Theorem 2:} In each time step of \textbf{Algorithm 1}, the CoMP beamforming subroutine converges within a finite number of iterations. Moreover, the action semantics actor network converges to a local optimal policy, i.e., 
	\begin{align}
		\underset{E\to \infty }{\mathop{\lim }}\,\left\| {{\nabla }_{{{\bm{\varphi} }_{n}}}}{{L}^{\text{a}}}\left( {{\bm{\varphi} }_{n}} \right) \right\|=0,\ \forall n. \label{eq:36}
	\end{align}

	\textit{Proof:} Please refer to Appendix B. $\qquad\qquad\qquad\qquad\quad \blacksquare$
	
	\subsection{Computational Complexity and Signaling Cost Analysis}
	
	The computational complexity of \textbf{Algorithm 1} is analyzed below. Typically, the training complexity of a $L$-layer neural network is $\mathcal{O}\left( \sum\nolimits_{l=1}^{L-1}{2\left( {{\Upsilon }_{l}}-1 \right){{\Upsilon }_{l+1}}} \right)$, where ${{\Upsilon }_{l}}$ denotes the number of neurons in the $l$-th layer. In Appendix C, we present the detailed architectures of actor network, critic network, and ESN, which can be substituted into this formula to calculate the corresponding complexities $\mathcal{O}\left( C{{o}_{\text{act}}} \right)$, $\mathcal{O}\left( C{{o}_{\text{cri}}} \right)$, and $\mathcal{O}\left( C{{o}_{\text{ESN}}} \right)$, respectively. Additionally, suppose that the optimization sub-routine needs $I$ iterations to converge, the beamforming design in each step has a complexity of $\mathcal{O}\left( I{{\left( {{M}^{2}}{{N^{2}_{\text{par}}}} \right)}^{3.5}} \right)$ \cite{34-1}, where $N_{\text{par}}$ denotes the number of participating edge nodes in broadcasting each PB. Note that $N_{\text{par}}\le N$ since the excessive PB migration among edge nodes is prohibited to reduce backhaul link delay. As a consequence, the overall training complexity for $K$ steps and $E$ episodes is
	\begin{align}
		\mathcal{O}\left(\! EK\!\left[ N\left( C{{o}_{\text{act}}}\!+\!C{{o}_{\text{cri}}} \right)\!+\!C{{o}_{\text{ESN}}}\!+\! I{{\left( {{M}^{2}}{{N^{2}_{\text{par}}}} \right)}^{3.5}} \right] \right). \label{eq:37}
	\end{align}
    
    After training, the implementation stage merely requires executing the actor network and the optimization subroutine, leading to a computational complexity of $\mathcal{O}(K[N{\Upsilon}_{\text{act},1}{\Upsilon }_{\text{act},L}+I{{\left( {{M}^{2}}{{N^{2}_{\text{par}}}} \right)}^{3.5}}])$, where ${\Upsilon}_{\text{act},l}$ is the number of neurons in the $l$-th layer of the actor. To form the input observations, edge nodes exchange backhaul rates, user requests, and remaining storage capacities, as specified in $\mathbf{o}_{n,m}^{\text{oth}}\left( k \right)$ in (\ref{eq:10b}). The resultant signaling cost is $\mathcal{O}(K\sum_{n\in\mathcal{N}}\sum_{m\in\mathcal{N}\backslash \left\{ n \right\}}{{\varpi }_{n,m}}(|\mathcal{U}_m|+2))$. 
    
	\section{Performance Evaluation} \label{sec:evaluation}
	
	\subsection{Simulation Setting and AI Model Repository Construction}
	
	In this section, we conduct simulations to evaluate the performance as well as glean insights of our proposed FGAMCD system. We consider a square area of 1 km$^\text{2}$ with 4-8 edge nodes and 18-48 randomly distributed users. The edge-user wireless channel is generated based on (\ref{eq:4}), and the small-scale fading ${{\mathbf{\bar{h}}}_{n,u}}\left( k \right)$ follows Rician channel model with factor of 3. The default parameter setting is listed in Table \ref{tab:I}, we set ${{\varpi }_{n,m}}$ by allowing information exchange between edge nodes less than 500 m away, and the neural network architectures of MAASN-DA are given in Appendix C. 
	
	\begin{table}[t]\footnotesize
		\centering
		\setlength{\abovecaptionskip}{0pt}    
		\setlength{\belowcaptionskip}{10pt}
		\caption{Simulation Parameters} \label{tab:I}
		\begin{threeparttable}
			\begin{tabular}{c|c|c|c}
				\hline
				\textbf{Parameter}&\textbf{Value}&\textbf{Parameter}&\textbf{Value}\\
				\hline
				$N$&6&$U$&30\\
				$J$&60&$K$&450\\
				${{Q}_{u}}$&[5, 7] Gbps&${{C}_{n}}$&1.25 GB\\
				$M$&20&$\upsilon$&-30 dB\\
				$\alpha $&3&$\sigma _{u}^{2}$&-80 dBm\\
				${{\mathbf{C}}_{n,u}}$&${{10}^{10}}{{\mathbf{I}}_{M}}$&$B$&400 MHz\\
				$R_{n,m}^{\text{bac}}\left( k \right)$&[8, 12] Gbps&${{P}^{\max }}$&43 dBm\\
				${{r}_{1}},{{r}_{2}}$&10, 10&$\gamma$&0.95\\
				${{\tau }_{0}},\xi$&0.8, 1.12&$\Lambda$&0.8\\
				$E,\bar{E}$&600, 10&$\mu$&1\\
				\hline
			\end{tabular}
		\end{threeparttable}			
	\end{table} 
    
    The AI model repository is constructed based on 3 basic AI models, i.e., Inception-v3, ResNet-18, and MobileNet-v2, per-trained on ImageNet. We fine-tune these models on the CIFAR-100 dataset to generate 60 task-specific models. In particular, each basic model is fine-tuned on 20 sub-datasets, each corresponding to one of the 20 superclasses in CIFAR-100\cite{35}. Therefore, the fine-tuned models are responsible for classifying the images within their respective superclass, e.g., the Inception-v3 model for “people” superclass is used to distinguish “baby”, “boy”, “girl”, “man”, and “woman”. Users request the 60 models in the repository according to a Zipf distribution, where the request probability of model $j$ is given by ${{{j}^{-\iota}}}/{\sum\nolimits_{i=1}^{60}{{{i}^{-\iota}}}}$, where distribution parameter $\iota=\text{0.5}$ in default with sensitivity analysis provided later.
    
    To capture the parameter reusability, we adopt a two-stage AI model fine-tuning. In the first stage, all parameters of the basic models are fine-tuned on several selected superclasses. In the second stage, we freeze a portion of the parameters, and further fine-tune the models on the remaining superclasses. For instance, all parameters of the basic ResNet-18 are first fine-tuned on “fruit and vegetables” superclass, the resulting model is then fine-tuned on “flower” superclass, with bottom layers frozen during this process. The frozen layers correspond to the reusable PBs in the AI model repository. 
    
    \subsection{Effectiveness of FGAMCD}
    
\begin{figure}[t]
		\begin{tabular}[t]{cc}
			\begin{minipage}[t]{0.5\linewidth}\centering
				\includegraphics[width = 4.3cm]{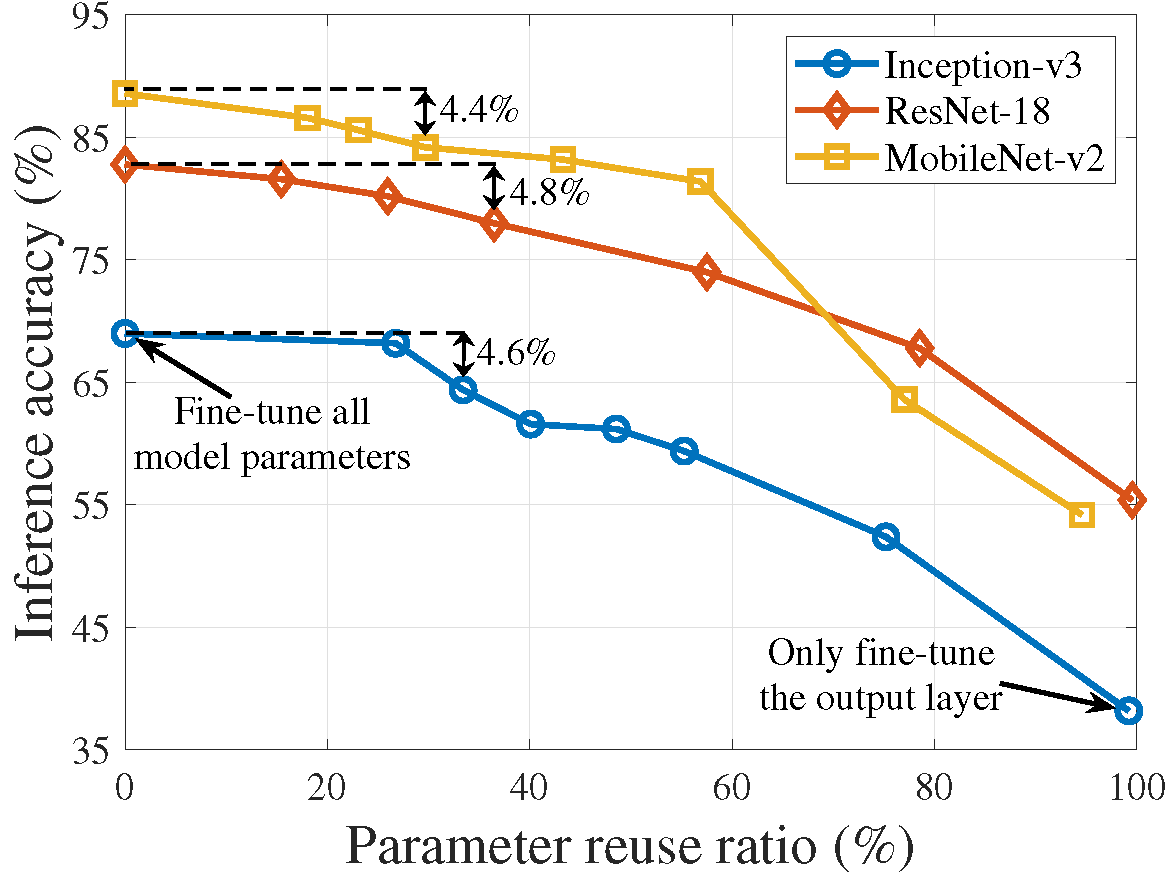}
				\vspace{1mm}
				\caption{Inference accuracy versus AI model parameter reuse ratio. }  
				\label{fig:3}  
			\end{minipage}
			\begin{minipage}[b]{0.46\linewidth}\centering
				\includegraphics[width = 3.9cm]{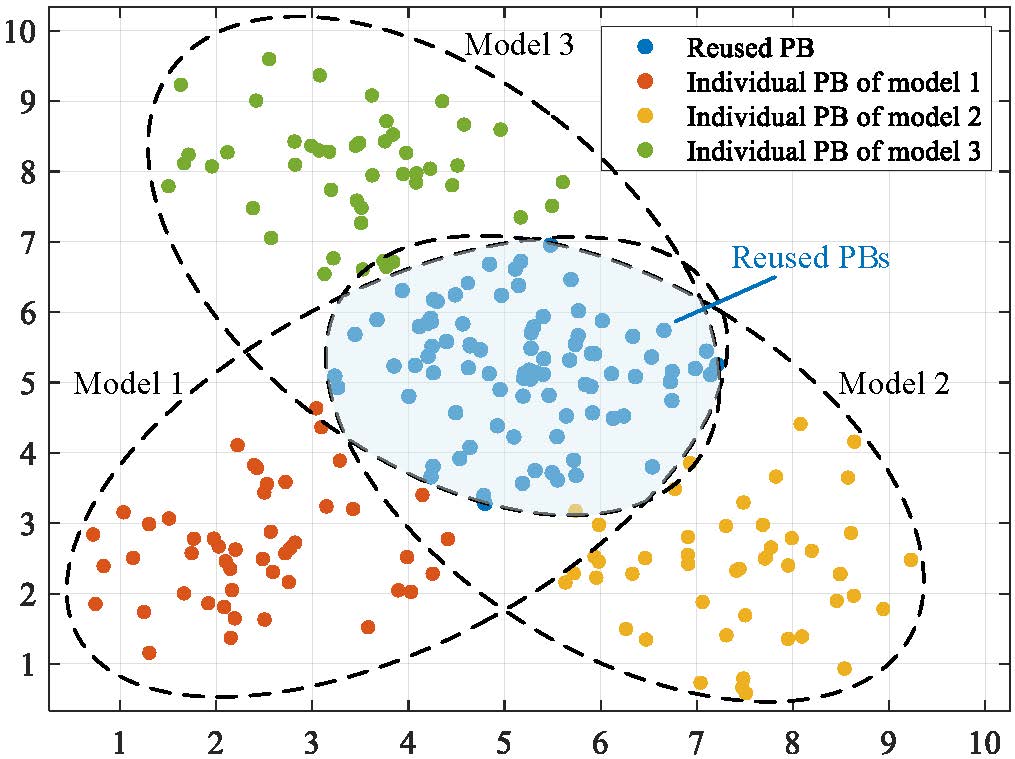}
				\vspace{1mm}
				\caption{A visualization of AI model parameter reuse, where different points indicate different PBs.}
				\label{fig:4}  
			\end{minipage}
			\vspace{0mm}	
		\end{tabular}
	\end{figure}

    Fig. \ref{fig:3} illustrates the impact of parameter reuse on the inference accuracy of task-specific models, where the parameter reuse ratio is controlled by adjusting the number of frozen layers during the second stage of fine-tuning. As can be observed, when the parameter reuse ratio is moderate, the degradation in inference accuracy is slight compared to full-parameter fine-tuning (i.e., no frozen layer and a reuse ratio of 0). This phenomenon demonstrates that the AI models fine-tuned for different downstream tasks can share a proportion of PBs, validating the effectiveness of our proposed FGAMCD. In the subsequent experiments, we select appropriate parameter reuse ratios for different basic models to ensure that the accuracy loss remains below 5\%. 
	
	\nop{\begin{figure}[t]\centering
	    \includegraphics[width = 5cm]{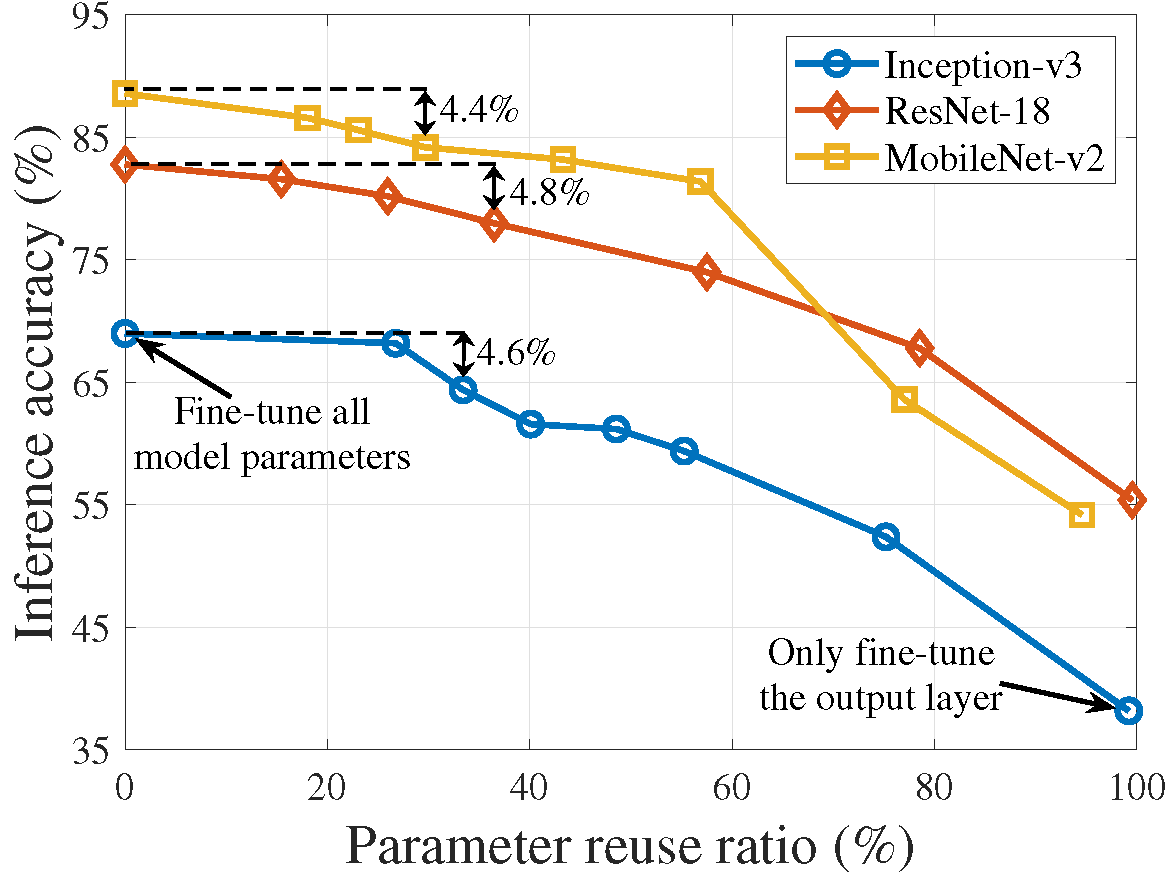}
	    \vspace{1mm}
	    \caption{Inference accuracy versus AI model parameter reuse ratio. }
    \end{figure}
    
    \begin{figure}[t]\centering
    	\includegraphics[width = 5cm]{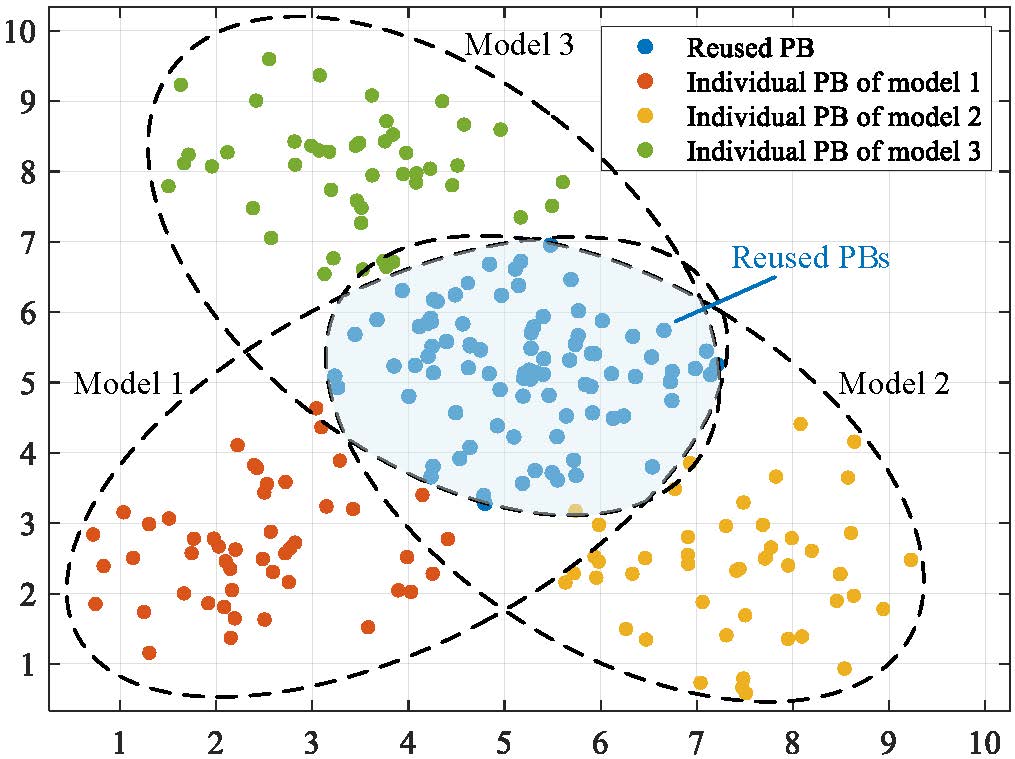}
    	\vspace{1mm}
    	\caption{A visualization of AI model parameter reuse, where different points indicate different PBs. }
    \end{figure}}
    
    Fig. \ref{fig:4} provides a visualization of AI model parameter reuse, employing models 1-3 in the repository as an example. These models are fine-tuned from the basic Inception-v3 model with parameter reuse ratio of 33.41\%, resulting in both reused and individual PBs. If an edge node needs to cache models 1-3, only one copy of the reused PBs is stored. Moreover, we adopt the actual sizes of PBs (ranging from 3.71 KB to 24.31 MB) to derive the numerical results, capturing their distinct storage requirements on the edge nodes. 
    
    \subsection{Learning Performance Evaluation}
    
\begin{figure}[t]\centering
    	\includegraphics[width = 3.9cm]{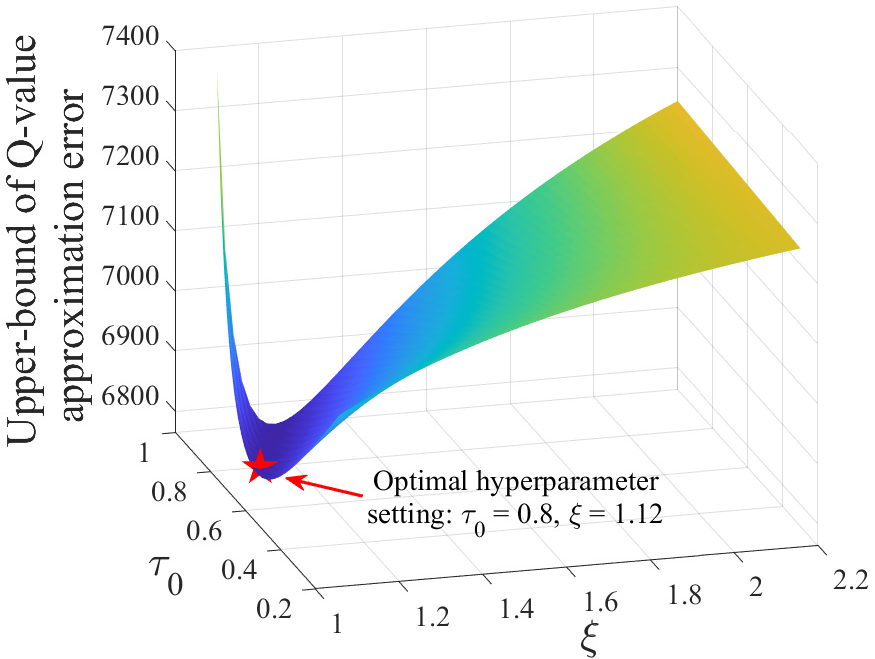}
    	\vspace{1mm}
    	\caption{Upper-bound of Q-value approximation error versus $\tau_0$ and $\xi$. }  
    	\label{fig:5}   
    \end{figure}

    According to Theorem 1, Fig. \ref{fig:5} depicts the upper bound of Q-value approximation error with different ${{\tau }_{0}}$ and $\xi$ . We set ${{B}_{\mathbf{s}}}=\text{1}$, ${{B}_{\mathbf{d}}}=\text{1}$, ${{B}_{r}}=\text{1}$ due to state/action/reward normalization, $\psi _{\text{in}}^{\max }=\text{0.5}$, $\psi _{\text{re}}^{\max }=\text{0.5}$, $\psi _{\text{out}}^{\max }=\text{0.5}$ to satisfy the echo state property, and $\varsigma =\text{0.1}$, $\phi _{\text{in}}^{\max }=\text{0.5}$, $\phi _{\text{out}}^{\max }=\text{10}$, $U_{\varsigma }^{\text{DRQN}}=\text{201}$, $L_{\varsigma }^{\text{DRQN}}=\text{46.2}$ based on \cite{30}. To minimize the Q-value approximation error, we select the optimal hyperparameters ${{\tau }_{0}}=\text{0.8}$ and $\xi =\text{1.12}$ for data augmentation experience replay. 
    
\begin{figure}[t] \centering
		\subfigure[Comparison of different training algorithms.] { 
			{\includegraphics[width=4.1cm]{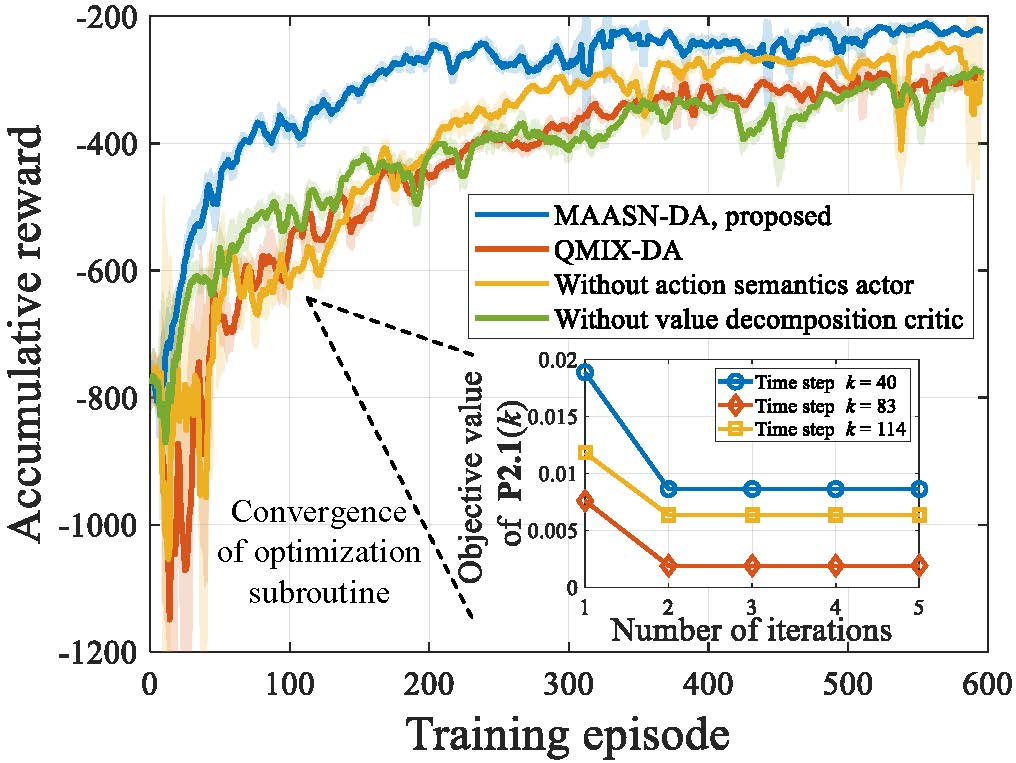}} 
		}     
		\subfigure[Comparison of different data augmentation strategies.] { 
			{\includegraphics[width=4.1cm]{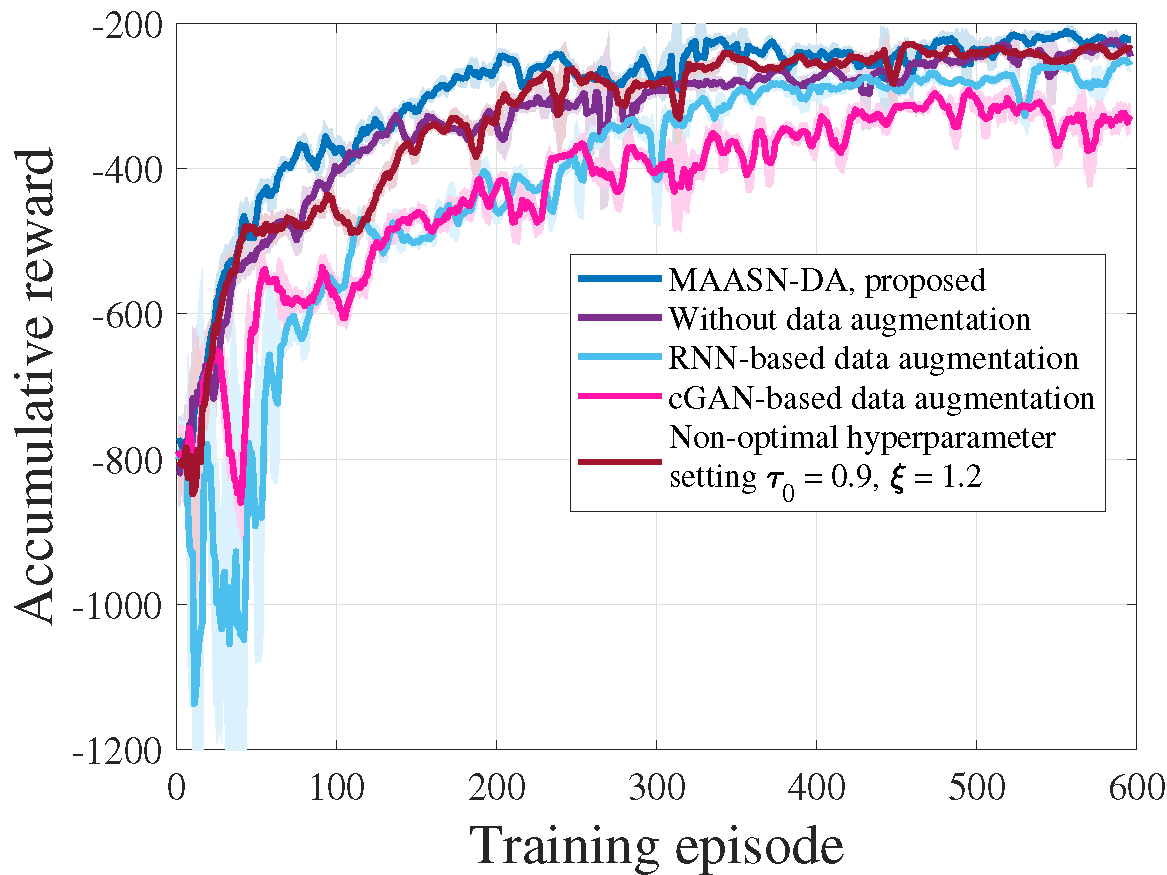}} 
		}
		\vspace{0mm}
		\caption{Accumulative reward versus training episode.}    
		\label{fig:6}     
	\end{figure}

	Fig. \ref{fig:6} assesses the learning performance of our proposed MAASN-DA through an ablation study. The average values (represented by the curves) and standard deviations (illustrated by the shaded regions) of the cumulative reward are used to evaluate convergence speed and stability, respectively. Baseline learning algorithms are outlined below. 
	
    \textit{1) QMIX-DA \cite{19,26}:} The actor-critic structure of MAASN-DA is replaced by QMIX, which employs a DRQN to estimate each agent’s Q-value for discrete actions, and combines these via a mixing network to generate the global Q-value. For a fair comparison, we also introduce the data augmentation experience replay into QMIX.
    
    \textit{2) Without Action Semantics Actor:} The actor network for each agent is replaced by a traditional fully-connected neural network with two hidden layers, each containing 256 neurons. 
    
    \textit{3) Without Value Decomposition Critic:} Each agent uses an independent critic network to estimate its individual Q-value without further aggregation or value decomposition.
     
    \textit{4) Without Data Augmentation:} We remove the ESN for generating synthetic samples, and the neural networks are trained solely using data from environment interactions. 
    
    \textit{5) RNN-Based Data Augmentation:} The ESN is replaced by a RNN with an identical architecture, where all weight parameters are tunable. 
    
    \textit{6) Conditional Generative Adversarial Network (cGAN)-Based Data Augmentation:} We train a cGAN that leverages $\mathbf{v}\left( k \right)$ as conditional information to generate $\left( \bar{r}\left( k \right),\mathbf{\bar{s}}\left( k+1 \right) \right)$, while a discriminator is employed to distinguish real experience data from generated samples.
    
    \textit{7) Non-Optimal Hyperparameter Setting:} We set ${{\tau }_{0}}=\text{0.9}$ and $\xi=\text{1.2}$ which are different from their optimal setting to validate the theoretical convergence bound analysis. 
    
    In Fig. \ref{fig:6} (a), we observe that MAASN-DA achieves the highest reward value with the fastest convergence speed. Compared to \textit{QMIX-DA}, the proposed approach utilizes an actor-critic structure and invokes Gumbel-Softmax reparameterization to ensure the discrete constraint, thereby enhancing the model’s ability to learn high-dimensional actions. MAASN-DA outperforms \textit{without action semantics actor} owing to its explicit characterization of actions’ mutual influence, which enables edge nodes to discern the diverse impact of different action dimensions on other agents. For \textit{without value decomposition critic}, it is difficult for each agent to evaluate the contribution of its own observation and action on the overall system, leading to lower performance than MAASN-DA. Additionally, the lower right corner of Fig. \ref{fig:6} (a) showcases the convergence of the beamforming optimization subroutine. It is evident that the subroutine converges within a few iterations (on average 2-3) for all time steps, demonstrating its efficiency for embedding into MAASN-DA.
    
    Fig. \ref{fig:6} (b) illustrates that MAASN-DA outperforms baseline data augmentation strategies, verifying the effectiveness of incorporating ESN to boost sample efficiency. We can see that the convergence performance of \textit{RNN-based data augmentation} is even worse than \textit{without data augmentation}. This is because tuning all parameters of the RNN makes it difficult to find optimal weights, thereby slowing down overall convergence. In contrast, ESN focuses exclusively on tuning the output weights, and the weights can be efficiently updated via ridge regression. \textit{cGAN-based data augmentation} fails to achieve satisfactory performance, as it struggles to capture the time-dependent nature of state transitions, thereby exhibiting limited compatibility in producing effective synthetic samples for MADRL. Moreover, when using different ${{\tau }_{0}}$ and $\xi$ in \textit{non-optimal hyperparameter setting}, the degradation in convergence performance aligns with the numerical result in Fig. \ref{fig:5}. To conclude, the ablation study corroborates the indispensability of all proposed components in MAASN-DA. 
    
    \begin{table}[t]\footnotesize
    	\centering
    	\caption{Evaluation of Algorithm Running Time Per Step}\label{tab:time}
    	\vspace{2.5 mm}
    	\begin{tabular}{p{1.6cm}<{\centering} p{1.6cm}<{\centering} p{2cm}<{\centering} p{2cm}<{\centering}}
    			\Xhline{1\arrayrulewidth}
    			\textbf{Parameter}& \textbf{Optimization}& \multicolumn{2}{c}{\textbf{Methods}}\\ \cline{3-4}
    			\textbf{settings}& \textbf{stages}&MAASN-DA& Full CoMP\\
    			\hline
    			$N=6$, &  MADRL & 0.01562 s & / \\\cline{2-4}
    			$M=20$ &  Subroutine & 0.6501 s & 3.4136 s\\
    			\hline
    			$N=6$, &  MADRL & 0.01606 s & / \\\cline{2-4}
    			$M=60$ &  Subroutine & 3.0884 s & 60.321 s\\
    			\hline
    			$N=12$, &  MADRL & 0.02013 s & / \\\cline{2-4}
    			$M=60$ &  Subroutine & 8.3760 s & 366.083 s\\
    			\Xhline{1\arrayrulewidth}   	
    	\end{tabular}
    \end{table}
    
    Table \ref{tab:time} reports the running time of different optimization stages (i.e., MADRL for caching and migration design, and subroutine for beamforming optimization) under varying $N$ and $M$, where \textit{full CoMP} refers to the traditional setting in which all edge nodes participate in the delivery of each PB. It can be observed that the well-trained MADRL agents incur negligible delays for making a decision, while the running time of the proposed optimization subroutine remains moderate. In particular, when $N=12$ and $M=60$, our approach reduces the time complexity by nearly two orders of magnitude compared with \textit{full CoMP}. The reason is that MAASN-DA strategically designs PB migration to manage backhaul latency, ensuring that the number of participating edge nodes in broadcasting each PB does not scale monotonically with $N$, hence the dimensionality of the beamforming variables remains moderate.
    
    \subsection{System Performance Evaluation }

    In this subsection, we evaluate the performance of the FGAMCD system. In addition to the previously discussed QMIX-DA, the following benchmark methods are employed for comparison. 
    
    \textit{1) TrimCaching \cite{24}:} This method optimizes AI model caching at edge nodes by exploiting parameter shareability to maximize the cache hit ratio, i.e., the satisfaction of user requests. The caching decisions are optimized using a greedy algorithm. As the original method does not account for active PB migration and CoMP beamforming, the corresponding variables are optimized through our proposed approach. 
    
    \textit{2) Without Edge Cooperation \cite{25}:} In this method, each edge node independently optimizes its model caching based on the requests of its associated users. During the downloading phase, the edge node broadcasts its cached PBs to the associated users, while the PB migration and CoMP are omitted. 
    
    \textit{3) Time Division Multiple Access (TDMA)-Based Unicasting:} The PB caching and migration are optimized by our approach. Then, edge nodes unicast the requested PBs to each user via TDMA. Since the PB delivery for each user is orthogonal, the CoMP beamforming is determined using maximum ratio transmission.
    
    \textit{4) Coarse-Grained Caching \cite{10,11}:} Similar to conventional content caching, edge nodes directly select the entire AI models for caching, disregarding the parameter reusability among different models. Afterwards, whole AI models are delivered to users during the downloading phase. 
    
\begin{figure}[t]
    	\begin{tabular}{cc}
    		\begin{minipage}[t]{0.48\linewidth}\centering
    			\includegraphics[width = 4.1cm]{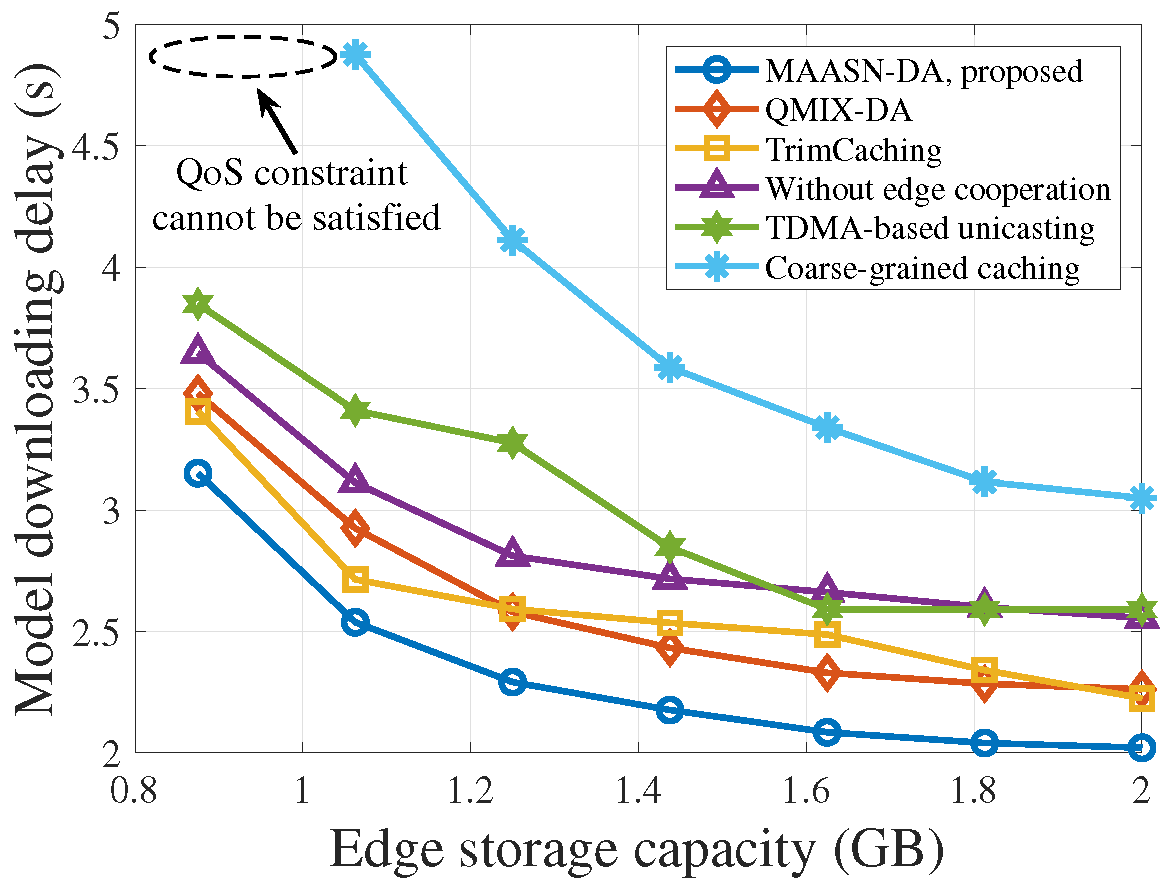}
    			\vspace{1mm}
    			\caption{Model downloading delay versus edge storage capacity.}  
    			\label{fig:7}  
    		\end{minipage}
    		\begin{minipage}[t]{0.48\linewidth}\centering
    			\includegraphics[width = 4.1cm]{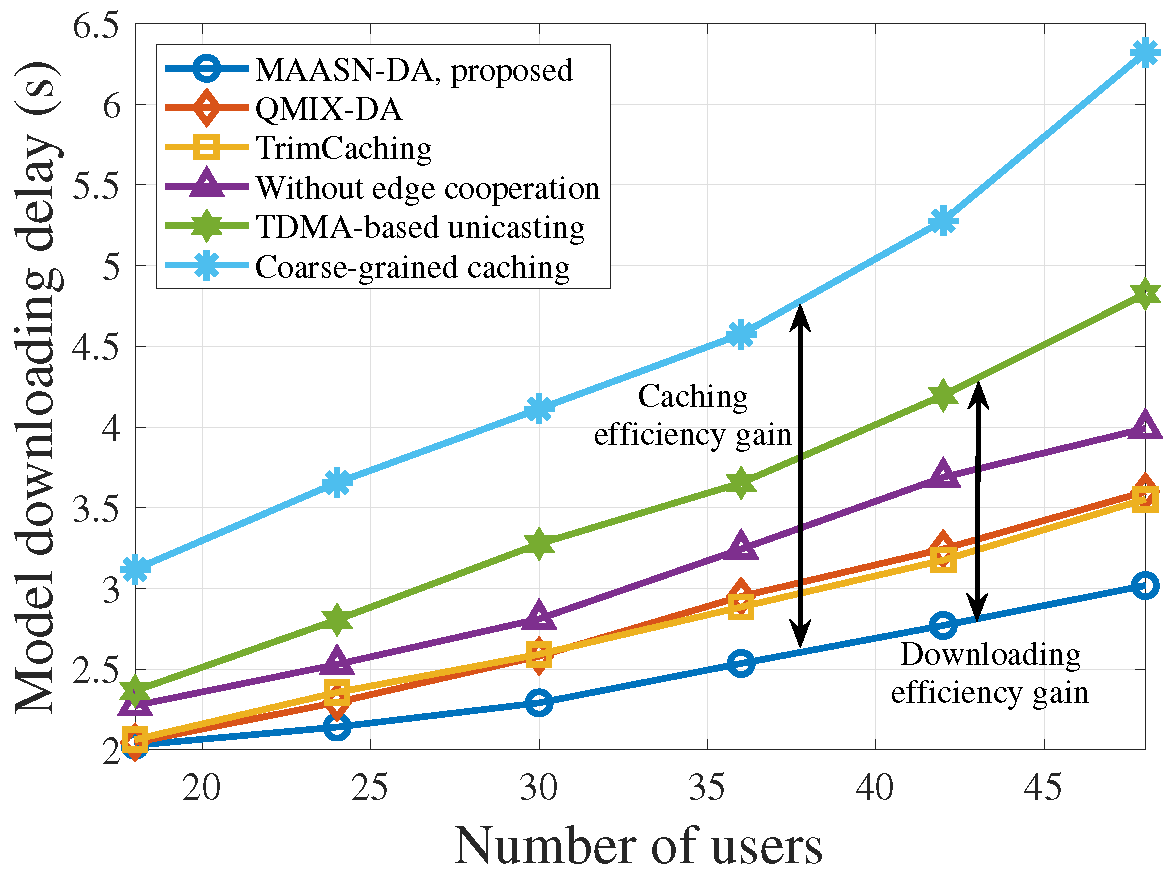}
    			\vspace{1mm}
    			\caption{Model downloading delay versus number of users. } 
    			\label{fig:8}  
    		\end{minipage}
    		\vspace{0mm}	
    	\end{tabular}
    \end{figure}

    Fig. \ref{fig:7} shows the relationship between model downloading delay and edge storage capacity ${{C}_{n}}$. As ${{C}_{n}}$ increases, all curves exhibit a downward trend. The rationale is that larger storage capacity allows edge nodes to cache more PBs or models, which facilitates the execution of CoMP broadcasting and improves the downloading channel conditions. Furthermore, the proposed MAASN-DA reduces model downloading delay by 12.20\%, 12.24\%, 23.30\%, 29.74\%, and 67.86\%, respectively, compared to the five benchmark methods. \textit{TrimCaching} focuses on maximizing the cache hit ratio but overlooks the interplay between edge model placement and delivery, the resultant caching decisions do not effectively reduce model downloading delay. For \textit{without edge cooperation}, each edge node only serves its associated users and PB migration is not considered, thus missing the performance improvement introduced by CoMP broadcasting. Both \textit{TDMA-based unicasting} and \textit{coarse-grained caching} exhibit high model downloading delays since neither method leverages parameter reusability during the caching and downloading phases, leading to inefficient use of storage and communication resources. Notably, \textit{coarse-grained caching} fails to meet users’ QoS requirements when edge storage capacity is limited.
    
    In Fig. \ref{fig:8}, one can observe that the model downloading delay monotonically increases as the number of users grows. This is because a larger variety of model requests necessitates edge nodes to store more diverse PBs so as to meet QoS requirements, which limits the performance improvement of CoMP. Nonetheless, the reduction in model downloading delay achieved by MAASN-DA remains significant, with improvements of 13.06\%, 12.44\%, 25.38\%, 42.85\%, and 82.96\% over the five benchmark methods, respectively. More insightful, it is found that the proposed FGAMCD system achieves a two-fold performance gain compared to traditional systems, owing to the exploitation of parameter reusability. The \textit{caching efficiency gain} is realized by eliminating redundant caching of reusable PBs at edge nodes, while the \textit{downloading efficiency gain} is achieved by broadcasting overlapping PBs to simultaneously fulfill the requests of multiple users. In addition, fully unlocking this two-fold performance gain requires dedicated PB caching, migration, and beamforming design, which is effectively implemented by the proposed MAASN-DA.  
    
    \begin{figure}[t]
    	\begin{tabular}{cc}
    		\begin{minipage}[t]{0.48\linewidth}\centering
    			\includegraphics[width = 4.1cm]{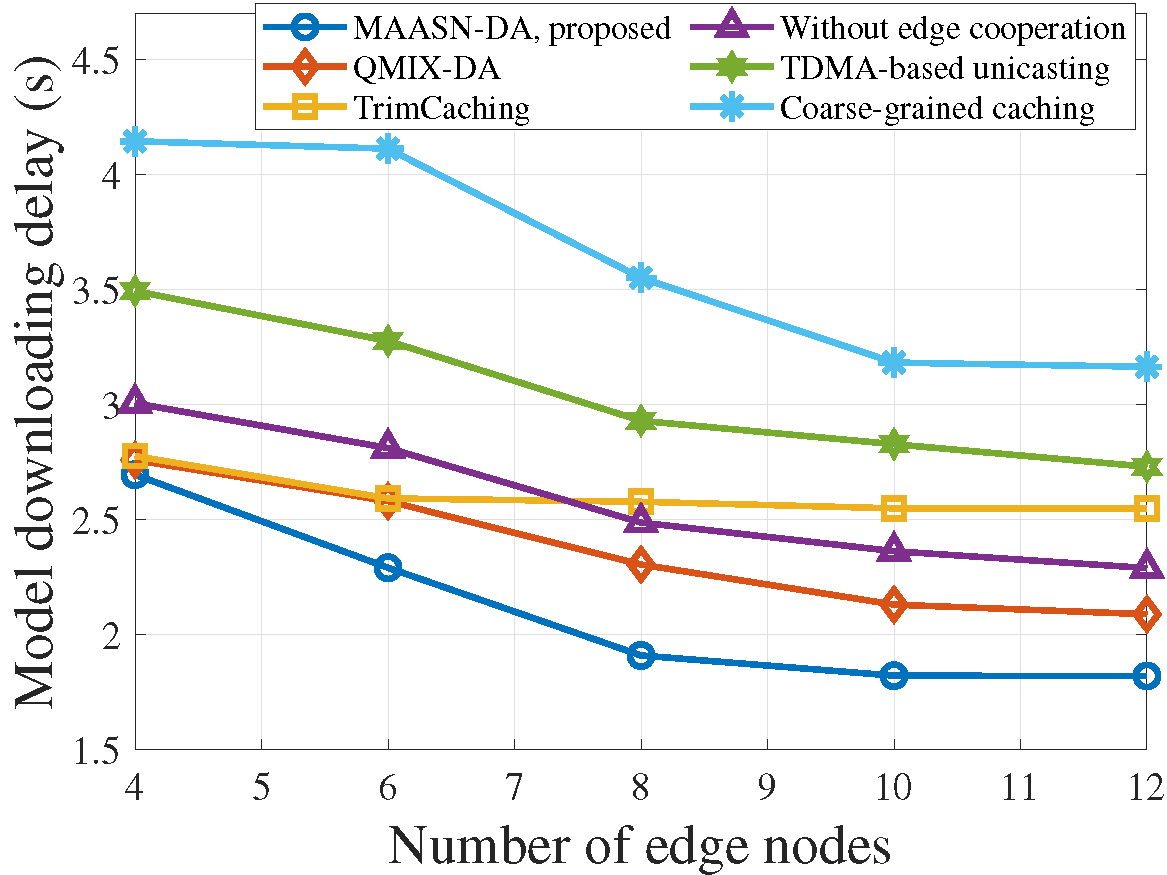}
    			\vspace{1mm}
    			\caption{Model downloading delay versus number of edge nodes. } 
    			\label{fig:edge_num}  
    		\end{minipage}
    		\begin{minipage}[t]{0.48\linewidth}\centering
    			\includegraphics[width = 4.1cm]{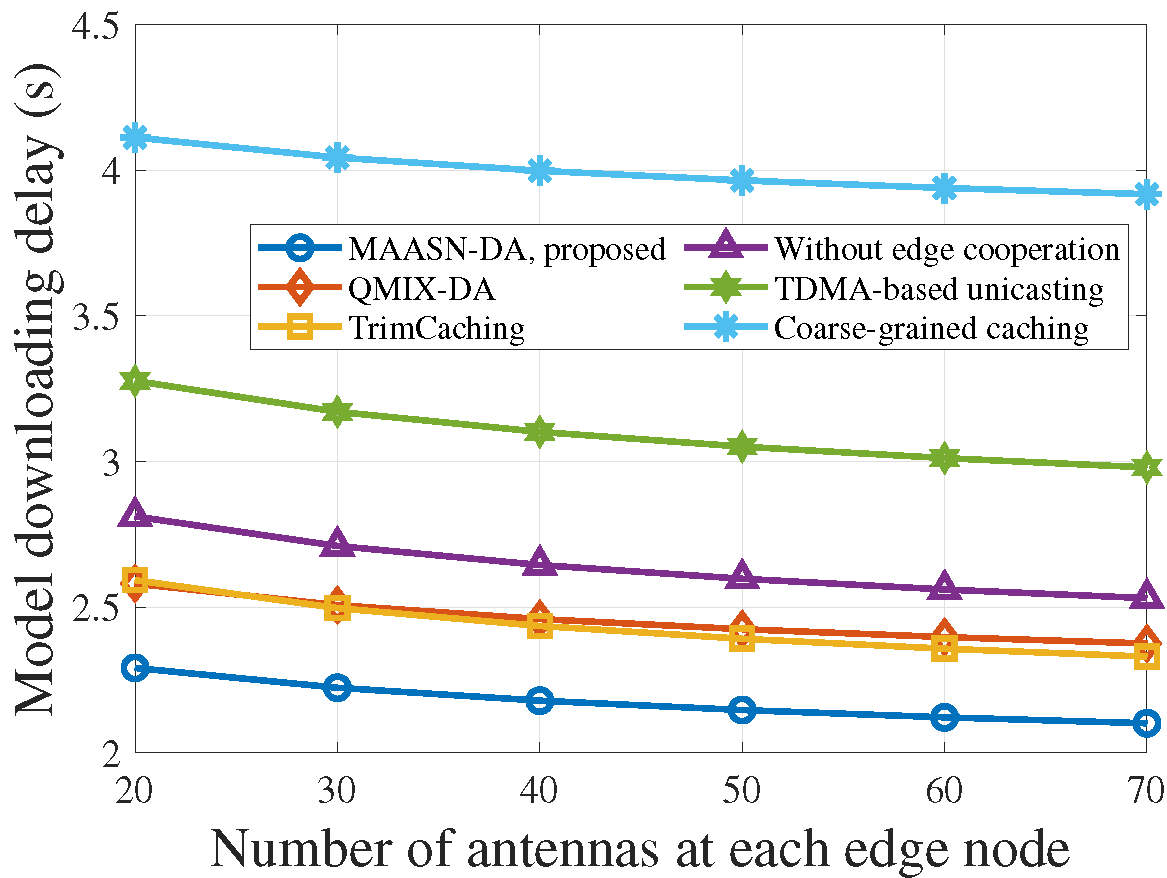}
    			\vspace{1mm}
    			\caption{Model downloading delay versus number of antennas.}  
    			\label{fig:edge_antennas}  
    		\end{minipage}
    		\vspace{0mm}	
    	\end{tabular}
    \end{figure}
    
    Figs. \ref{fig:edge_num}-\ref{fig:edge_antennas} evaluate the model downloading delay under different network scales. On the one hand, increasing the number of edge nodes $N$ enhances the caching diversity and CoMP gains, while the reduction in downloading delay gradually diminishes when $N$ becomes sufficiently large owing to the limitation of backhaul rates. On the other hand, equipping edge nodes with more antennas provides higher spatial multiplexing gains, thereby enhancing downlink rates for PB broadcasting. Furthermore, our MAASN-DA consistently outperforms the baseline methods under large-scale network settings. 
    
    \begin{figure}[t]
    	\begin{tabular}{cc}
    		\begin{minipage}[t]{0.48\linewidth}\centering
    			\includegraphics[width = 4.1cm]{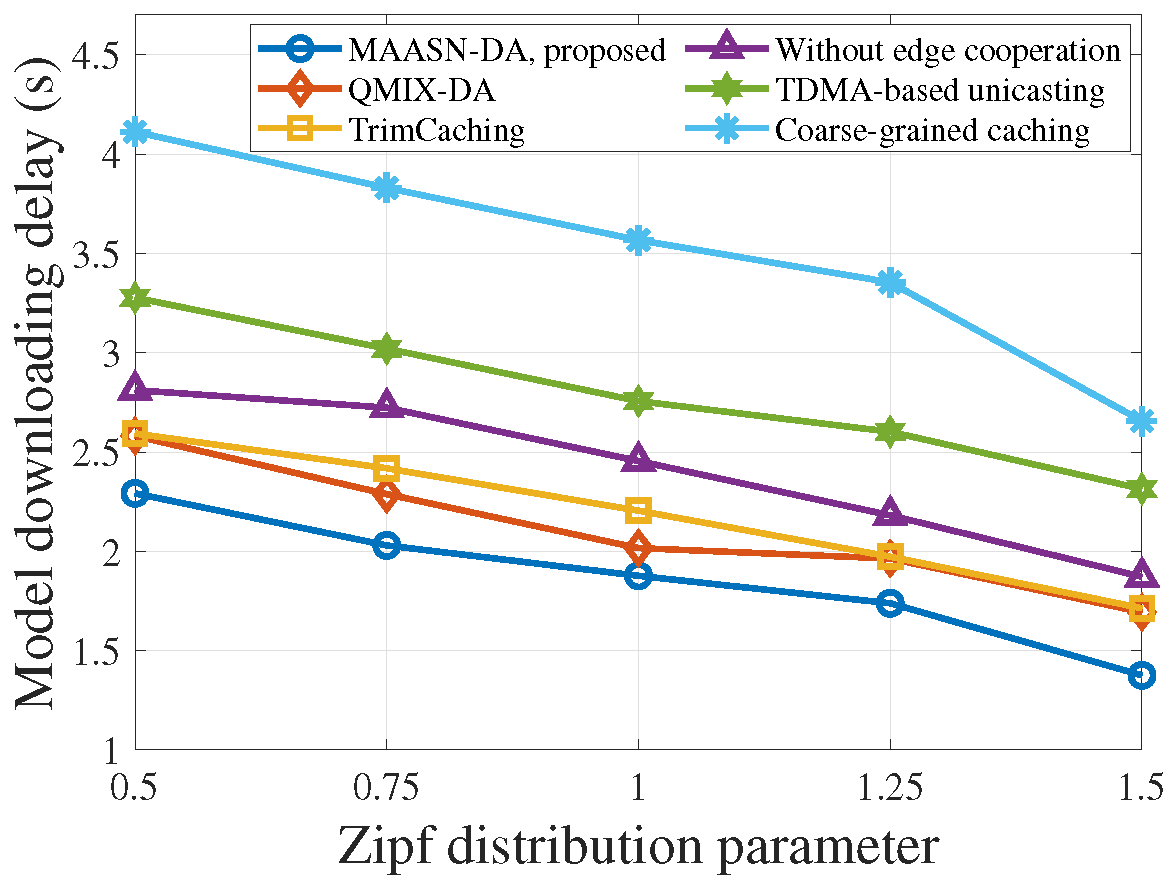}
    			\vspace{1mm}
    			\caption{Model downloading delay versus Zipf distribution parameter. }
    			\label{fig:Zipf}  
    		\end{minipage}
    		\begin{minipage}[t]{0.48\linewidth}\centering
    			\includegraphics[width = 4.1cm]{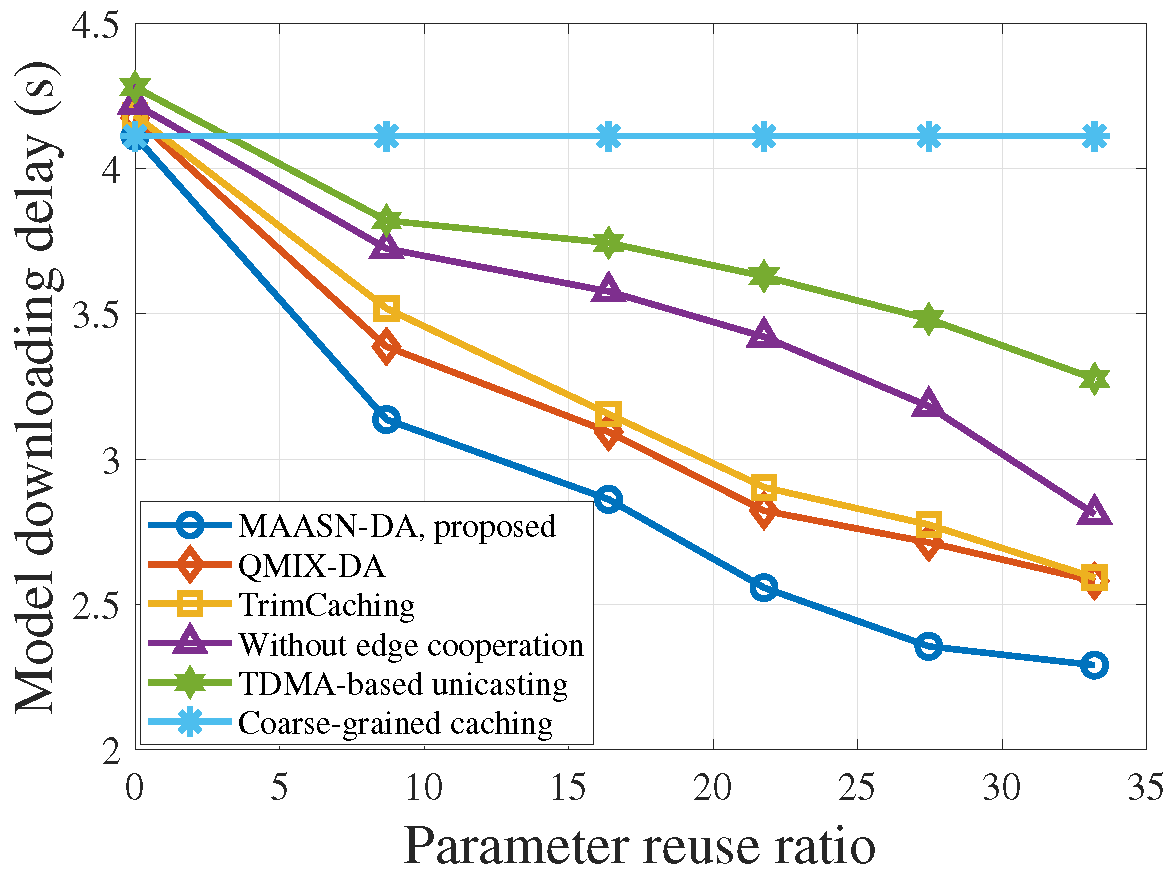}
    			\vspace{1mm}
    			\caption{Model downloading delay versus parameter reuse ratio.}
    			\label{fig:reuse_ratio}  
    		\end{minipage}
    		\vspace{0mm}	
    	\end{tabular}
    \end{figure}
    
    Fig. \ref{fig:Zipf} presents a sensitivity analysis with different Zipf distribution parameters $\iota$, showing that the model downloading delay decreases as $\iota$ increases. The rationale is that a larger $\iota$ concentrates user requests on hotspot AI models, enabling the PBs constituting these models to be cooperatively cached across multiple edge nodes, thereby facilitating CoMP. This result confirms the robustness of MAASN-DA in adapting to diverse user request patterns.
    
    Fig. \ref{fig:reuse_ratio} illustrates the delay performance of different schemes under varying parameter reuse ratios across AI models. Notably, the proposed FGAMCD framework inherently accommodates different reuse ratios by adjusting the input information $\{\mathcal{K},\mathcal{K}_j:\forall j\}$, without requiring any modification to the decision-making process. As expected, low parameter reuse leads to higher model downloading delay due to reduced caching and broadcasting gains, while MAASN-DA degenerates to \textit{coarse-grained caching} when the reuse ratio is zero. Moreover, our approach performs well even under small reuse ratios. For instance, when the reuse ratio is 8.70\%, the downloading delay is reduced by 8.00\%, 12.19\%, 18.75\%, 21.83\%, and 31.09\% compared with the five benchmark schemes. 
    
    \begin{figure}[t]
    	\begin{tabular}{cc}
    		\begin{minipage}[t]{0.48\linewidth}\centering
    			\includegraphics[width = 4.1cm]{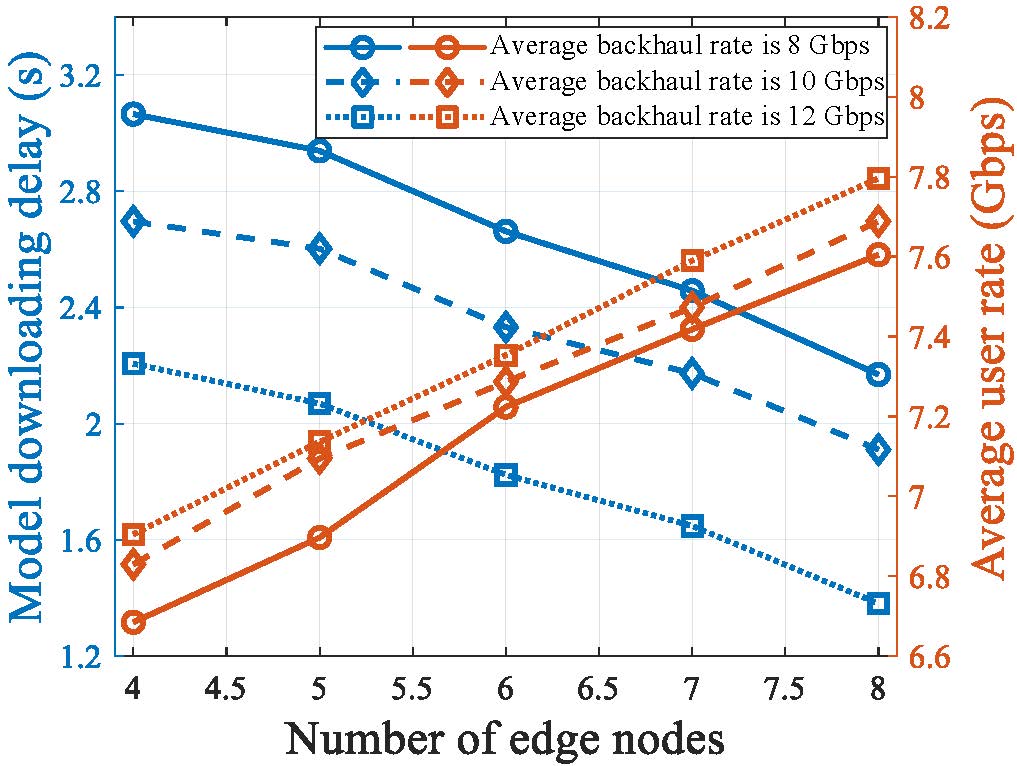}
    			\vspace{1mm}
    			\caption{Impact of number of edge nodes and average backhaul rate.}  
    			\label{fig:9}  
    		\end{minipage}
    		\begin{minipage}[t]{0.48\linewidth}\centering
    			\includegraphics[width = 4.1cm]{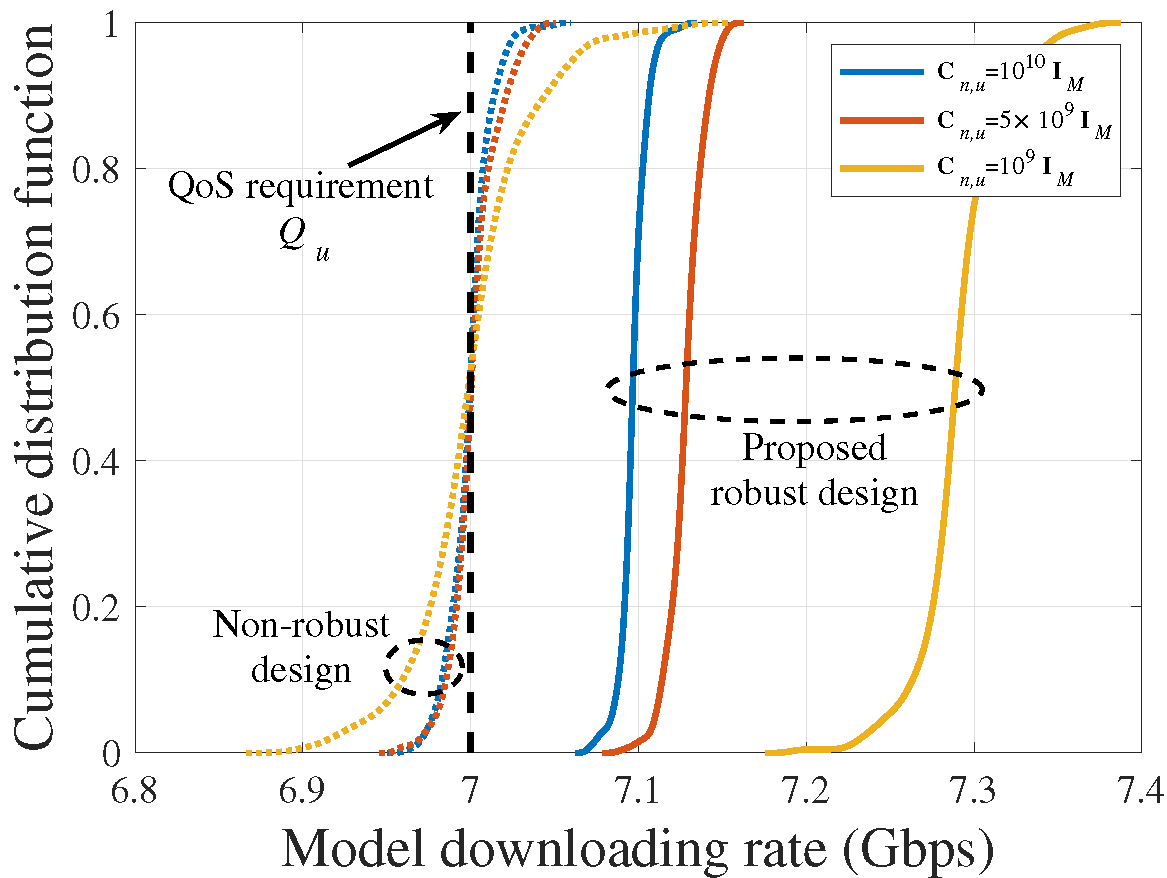}
    			\vspace{1mm}
    			\caption{Performance of the proposed robust CoMP beamforming design. } 
    			\label{fig:10}  
    		\end{minipage}
    		\vspace{0mm}	
    	\end{tabular}
    \end{figure}
    
    Fig. \ref{fig:9} delineates the impact of number of edge nodes and average backhaul rate on the system performance. As we observe, an increase in both the number of edge nodes and the backhaul rate enhances PB migration and CoMP broadcasting, thereby improving the average user rate and reducing model downloading delay. This result further validates the scalability of the proposed MAASN-DA approach, corroborating its ability to learn the cooperative caching and migration policies, while adapting to varying environmental factors such as the number of edge nodes and the backhaul link conditions.
    
    \begin{figure}[t] \centering
    	\subfigure[$n=2$, $k=40$.] { 
    		{\includegraphics[width=4.1cm]{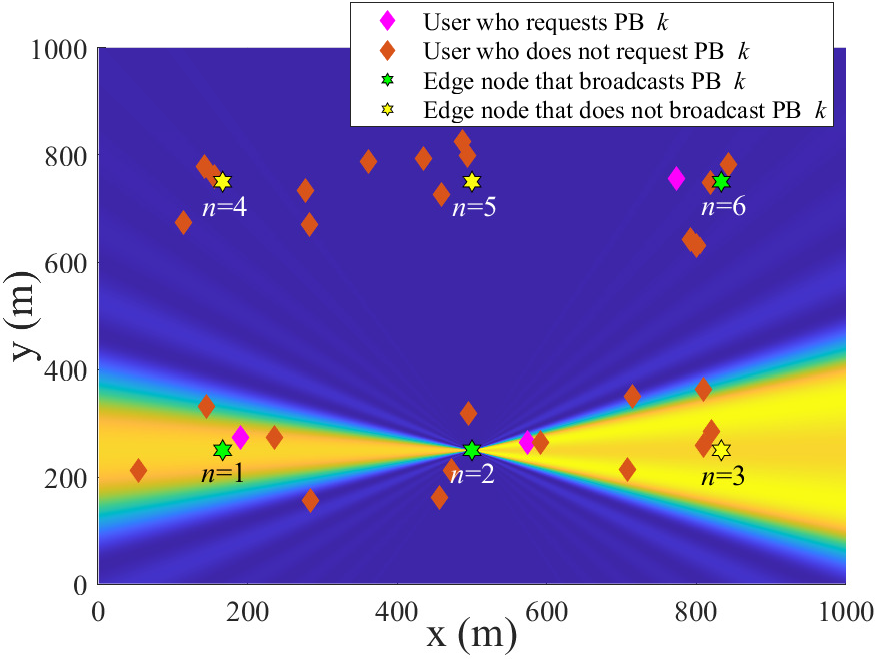}} 
    	}     
    	\subfigure[$n=5$, $k=370$.] { 
    		{\includegraphics[width=4.1cm]{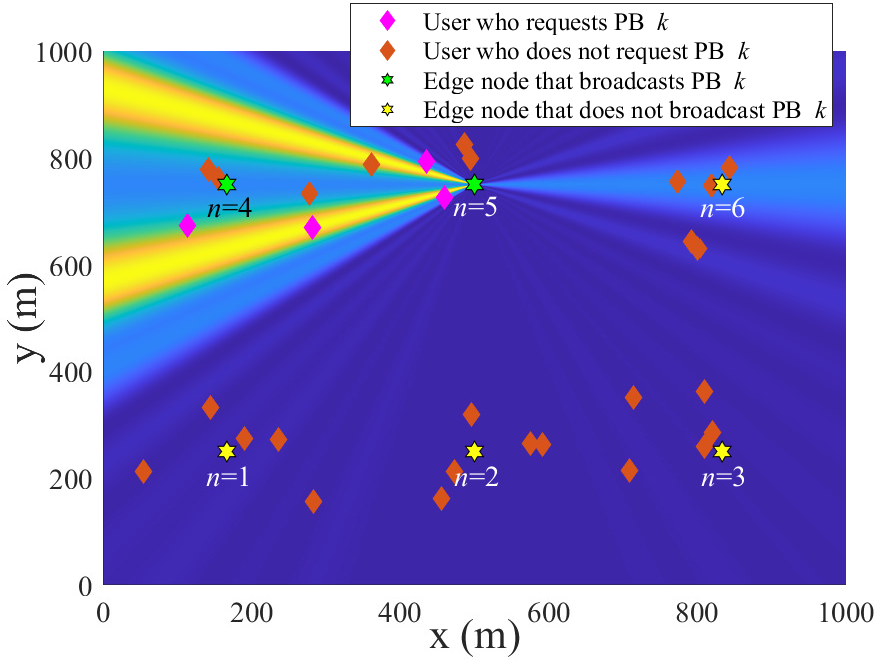}} 
    	}
    	\vspace{0mm}
    	\caption{Beampatterns of edge nodes for broadcasting different PBs.}     
    	\label{fig:11}     
    \end{figure}

    In Fig. \ref{fig:10}, we plot the cumulative distribution function (CDF) of model downloading rates across 200 channel error realizations. To emphasize the advantages of the proposed robust beamforming approach, we compare it with a non-robust design, in which the QoS constraint in (\ref{eq:9b}) is simplified as ${{\tilde{R}}_{u}}\left( k \right)\ge \mathbb{I}\left\{ k\in {{\mathcal{K}}_{{{r}_{u}}}} \right\}{{Q}_{u}},\text{ }\forall k,u$, with ${{\tilde{R}}_{u}}\left( k \right)$ being computed based on the estimated CSI. As shown in this figure, the downloading rates achieved by our robust design consistently exceed ${{Q}_{u}}$ under different channel error settings ${{\mathbf{C}}_{n,u}}$. In contrast, the non-robust design may violate the QoS constraints due to the fluctuations in channel errors. This highlights the critical importance of robust beamforming in ensuring QoS requirements during AI model downloading, particularly in practical edge networks with CSI uncertainty. 
    
    Fig. \ref{fig:11} portrays the beampatterns obtained by the proposed robust CoMP beamforming subroutine, where the beampattern of each edge node $n$ is calculated by ${{\left| {{\mathbf{z}}^{\text{H}}}\left( \bm{\theta}  \right){{\mathbf{w}}_{n}}\left( k \right) \right|}^{2}}$ with $\mathbf{z}\left( \bm{\theta}  \right)$ being the steering vector towards direction $\bm{\theta}$. When broadcasting PB $k=40$, we find that edge node $n=2$ not only serves its associated user but also radiates power towards the user associated with a neighboring edge node. This is owing to the edge nodes share the PB requested by users, hence CoMP is executed to boost the downloading rate. Analogously, two edge nodes participate in the broadcasting of PB $k=370$, and the beampattern is appropriately steered towards the requesting users. 
    
    \begin{figure}[t]\centering
    	\includegraphics[width = 3.9cm]{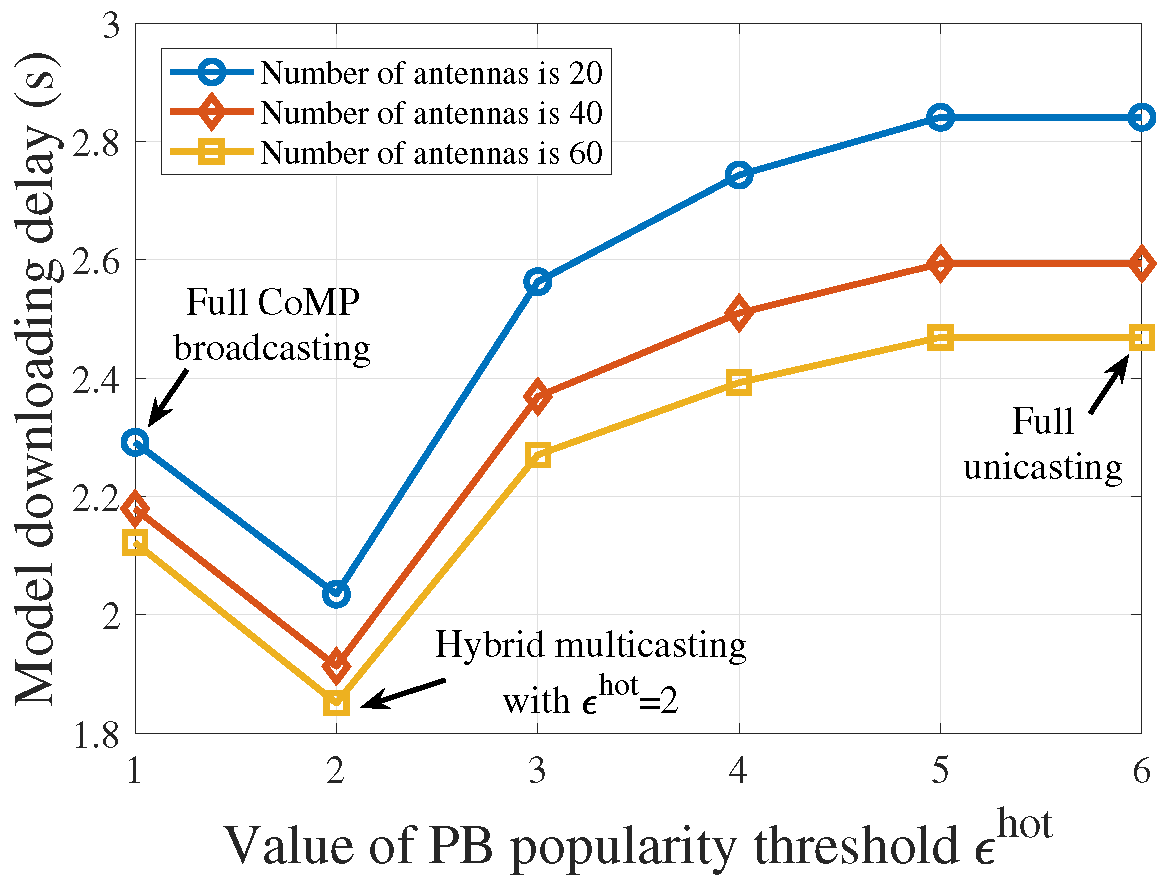}
    	\vspace{1mm}
    	\caption{Integration of FGAMCD with hybrid multicasting.}  
    	\label{fig:hybrid_multicast}   
    \end{figure}
    
    In Fig. \ref{fig:hybrid_multicast}, we examine the integration of FGAMCD with hybrid multicasting. To be specific, we introduce a PB popularity threshold $\epsilon^{\text{hot}}$. When the number of users requesting a PB exceeds $\epsilon^{\text{hot}}$, CoMP broadcasting across multiple edge nodes is employed; otherwise, the PB is delivered via unicasting from the associated or nearby edge node. As can be observed, for different edge antennas settings, hybrid multicasting with $\epsilon^{\text{hot}}=\text{2}$ achieves lower model downloading delay compared to full CoMP broadcasting. The reason is that the transmission mode is adaptively selected based on the popularity of each PB, thereby enhancing the flexibility of FGAMCD and highlighting a promising research direction for future exploration.
    
    \subsection{Extension to LLM Caching and Downloading}
    
    \begin{figure}[t] \centering
    	\subfigure[Varying edge storage capacity.] { 
    		{\includegraphics[width=4.1cm]{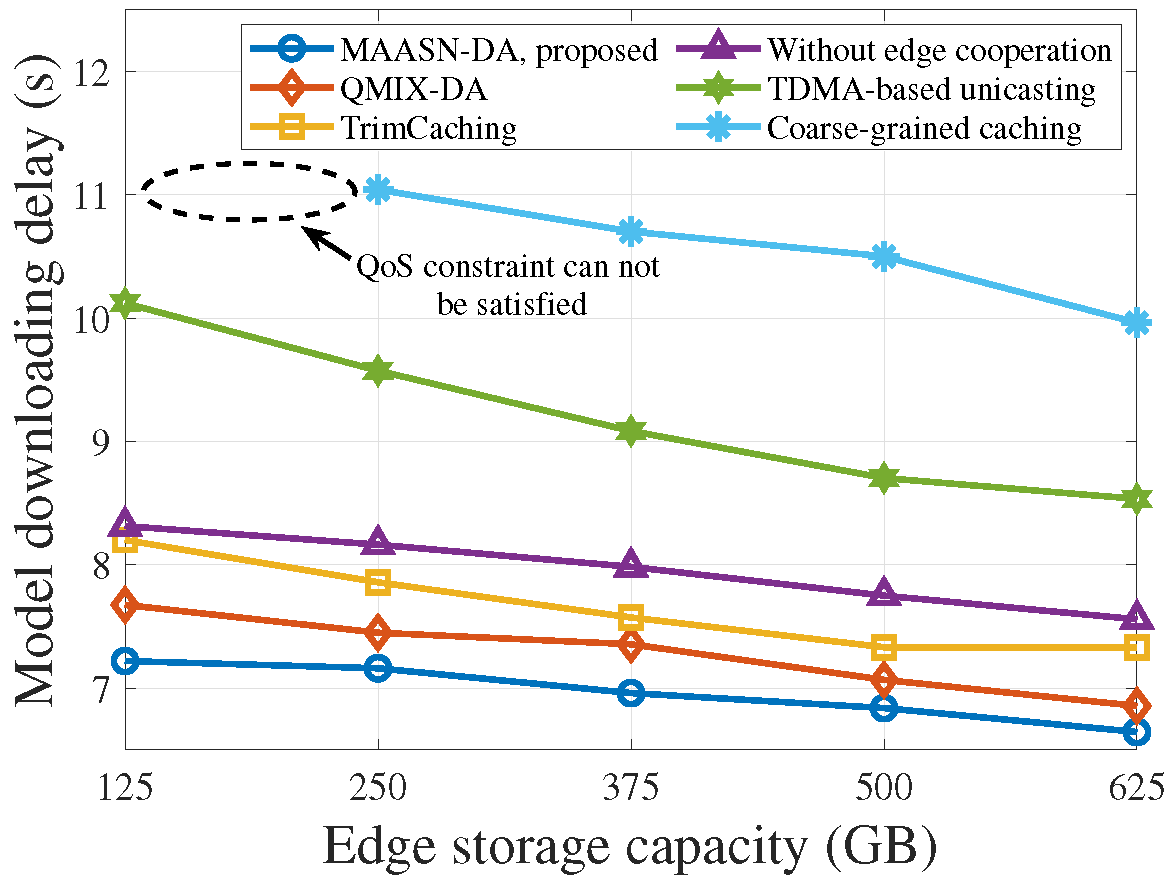}} 
    	}     
    	\subfigure[Varying number of edge nodes.] { 
    		{\includegraphics[width=4.1cm]{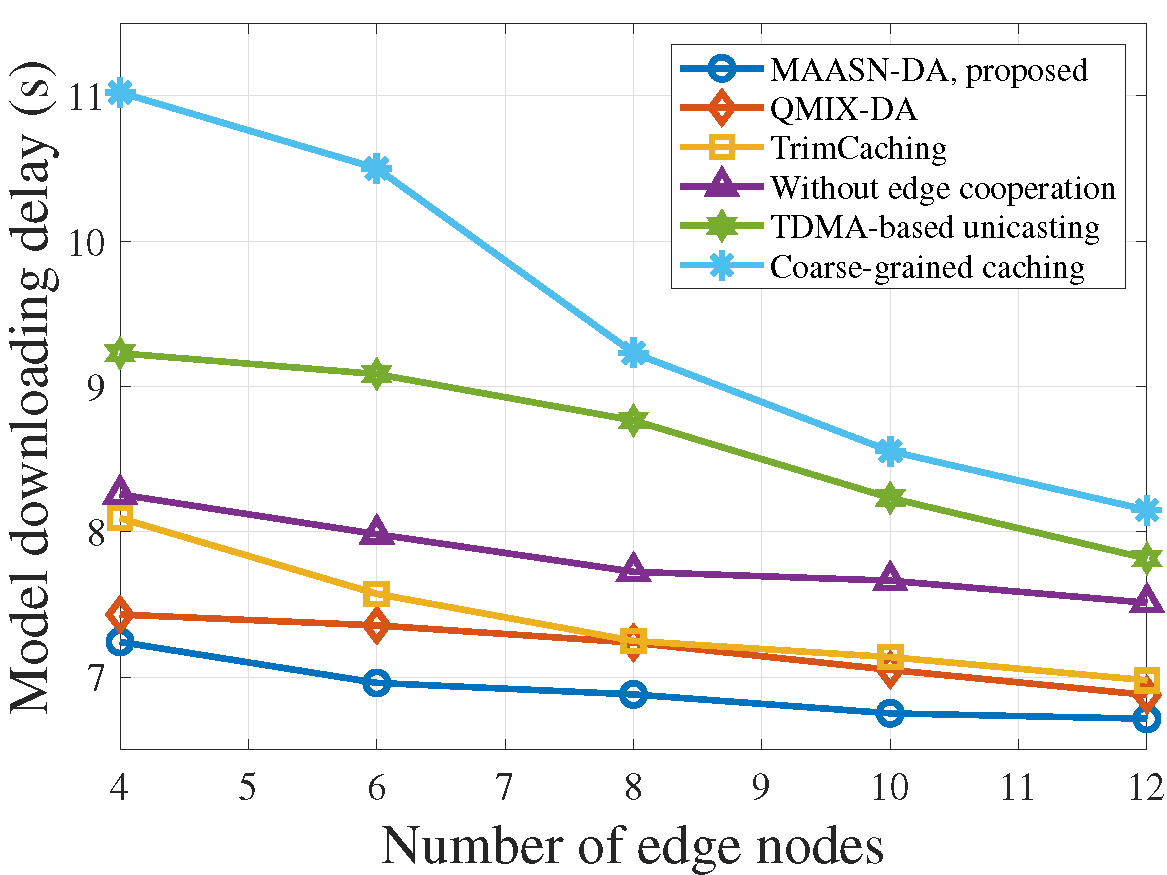}} 
    	}
    	\vspace{0mm}
    	\caption{Model downloading delay evaluation using fine-tuned Llama2-7B and Llama2-13B models.}     
    	\label{fig:LLM_test}     
    \end{figure}

    In this subsection, we extend the evaluation of MAASN-DA to billion-parameter LLMs. Specifically, we fine-tune $J=20$ LLMs based on Llama2-7B and Llama2-13B using the LLaMA-Factory platform \cite{zheng2024llamafactory}. Starting from the pre-trained models on HuggingFace, we freeze the embedding layers and a certain number of initial decoder layers, while fine-tuning the remaining decoders on different subsets of the LIMA \cite{zhou2023lima} and Dolly 15k \cite{databricks2023dolly15k} datasets. The perplexity (PPL) evaluation \cite{zheng2024llamafactory} provided by LLaMA-Factory is employed to measure the performance of the fine-tuned models. Experiments indicate that freezing 28 decoder layers (corresponding to reusable PBs) in Llama2-7B and 35 decoder layers in Llama2-13B leads to a PPL increase of less than 5, which is considered acceptable. To accommodate the incremental storage requirements of LLMs, we adjust the default parameter setting to $K=\text{285},C_n=\text{375 GB},B=\text{4}\times\text{10}^{\text{4}}\text{ MHz},R_{n,m}^{\text{bac}}\left( k \right)\in[\text{3.2}, \text{4.8}]\times\text{10}^{\text{3}}\text{ Gbps}$, thereby ensuring the feasibility of the solution.
    
    Under the LLM setup, Fig. \ref{fig:LLM_test} (a)-(b) illustrate the model downloading delays with respect to edge storage capacity and the number of edge nodes, respectively, which follow consistent trends with those observed in Figs. \ref{fig:7} and \ref{fig:edge_num}. Numerical results demonstrate that the proposed MAASN-DA outperforms \textit{QMIX-DA}, \textit{TrimCaching}, \textit{without edge cooperation}, \textit{TDMA-based unicasting}, and \textit{coarse-grained caching} by 5.68\%, 8.80\%, 14.7\%, 30.52\%, and 50.86\%, respectively, when $N=\text{6}$ and $C_n=\text{375 GB}$. Owing to the pronounced parameter reusability of LLMs, these results corroborate the applicability of MAASN-DA in improving the caching and downloading efficiency for edge LLM deployment.
    
	\section{Conclusion} \label{sec:conclusion}	
	
	In this work, we proposed the FGAMCD system to achieve adaptive and low-latency model downloading for on-device AI inference. To unleash the full potential of this system, PB caching, migration, and broadcasting beamforming were jointly optimized to minimize the total model downloading delay, satisfying user QoS requirements, edge storage capacity and transmission power constraints. We developed MAASN-DA, an improved MADRL framework, to enable this joint optimization in a distributed manner. Specifically, MAASN-DA facilitated cooperative decision-making for PB caching and migration via action semantics actor as well as value decomposition critic networks. Moreover, it leveraged an ESN to generate synthetic training samples, thereby enriching the replay buffer and accelerating policy training. We also designed a robust CoMP beamforming subroutine to optimize broadcasting beamformers under CSI uncertainty, which was integrated into the reward calculation to complete the MAASN-DA training loop. Our theoretical analysis showed the convergence of both actor and critic networks, while providing guidelines for configuring learning hyperparameter. In our simulations, the ablation study confirmed the superior learning performance of the MAASN-DA compared to existing MADRL baselines. Furthermore, the FGAMCD system achieved a two-fold performance gain over conventional coarse-grained caching and TDMA-based unicasting schemes, effectively reducing model downloading delay by 29.74\% to 67.86\%. 
	
	\begin{appendices} 
		\section{Proof of Theorem 1}
		
		According to Theorem 4.5 in \cite{34}, when the training process only incorporates real samples, the approximation error between ${{Q}_{{{\pi }_{E}}}}$ and ${{Q}^{*}}$ is upper-bounded by 
		\begin{align}
			{{\mathbb{E}}_{\Xi }}\left[ \left| {{Q}_{{{\pi }_{E}}}}-{{Q}^{*}} \right| \right]\le \frac{2{{\vartheta }_{\Xi ,\Omega }}\gamma }{{{\left( 1-\gamma  \right)}^{2}}}\sqrt{{{\nu }^{\max }}}+\frac{4{{\gamma }^{E+1}}}{{{\left( 1-\gamma  \right)}^{2}}}{{B}_{r}}, \label{eq:38}
		\end{align}
		where $\Xi $ signifies a fixed distribution of the Dec-POMDP in Section III-B, ${{\vartheta }_{\Xi ,\Omega }}$ denotes the concentration coefficient determined by $\Xi $ and $\Omega $. ${{\nu }^{\max }}=\underset{e\in \mathcal{E}}{\mathop{\max }}\,{{\mathbb{E}}_{\Omega }}\left[ \left| {{{\tilde{Q}}}_{e}}-T{{{\tilde{Q}}}_{e-1}} \right| \right]$ represents the maximum one-step error, ${{\tilde{Q}}_{e}}$ is the Q-value in the $e$-th episode, $T$ is the Bellman optimality operator. Since our work adopts the ESN to generate synthetic samples, we define the optimal Q-value with respect to both the real and synthetic samples as ${{\bar{Q}}^{*}}$, then we modify (\ref{eq:38}) into  
		\begin{align}
			&{{\mathbb{E}}_{\Xi }}\left[ \left| {{Q}_{{{\pi }_{E}}}}-{{Q}^{*}} \right| \right]={{\mathbb{E}}_{\Xi }}\left[ \left| {{Q}_{{{\pi }_{E}}}}-{{{\bar{Q}}}^{*}}+{{{\bar{Q}}}^{*}}-{{Q}^{*}} \right| \right]\nonumber\\
			&\ \le {{\mathbb{E}}_{\Xi }}\left[ \left| {{Q}_{{{\pi }_{E}}}}-{{{\bar{Q}}}^{*}} \right| \right]+{{\mathbb{E}}_{\Xi }}\left[ \left| {{{\bar{Q}}}^{*}}-{{Q}^{*}} \right| \right]\nonumber \\ 
			&\  \le \frac{2{{\vartheta }_{\Xi ,\Omega }}\gamma }{{{\left( 1-\gamma  \right)}^{2}}}{{\nu }^{\max }}+\frac{4{{\gamma }^{E+1}}}{{{\left( 1-\gamma  \right)}^{2}}}{{B}_{r}}+{{\mathbb{E}}_{\Xi }}\left[ \left| {{{\bar{Q}}}^{*}}-{{Q}^{*}} \right| \right].\label{eq:39}
		\end{align}
		
		Thereafter, we further express $\left| {{{\bar{Q}}}^{*}}-{{Q}^{*}} \right|$ in (\ref{eq:40}),
		\begin{figure*}[t]
		\begin{align}
			&\left| {{{\bar{Q}}}^{*}}-{{Q}^{*}} \right|\overset{\left( a \right)}{\mathop{=}}\,\left| \bar{r}\left( k \right)-r\left( k \right)+\gamma \left[ Q\left( \mathbf{\bar{s}}\left( k+1 \right),\mathbf{\bar{d}}\left( k+1 \right) \right)-Q\left( \mathbf{s}\left( k+1 \right),\mathbf{d}\left( k+1 \right) \right) \right] \right| \nonumber\\ 
			&\le \left| \bar{r}\left( k \right)-r\left( k \right) \right|+\gamma \left| Q\left( \mathbf{\bar{s}}\left( k+1 \right),\mathbf{d}\left( k+1 \right) \right)-Q\left( \mathbf{s}\left( k+1 \right),\mathbf{d}\left( k+1 \right) \right) \right| \nonumber\\ 
			&=\left| \bar{r}\left( k \right)-r\left( k \right) \right|+\gamma \left| {{\bm{\omega} }_{\text{out}}}\left[ \tanh \left( {{\bm{\omega} }_{\text{in}}}\mathbf{\bar{v}}\left( k+1 \right)+{{\bm{\omega} }_{\text{re}}}\mathbf{p}\left( k \right) \right)-\tanh \left( {{\bm{\omega} }_{\text{in}}}\mathbf{v}\left( k+1 \right)+{{\bm{\omega} }_{\text{re}}}\mathbf{p}\left( k \right) \right) \right] \right| \nonumber\\ 
			&\le \left| \bar{r}\left( k \right)-r\left( k \right) \right|+\gamma \left| {{\bm{\omega} }_{\text{out}}}\left[ {{\bm{\omega} }_{\text{in}}}\left( \mathbf{\bar{s}}\left( k+1 \right)-\mathbf{s}\left( k+1 \right) \right) \right] \right| \nonumber\\ 
			&\le \left| \bar{r}\left( k \right)-r\left( k \right) \right|+\gamma \phi _{\text{out}}^{\max }\phi _{\text{in}}^{\max }\left\| \mathbf{\bar{s}}\left( k+1 \right)-\mathbf{s}\left( k+1 \right) \right\| \overset{\left( b \right)}{\mathop{\le }}\,\xi \left( 1+\gamma \phi _{\text{out}}^{\max }\phi _{\text{in}}^{\max } \right)
			\label{eq:40}
		\end{align}
     	\end{figure*}
		where $\left( a \right)$ follows the Bellman equation. $\mathbf{\bar{d}}\left( k+1 \right)$ is the action output by the actor under $\mathbf{\bar{s}}\left( k+1 \right)$, whose Q-value is higher than that of taking $\mathbf{d}\left( k+1 \right)$ under $\mathbf{\bar{s}}\left( k+1 \right)$. $\phi _{\text{in}}^{\max }$ and $\phi _{\text{out}}^{\max }$ are the bounds of $\left\| {{\bm{\omega} }_{\text{in}}} \right\|$ and $\left\| {{\bm{\omega} }_{\text{out}}} \right\|$, respectively. $\left( b \right)$ is owing to the data quality selection in (\ref{eq:17}). By substituting (\ref{eq:40}) into (\ref{eq:39}), the error bound is given by
		\begin{align}
			&{{\mathbb{E}}_{\Xi }}\left[ \left| {{Q}_{{{\pi }_{E}}}}-{{Q}^{*}} \right| \right]\nonumber\\
			&\le\!{{\vartheta }_{\Xi ,\Omega }}\bigg[ \frac{2\gamma }{{{\left(1\!-\!\gamma  \right)}^{2}}}\!\sqrt{{{\nu }^{\max }}}\!+\!\xi \left( 1\!+\!\gamma \phi _{\text{out}}^{\max }\phi _{\text{in}}^{\max } \right) \bigg]\!+\!\frac{4{{\gamma }^{E+1}}}{{{\left( 1\!-\!\gamma  \right)}^{2}}}{{B}_{r}}.  \label{eq:41}
		\end{align}
		
		To proceed with, we derive an upper-bound of ${{\nu }^{\max }}$. Specifically, Theorem 2 in \cite{30} is extended as follows
		\begin{align}
			{{\nu }^{\max }}\le &4\max {{\left\{ {{{B}_{r}}}/{\left( 1-\gamma  \right)}\;-\varsigma L_{\varsigma }^{\text{DRQN}},0 \right\}}^{2}}\nonumber\\
			&+{{{D}_{1}}{{V}^{2}}\ln U_{\varsigma }^{\text{DRQN}}}/{{{K}'}}\;+{{D}_{2}}{{V}^{2}}\varsigma , \label{eq:42}
		\end{align}
		where $\varsigma $ is a positive constant, ${K}'$ denotes the number of training samples in each episode. ${{D}_{1}}=8\sqrt{2{K}'}+{256}/{V}$, ${{D}_{2}}=4\sqrt{2{K}'}+52$, $V={{{B}_{r}}}/{\left( 1-\gamma  \right)}$. $L_{\varsigma }^{\text{DRQN}}$ and $U_{\varsigma }^{\text{DRQN}}$ are the upper- and lower-bounds of the exterior $\varsigma$-number of the DRQN, respectively, whose specific expressions can be seen in Appendix B of \cite{30}.
		
		Our subsequent aim is to characterize ${K}'$ by taking into account both real and synthetic samples. For tractable analysis, we only consider the effect of the maximum initial number of synthetic samples ${{\tau }_{0}}K$ and neglect its decay with training episodes. Therefore, ${K}'$ is given by
		\begin{align}
			&{K}'\!=\! K\! +\! {{\tau }_{0}}K\Pr \left\{ \left\| \left( \bar{r}\left( k \right),\mathbf{\bar{s}}\left( k\! +\! 1 \right) \right)\!-\!\left( r\left( k \right),\mathbf{s}\left( k\! +\!1 \right) \right) \right\| \!\le\! \xi  \right\} \nonumber\\ 
			& \overset{\left( a \right)}{\mathop{\ge }}K\!+\!{{\tau }_{0}}K\left( 1\!-\!{\mathbb{E}\left[ \left\| \left( \bar{r}\left( k \right),\mathbf{\bar{s}}\left( k\!+\! 1 \right) \right)\!-\! \left( r\left( k \right),\mathbf{s}\left( k\!+\! 1 \right) \right) \right\| \right]}/{\xi }\right) \nonumber\\ 
			&  \ge K\big( 1+{{\tau }_{0}}-\frac{{{\tau }_{0}}}{\xi }\{ \mathbb{E}\left[ \left\| \left( \bar{r}\left( k \right),\mathbf{\bar{s}}\left( k+1 \right) \right) \right\| \right]\nonumber\\
			&\qquad\qquad\qquad\qquad\qquad+\mathbb{E}\left[ \left\| \left( r\left( k \right),\mathbf{s}\left( k+1 \right) \right) \right\| \right] \} \big), \label{eq:43}
		\end{align}
		where $\left( a \right)$ is due to the Markov inequality. Since $\left( \bar{r}\left( k \right),\mathbf{\bar{s}}\left( k+1 \right) \right)$ is output by the ESN, we can bound it as below
		\begin{align}
			\left\| \mathbf{q}\left( k \right) \right\|&=\!\left\| \tanh \left( {{\eta }_{\text{in}}}\mathbf{v}\left( k \right)\!+\!{{\eta }_{\text{re}}}\mathbf{q}\left( k\!-\! 1 \right) \right) \right\|\nonumber\\
			& \le\! \left\| {{\eta }_{\text{in}}}\mathbf{v}\left( k \right)\!+\!{{\eta }_{\text{re}}}\mathbf{q}\left( k\!-\! 1 \right) \right\| \nonumber\\ 
			&\le \left\| {{\eta }_{\text{in}}} \right\|\left\| \mathbf{v}\left( k \right) \right\|+\left\| {{\eta }_{\text{re}}} \right\|\left\| \mathbf{q}\left( k-1 \right)\right\|\nonumber\\
			&\overset{\left( a \right)}{\mathop{\le }}\,\psi _{\text{in}}^{\max }\sqrt{B_{\mathbf{s}}^{2}+B_{\mathbf{d}}^{2}}+\psi _{\text{re}}^{\max }\left\| \mathbf{q}\left( k-1 \right) \right\|, \label{eq:44}
		\end{align}
		where $\left( a \right)$ follows Assumption 1. By calculating the sum of geometric series with $\left\| \mathbf{q}\left( 0 \right) \right\|=0$, we have
		\begin{align}
			\left\| \mathbf{q}\left( k \right) \right\|&\le \psi _{\text{in}}^{\max }\sqrt{{{B}_{\mathbf{s}}}+{{B}_{\mathbf{d}}}}\sum\limits_{i=0}^{k-1}{{{\left( \psi _{\text{re}}^{\max } \right)}^{i}}}\nonumber\\
			&=\psi _{\text{in}}^{\max }\sqrt{B_{\mathbf{s}}^{2}+B_{\mathbf{d}}^{2}}\frac{1-{{\left( \psi _{\text{re}}^{\max } \right)}^{k}}}{1-\psi _{\text{re}}^{\max }}. \label{eq:45}
		\end{align}
		Thus, the output of the ESN is bounded by
		\begin{align}
			&\left\| \left( \bar{r}\left( k \right),\mathbf{\bar{s}}\left( k+1 \right) \right) \right\|\le \left\| {{\eta }_{\text{out}}}\mathbf{q}\left( k \right) \right\|\le \left\| {{\eta }_{\text{out}}}\mathbf{q}\left( K \right) \right\|\nonumber\\
			&\qquad\le \psi _{\text{out}}^{\max }\psi _{\text{in}}^{\max }\sqrt{B_{\mathbf{s}}^{2}+B_{\mathbf{d}}^{2}}\frac{1-{{\left( \psi _{\text{re}}^{\max } \right)}^{K}}}{1-\psi _{\text{re}}^{\max }}. \label{eq:46}
		\end{align}
		Besides, we can obtain from Assumption 1 that $\left\| \left( r\left( k \right),\mathbf{s}\left( k+1 \right) \right) \right\|\le \sqrt{B_{r}^{2}+B_{\mathbf{s}}^{2}}$, then ${K}'$ is further derived in (\ref{eq:47}).
		\begin{figure*}
		\begin{align}
			{K}'\ge K\left( 1+{{\tau }_{0}}-\frac{{{\tau }_{0}}}{\xi }\left\{ \psi _{\text{out}}^{\max }\psi _{\text{in}}^{\max }\sqrt{B_{\mathbf{s}}^{2}+B_{\mathbf{d}}^{2}}\frac{1-{{\left( \psi _{\text{re}}^{\max } \right)}^{K}}}{1-\psi _{\text{re}}^{\max }}+\sqrt{B_{r}^{2}+B_{\mathbf{s}}^{2}} \right\} \right). \label{eq:47}
		\end{align}
	    \end{figure*}
		Plugging (\ref{eq:47}) into (\ref{eq:42}), and (\ref{eq:42}) into (\ref{eq:41}), we can derive Theorem 1. This completes the proof.
		
		\section{Proof of Theorem 2}
		
		Since the CoMP beamforming subroutine is executed in each time step $k$ of \textbf{Algorithm 1}, we first prove that the beamforming problem \textbf{P2}$\left( k \right)$ can be solved within a finite number of iterations. Denote $f\left( \mathbf{W} \right)$ and ${{f}^{\text{up}}}\left( \mathbf{W} \right)$ as the objective value in (\ref{eq:32}) and (\ref{eq:33}) for a feasible solution $\mathbf{W}$, respectively. Given $\mathbf{\bar{W}}\left( k \right)$ from the previous iteration, by solving the \textbf{P2.2}$\left( k \right)$, we have
		\begin{align}
			f\left( \mathbf{\bar{W}}\left( k \right) \right)\overset{\left( a \right)}{\mathop{=}}\,{{f}^{\text{up}}}\left( \mathbf{\bar{W}}\left( k \right) \right)\overset{\left( b \right)}{\mathop{\ge }}\,{{f}^{\text{up}}}\left( \mathbf{W}\left( k \right) \right)\overset{\left( c \right)}{\mathop{\ge }}\,f\left( \mathbf{W}\left( k \right) \right), \label{eq:48}
		\end{align}
		where $\left( a \right)$ holds owing to the approximation in (\ref{eq:20}) is tight with given point $\mathbf{\bar{W}}\left( k \right)$, $\left( b \right)$ is attributed to that \textbf{P2.2}$\left( k \right)$ is addressed optimally with solution $\mathbf{W}\left( k \right)$, $\left( c \right)$ follows that ${{f}^{\text{up}}}$ is a upper bound of $f$ due to the linearization. (\ref{eq:48}) indicates that $f$ is non-increasing after each optimization. In addition, according to \cite{33}, when \textbf{P2.1}$\left( k \right)$ with regularization term is solved, the attained $\mathbf{W}\left( k \right)$ is a KKT point of \textbf{P2}$\left( k \right)$. This ensures that the rank-one constraint (\ref{eq:23f}) is met, yielding a converged beamforming decision. 
		
		Subsequently, we show the convergence of the action semantics actor network. Given the actor of each agent $\pi \left( {{\mathbf{o}}_{n}}\left( k \right);{{\bm{\varphi} }_{n}} \right)$, the joint policy can be expressed as 	$\pi =\prod\limits_{n\in \mathcal{N}}{\pi \left( {{\mathbf{o}}_{n}}\left( k \right);{{\bm{\varphi} }_{n}} \right)}$. After performing gradient descent on the actor loss in (\ref{eq:22}) to generate new policy ${{\pi }_{\text{new}}}$, we have the following inequality: 
		\begin{align}
			{{Q}_{{{\pi }_{\text{new}}}}}\!\left( \mathbf{s}\left( k \right)\!,\! \mathbf{d}\left( k \right) \right)\!\ge\! {{Q}_{\pi }}\!\left( \mathbf{s}\left( k \right)\!,\!\mathbf{d}\left( k \right) \right),\forall\!\left( \mathbf{s}\left( k \right),\!\mathbf{d}\left( k \right) \right)\!\in\! \mathcal{S}\!\times\! \mathcal{D}. \label{eq:49}
		\end{align}
		Let $E\to \infty $, we get ${{Q}_{{{\pi }_{\text{E}}}}}>{{Q}_{\pi }}$ for any ${{\pi }_{E}}\ne \pi $. According to Theorem 1, ${{Q}_{{{\pi }_{\text{E}}}}}$ can converge to ${{Q}^{*}}$ with an error bound, then the joint policy ${{\pi }_{E}}$ that maximize ${{Q}_{{{\pi }_{\text{E}}}}}$ is local optimal. Therefore, the minimum loss value characterized by (\ref{eq:36}) is achieved. This completes the proof.
		
		\section{Neural Network Architectures of MAASN-DA}
		
\begin{figure*}[t]\centering
			\includegraphics[width = 15cm]{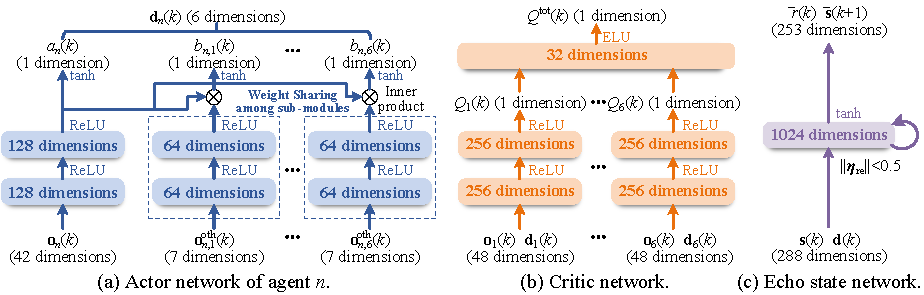}
			\vspace{1mm}
			\caption{Neural network architectures of MAASN-DA. }  
			\label{fig:19}   
		\end{figure*}

		With default parameter setting in Section V, the neural network architectures of MAASN-DA are shown in Fig. \ref{fig:19}. 
		
	\end{appendices}

	\bibliographystyle{IEEEtran}
	\bibliography{reference}

\begin{thebibliography}{10}
\providecommand{\url}[1]{#1}
\csname url@samestyle\endcsname
\providecommand{\newblock}{\relax}
\providecommand{\bibinfo}[2]{#2}
\providecommand{\BIBentrySTDinterwordspacing}{\spaceskip=0pt\relax}
\providecommand{\BIBentryALTinterwordstretchfactor}{4}
\providecommand{\BIBentryALTinterwordspacing}{\spaceskip=\fontdimen2\font plus
\BIBentryALTinterwordstretchfactor\fontdimen3\font minus
  \fontdimen4\font\relax}
\providecommand{\BIBforeignlanguage}[2]{{%
\expandafter\ifx\csname l@#1\endcsname\relax
\typeout{** WARNING: IEEEtran.bst: No hyphenation pattern has been}%
\typeout{** loaded for the language `#1'. Using the pattern for}%
\typeout{** the default language instead.}%
\else
\language=\csname l@#1\endcsname
\fi
#2}}
\providecommand{\BIBdecl}{\relax}
\BIBdecl

\bibitem{1}
Y.~Mao, X.~Yu, K.~Huang \emph{et~al.}, ``Green edge {AI}: {A} contemporary
  survey,'' \emph{Proc. IEEE}, vol. 112, no.~7, pp. 880--911, 2024.

\bibitem{2}
\BIBentryALTinterwordspacing
ITU-R, ``Framework and overall objectives of the future development of {IMT}
  for 2030 and beyond,'' Nov. 2023. [Online]. Available:
  \url{https://www.itu.int/en/ITU-R/study-groups/rsg5/rwp5d/imt-2030/Pages/default.aspx}
\BIBentrySTDinterwordspacing

\bibitem{3}
Y.~Fu, P.~Qin, J.~Zhang \emph{et~al.}, ``Joint {AI} inference and target
  tracking at network edge: {A} hybrid offline-online design for {UAV}-enabled
  network,'' \emph{IEEE Trans. Wireless Commun.}, vol.~23, no.~12, pp.
  17\,959--17\,973, 2024.

\bibitem{4}
M.~A. Rahman, ``A survey on security and privacy of multimodal {LLM}s -
  connected healthcare perspective,'' in \emph{Proc. IEEE Glob. Commun. Conf.
  Workshops (GC Wkshps)}, 2023, pp. 1807--1812.

\bibitem{5}
\BIBentryALTinterwordspacing
S.~Mehta, M.~H. Sekhavat, Q.~Cao \emph{et~al.}, ``Open{ELM}: {A}n efficient
  language model family with open training and inference framework,'' 2024.
  [Online]. Available: \url{https://arxiv.org/abs/2404.14619}
\BIBentrySTDinterwordspacing

\bibitem{6}
K.~Huang, H.~Wu, Z.~Liu \emph{et~al.}, ``In-situ model downloading to realize
  versatile edge {AI} in {6G} mobile networks,'' \emph{IEEE Wireless Commun.},
  vol.~30, no.~3, pp. 96--102, 2023.

\bibitem{7}
\BIBentryALTinterwordspacing
{3GPP, document TS 22.874}, ``Study on traffic characteristics and performance
  requirements for {AI/ML} model transfer,'' Jun. 2021. [Online]. Available:
  \url{https://portal.3gpp.org/desktopmodules/Specifications/SpecificationDetails.aspx?specificationId=3721}
\BIBentrySTDinterwordspacing

\bibitem{8}
\BIBentryALTinterwordspacing
{3GPP, document TS 22.261}, ``Service requirements for the {5G} system,'' Mar.
  2025. [Online]. Available:
  \url{https://portal.3gpp.org/ChangeRequests.aspx?q=1&versionId=92077&release=195}
\BIBentrySTDinterwordspacing

\bibitem{9}
G.~Qu, Q.~Chen, W.~Wei \emph{et~al.}, ``Mobile edge intelligence for large
  language models: {A} contemporary survey,'' \emph{IEEE Commun. Surveys
  Tuts.}, pp. 1--1, 2025.

\bibitem{10}
W.~Fan, Z.~Chen, Z.~Hao \emph{et~al.}, ``{DNN} deployment, task offloading, and
  resource allocation for joint task inference in {II}o{T},'' \emph{IEEE Trans
  Ind. Informat.}, vol.~19, no.~2, pp. 1634--1646, 2023.

\bibitem{11}
K.~Zhao, Z.~Zhou, X.~Chen \emph{et~al.}, ``Edge{A}daptor: {O}nline
  configuration adaption, model selection and resource provisioning for edge
  {DNN} inference serving at scale,'' \emph{IEEE Trans. Mobile Comput.},
  vol.~22, no.~10, pp. 5870--5886, 2023.

\bibitem{12}
P.~Qin, M.~Fu, Y.~Fu \emph{et~al.}, ``Collaborative edge computing and program
  caching with routing plan in {C-NOMA}-enabled space-air-ground network,''
  \emph{IEEE Trans. Wireless Commun.}, vol.~23, no.~12, pp. 18\,302--18\,315,
  2024.

\bibitem{13}
P.~Qin, Y.~Fu, J.~Zhang \emph{et~al.}, ``{DRL}-based resource allocation and
  trajectory planning for {NOMA}-enabled multi-{UAV} collaborative caching {6G}
  network,'' \emph{IEEE Trans. Veh. Technol.}, vol.~73, no.~6, pp. 8750--8764,
  2024.

\bibitem{Fan2025Satellite}
W.~Fan, Q.~Meng, G.~Wang \emph{et~al.}, ``Satellite edge intelligence:
  {DRL}-based resource management for task inference in {LEO}-based
  satellite-ground collaborative networks,'' \emph{IEEE Trans. Mobile Comput.},
  pp. 1--18, 2025.

\bibitem{14}
J.~Yan, S.~Bi, and Y.-J.~A. Zhang, ``Optimal model placement and online model
  splitting for device-edge co-inference,'' \emph{IEEE Trans. Wireless
  Commun.}, vol.~21, no.~10, pp. 8354--8367, 2022.

\bibitem{15}
S.~Tuli, G.~Casale, and N.~R. Jennings, ``Split{P}lace: {AI} augmented
  splitting and placement of large-scale neural networks in mobile edge
  environments,'' \emph{IEEE Trans. Mobile Comput.}, vol.~22, no.~9, pp.
  5539--5554, 2023.

\bibitem{16}
J.~Chen, X.~Liang, J.~Xue \emph{et~al.}, ``Evolution of {RAN} architectures
  toward {6G}: {M}otivation, development, and enabling technologies,''
  \emph{IEEE Commun. Surveys Tuts.}, vol.~26, no.~3, pp. 1950--1988, 2024.

\bibitem{17}
M.~Jafri, S.~Srivastava, N.~K.~D. Venkategowda \emph{et~al.}, ``Cooperative
  hybrid transmit beamforming in cell-free mmwave {MIMO} networks,'' \emph{IEEE
  Trans. Veh. Technol.}, vol.~72, no.~5, pp. 6023--6038, 2023.

\bibitem{18}
J.-M. Liang, S.~Mishra, and I.-C. Chien, ``Enhanced cell clustering and
  multicast scheduling for energy-efficient {5G/B5G} {MBSFN} networks,''
  \emph{IEEE Internet Things J.}, pp. 1--1, 2025.

\bibitem{19}
J.~Chen, K.~Zhai, Z.~Wang \emph{et~al.}, ``Co{MP} and {RIS}-assisted multicast
  transmission in a multi-{UAV} communication system,'' \emph{IEEE Trans.
  Commun.}, vol.~72, no.~6, pp. 3602--3617, 2024.

\bibitem{20}
N.~Babu, C.~Masouros, C.~B. Papadias \emph{et~al.}, ``Precoding for multi-cell
  {ISAC}: {F}rom coordinated beamforming to coordinated multipoint and
  bi-static sensing,'' \emph{IEEE Trans. Wireless Commun.}, vol.~23, no.~10,
  pp. 14\,637--14\,651, 2024.

\bibitem{Xie2025Mixture}
G.~Xie, Z.~Xiong, R.~Xie \emph{et~al.}, ``Mixture of experts-enabled parallel
  scheduling and processing for vehicular generative {AI} services,''
  \emph{IEEE Trans. Cogn. Commun. Netw.}, pp. 1--1, 2025.

\bibitem{Li2025JointCommunication}
Y.~Li, Z.~Yi, D.~Guo \emph{et~al.}, ``Joint communication and offloading
  strategy of {CoMP} {UAV}-assisted {MEC} networks,'' \emph{IEEE Internet
  Things J.}, pp. 1--1, 2025.

\bibitem{21}
Z.~Lyu, Y.~Li, G.~Zhu \emph{et~al.}, ``Rethinking resource management in edge
  learning: {A} joint pre-training and fine-tuning design paradigm,''
  \emph{IEEE Trans. Wireless Commun.}, vol.~24, no.~2, pp. 1584--1601, 2025.

\bibitem{22}
H.~Wu, X.~Chen, and K.~Huang, ``Resource management for low-latency cooperative
  fine-tuning of foundation models at the network edge,'' \emph{IEEE Trans.
  Wireless Commun.}, pp. 1--1, 2025.

\bibitem{23}
E.~Hu, Y.~Shen, P.~Wallis \emph{et~al.}, ``Lo{RA}: {L}ow-rank adaptation of
  large language models,'' in \emph{Proc. Int. Conf. Learn. Represent. (ICLR)},
  2022.

\bibitem{24}
G.~Qu, Z.~Lin, F.~Liu \emph{et~al.}, ``Trimcaching: {P}arameter-sharing {AI}
  model caching in wireless edge networks,'' in \emph{IEEE International
  Conference on Distributed Computing Systems (ICDCS)}, 2024, pp. 36--46.

\bibitem{25}
H.~Wu, Q.~Zeng, and K.~Huang, ``Efficient multiuser {AI} downloading via
  reusable knowledge broadcasting,'' \emph{IEEE Trans. Wireless Commun.},
  vol.~23, no.~8, pp. 10\,459--10\,472, 2024.

\bibitem{Wang2025ModelOwnership}
R.~Wang, J.~Liang, C.~Feng \emph{et~al.}, ``Model ownership protection for
  healthcare consumer electronics in federated edge learning,'' \emph{IEEE
  Trans. Consumer Electron.}, vol.~71, no.~2, pp. 4616--4627, 2025.

\bibitem{Geng2025Layer}
H.~Geng, Y.~Li, S.~Wang \emph{et~al.}, ``Layer redundancy aware {DNN} model
  repository planning for fast model download in edge cloud,'' \emph{IEEE
  Trans. Cloud Comput.}, vol.~13, no.~3, pp. 1038--1049, 2025.

\bibitem{Gong2025Digital}
X.~Gong, X.~Liu, A.-A. Lu \emph{et~al.}, ``Digital twin of channel: {D}iffusion
  model for sensing-assisted statistical channel state information
  generation,'' \emph{IEEE Trans. Wireless Commun.}, vol.~24, no.~5, pp.
  3805--3821, 2025.

\bibitem{26}
P.~Qin, Y.~Fu, K.~Wu \emph{et~al.}, ``Packet routing and energy cooperation for
  {RTU} satellite-terrestrial multi-hop network in remote cyber-physical power
  system,'' \emph{IEEE Trans. Netw. Sci. Eng.}, vol.~11, no.~4, pp. 3585--3597,
  2024.

\bibitem{27}
P.~Qin, Y.~Fu, R.~Ding \emph{et~al.}, ``Competition-awareness partial task
  offloading and {UAV} deployment for multitier parallel computational internet
  of vehicles,'' \emph{IEEE Syst. J.}, vol.~18, no.~3, pp. 1753--1764, 2024.

\bibitem{Fan2025MADRL}
W.~Fan, P.~Chen, X.~Chun \emph{et~al.}, ``{MADRL}-based model partitioning,
  aggregation control, and resource allocation for cloud-edge-device
  collaborative split federated learning,'' \emph{IEEE Trans. Mobile Comput.},
  vol.~24, no.~6, pp. 5324--5341, 2025.

\bibitem{28}
T.~T. Nguyen, N.~D. Nguyen, and S.~Nahavandi, ``Deep reinforcement learning for
  multiagent systems: {A} review of challenges, solutions, and applications,''
  \emph{IEEE Trans. Cybern.}, vol.~50, no.~9, pp. 3826--3839, 2020.

\bibitem{29}
\BIBentryALTinterwordspacing
W.~Wang, T.~Yang, Y.~Liu \emph{et~al.}, ``Action semantics network:
  {C}onsidering the effects of actions in multiagent systems,'' 2020. [Online].
  Available: \url{https://arxiv.org/abs/1907.11461}
\BIBentrySTDinterwordspacing

\bibitem{30}
H.-H. Chang, N.~Mohammadi, R.~Safavinejad \emph{et~al.}, ``Dyna-{ESN}:
  {E}fficient deep reinforcement learning for partially observable dynamic
  spectrum access,'' \emph{IEEE Trans. Wireless Commun.}, pp. 1--1, 2024.

\bibitem{31}
\BIBentryALTinterwordspacing
T.~Rashid, M.~Samvelyan, C.~S. de~Witt \emph{et~al.}, ``{QMIX}: {M}onotonic
  value function factorisation for deep multi-agent reinforcement learning,''
  2018. [Online]. Available: \url{https://arxiv.org/abs/1803.11485}
\BIBentrySTDinterwordspacing

\bibitem{32}
Z.~Wei, X.~Yu, D.~W.~K. Ng \emph{et~al.}, ``Resource allocation for
  simultaneous wireless information and power transfer systems: {A} tutorial
  overview,'' \emph{Proc. IEEE}, vol. 110, no.~1, pp. 127--149, 2022.

\bibitem{33}
K.~Yang, Y.~Shi, W.~Yu \emph{et~al.}, ``Energy-efficient processing and robust
  wireless cooperative transmission for edge inference,'' \emph{IEEE Internet
  Things J.}, vol.~7, no.~10, pp. 9456--9470, 2020.

\bibitem{34}
Z.~Yang, Y.~Xie, and Z.~Wang, ``A theoretical analysis of deep {Q}-learning,''
  in \emph{Proceedings of the 2nd Conference on Learning for Dynamics and
  Control (L4DC)}, 2020, pp. 486--489.

\bibitem{34-1}
Y.~Fu, P.~Qin, G.~Tang \emph{et~al.}, ``Joint design of sensing, communication,
  and computation for multi-{UAV}-enabled over-the-air federated learning,''
  \emph{IEEE Trans. Veh. Technol.}, pp. 1--17, 2025.

\bibitem{35}
Z.~Zhang, Y.~Zhao, H.~Li \emph{et~al.}, ``{DVFO}: {L}earning-based {DVFS} for
  energy-efficient edge-cloud collaborative inference,'' \emph{IEEE Trans.
  Mobile Comput.}, vol.~23, no.~10, pp. 9042--9059, 2024.

\bibitem{zheng2024llamafactory}
\BIBentryALTinterwordspacing
Y.~Zheng, R.~Zhang, J.~Zhang \emph{et~al.}, ``{LlamaFactory}: {U}nified
  efficient fine-tuning of 100+ language models,'' in \emph{Proceedings of the
  62nd Annual Meeting of the Association for Computational Linguistics (Volume
  3: System Demonstrations)}.\hskip 1em plus 0.5em minus 0.4em\relax Bangkok,
  Thailand: Association for Computational Linguistics, 2024. [Online].
  Available: \url{http://arxiv.org/abs/2403.13372}
\BIBentrySTDinterwordspacing

\bibitem{zhou2023lima}
C.~Zhou, P.~Liu, P.~Xu \emph{et~al.}, ``{LIMA}: {L}ess is more for alignment,''
  \emph{NeurIPS}, vol.~36, 2024.

\bibitem{databricks2023dolly15k}
{Databricks, Inc.}, ``{databricks-dolly-15k}: An open instruction-following
  dataset,''
  \url{https://huggingface.co/datasets/databricks/databricks-dolly-15k}, 2023.

\end{thebibliography}
   
\end{document}